\documentclass{aa}

\usepackage{graphicx}

\usepackage{txfonts}
\usepackage[utf8]{inputenc}
\usepackage[table]{xcolor}
\usepackage{array}
\usepackage{rotating}
\usepackage{float}
\usepackage{subfigure}
\usepackage{multirow}
\usepackage[]{natbib}
\usepackage{epstopdf}
\usepackage[colorlinks=true,linkcolor=blue,citecolor=blue]{hyperref}
\newcommand\fref{$F_{\rm{r}}$ }

\newcommand\fscience{$F_{\rm{s}}$ }

\newcommand\iref{$I_{\rm{r}}$ }
\newcommand\irefnospace{$I_{\rm{r}}$}
\newcommand\iscience{$I_{\rm{s}}$ }
\newcommand\isciencenospace{$I_{\rm{s}}$}
\newcommand\fluxd{$F_{\rm{3b}}$ }

\newcommand\fad{$F_{\rm{13b}}$ }
\newcommand\fadnospace{$F_{\rm{13b}}$}

\newcommand\iab{$I_{\rm{12}}$ }

\newcommand\iacnospace{$I_{\rm{13a}}$}
\newcommand\iad{$I_{\rm{13b}}$ }
\newcommand\iadnospace{$I_{\rm{13b}}$}
\newcommand\ibc{$I_{\rm{23a}}$ }
\newcommand\ibcnospace{$I_{\rm{23a}}$}

\newcommand\ibdnospace{$I_{\rm{23b}}$}
\newcommand\mjnospace{$M_{\rm{J}}$}
\newcommand\mj{$M_{\rm{J}}$ }
\newcommand\teffnospace{$T_{\rm{eff}}$}
\newcommand\teff{$T_{\rm{eff}}$ }
\newcommand{\fluxratio}{$F_1/F_{\rm{3b}}$ }
\newcommand{\fluxrationospace}{$F_1/F_{\rm{3b}}$}
\def\degb{^{\circ}}

\begin{document}

\title{Search for cool giant exoplanets around young and nearby stars}
\subtitle{VLT/NaCo near-infrared phase-coronagraphic and differential imaging
\thanks{Based on observations collected at the European Southern
Observatory, Chile, ESO programs 085.C-0257A, 086.C-0164A, and 088.C-0893A.}}

\author{A.-L. Maire\inst{1,2}, A. Boccaletti\inst{3}, J. Rameau\inst{4}, G. Chauvin\inst{4}, A.-M. Lagrange\inst{4}, M. Bonnefoy\inst{5}, S. Desidera\inst{2}, M. Sylvestre\inst{3}, P. Baudoz\inst{3}, R. Galicher\inst{3}, and D. Mouillet\inst{4}}

   \institute{LUTH, Observatoire de Paris, CNRS and University Denis Diderot Paris 7, 5 place Jules Janssen, 92195 Meudon, France
   		  \and
          INAF - Osservatorio Astronomico di Padova, Vicolo dell'Osservatorio 5, 35122 Padova, Italy\\
          \email{annelise.maire@oapd.inaf.it}
          \and
          LESIA, Observatoire de Paris, CNRS, University Pierre et Marie Curie Paris 6 and University Denis Diderot Paris 7, 5 place Jules Janssen, 92195 Meudon, France 
          \and
          IPAG, Universit\'e Joseph Fourier, CNRS, BP 53, 38041 Grenoble, France
          \and
          Max-Planck-Institut f\"ur Astronomie, K\"onigstuhl 17, 69117 Heidelberg, Germany
             }
\date{Received 19 November 2013 / Accepted 12 April 2014}
 
  \abstract
   {Spectral differential imaging (SDI) is part of the observing strategy of current and future high-contrast imaging instruments. It aims to reduce the stellar speckles {that prevents} the detection of cool planets {by} using in/out methane-band images. It attenuates the signature of off-axis companions to the star, {such as} angular differential imaging (ADI). However, this attenuation {depends} on the spectral properties of the low-mass companions we are searching for. {The implications of this particularity on {estimating} the detection limits have been poorly explored so far.}}
   {{We perform an imaging survey to search for cool (T$_{\rm{eff}}$\,$<$\,1\,000--1\,300~K) giant planets at separations as close as 5--10~AU. We also aim to assess {the} sensitivity limits in SDI data taking {the photometric bias into account. This will lead to} a better view of the SDI performance.}}
   {{We {observed} a selected sample of 16 stars (age $<$ 200~Myr, distance $<$ 25~pc) with the phase-mask coronagraph, SDI, and ADI modes of VLT/NaCo.}}
   {We do not detect any companions. As for the estimation of the sensitivity limits, we argue that the SDI residual noise cannot be converted into mass limits because it represents a differential flux, {unlike} what is done for single-band images, in which fluxes are measured. This results in degeneracies for the mass limits, which may be {removed} with the use of single-band constraints. We {instead emply} a method {of directly determining} the mass limits and compare the results from a combined processing SDI-ADI (ASDI) and ADI. The SDI flux ratio of a planet is the critical parameter for the ASDI performance at close-in separations ($\lesssim$1$''$). {The survey is sensitive to cool giant planets beyond 10~AU {for 65\% and 30~AU for} 100\% of the sample.}}
   {{For close-in separations, the optimal regime for SDI corresponds to SDI flux ratios higher than $\sim$2. According to the BT-Settl model, this translates into T$_{\rm{eff}}$\,$\lesssim$\,800~K, which is significantly lower than the methane condensation temperature ($\sim$1300~K).} The methods described here can be applied to the data interpretation of SPHERE. In particular, we expect better performance with the dual-band imager IRDIS, thanks to more suitable filter characteristics and better image quality.}


\authorrunning{A.-L. Maire et al.}
\titlerunning{Imaging search for cool giant exoplanets combining coronagraphy and differential imaging}

\maketitle
% ------------------------------------------------------------------------------------
\section{Introduction} \label{sec:intro}
% ------------------------------------------------------------------------------------
The search for exoplanets by direct imaging is challenged by very large brightness ratios between stars and planets at short angular separations. Current facilities on large ground-based telescopes or in space allow {adequate contrasts to be reached}, and have revealed a few planetary-mass objects \citep{Marois2008c, Marois2010b, Lagrange2009, Rameau2013b, Kuzuhara2013} either massive ($>$3 Jupiter masses or \mjnospace) and young ($<$100--200~Myr) or with large angular separations ($>$1$''$). Even though some of them are questioned \citep{Kalas2008}, these objects {very} likely represent the top of the giant planet population at long periods. {They are therefore} very important {for understanding the planet's} formation mechanisms. These discoveries have been favored by {longstanding} instrumental developments {such as} adaptive optics (AO) and coronagraphy, but also {by} dedicated observing strategies and post-processing methods like differential imaging.

{The} purpose of differential imaging is to attenuate the {stellar speckles which prevent} the detection of faint planets around the star. More precisely, a reference image of the star is built and subtracted {from} the science images. Several kinds of differential imaging have been proposed in the {past} decade. Angular differential imaging {(ADI)} takes advantage of the field rotation occurring in an alt-az telescope \citep{Marois2006a}. Spectral differential imaging {(SDI)} exploits the natural wavelength dependence of a star image \citep{Racine1999}. An extension of this technique consists in using the spectral information in many spectral channels, provided for instance by an integral field spectrometer {\citep{Sparks2002, Thatte2007, Crepp2011, Pueyo2012}}. Polarimetric differential imaging uses differences between the polarimetric fluxes of the star and the planet {or} disk \citep{Kuhn2001}{, but has not permitted any planet detections so far}. Introducing a difference between star and planet properties allows {differentiating} the {unwanted stellar speckles from} the much fainter planet signals. Still, achieving high performance with these methods also {requires good} knowledge of the instrument behavior and biases.

Large direct imaging surveys have tentatively constrained the frequency of young giant planets at long periods ($\gtrsim$10--20~AU) to $\sim$10--20\% \citep[{e.g.},][]{Lafreniere2007b, Chauvin2010, Vigan2012, Rameau2013a, Wahhaj2013a, Nielsen2013, Biller2013}. {Typical contrasts} of 10 to 15 magnitudes {have been obtained,} but for separations beyond $\sim$1$''$. Consequently, the occurrence of young giant planets down to a few Jupiter masses was mostly investigated at physical separations from $\sim$10~AU to hundreds of AU, $\beta$~Pictoris~b {being the object detected with the closest separation} \citep[8~AU,][]{Lagrange2010b}. To analyze closer-in, colder, and {less-massive} giant planets, we need to push {the contrast performance further}. For this very purpose, several new-generation imaging instruments {are now ready to start operation{, such as} SPHERE \citep[Spectro-Polarimetric High-contrast Exoplanet REsearch,][]{Beuzit2008} and GPI \citep[Gemini Planet Imager,][]{Macintosh2008}. They} were built to take advantage of several high-contrast imaging techniques{, namely} extreme AO, advanced coronagraphy, {ADI, and SDI.}

The evolutionary models extrapolated from stellar mechanisms \citep{Burrows1997, Chabrier2000} predict that {Jovian} planets are very hot when formed; they cool over time and can be relatively bright at young ages\footnote{We note that ``cold-start'' models \citep{Marley2007, Fortney2008, Spiegel2012, Mordasini2012} also predict bright planets at young ages but fainter than those in ``hot-start'' models.}. {SDI \citep{Racine1999}} is intended to take advantage of the presence of a methane absorption band at $\sim$1.6~$\muup$m in the spectra of cool ($\lesssim$1\,300~K) giant planets \citep{Burrows1997, Chabrier2000}, while the star is not expected to contain this chemical element. Thus, this spectral feature provides an efficient tool {for disentangling} stellar speckles from planet signal(s). SDI offers the potential to reach the detection of planets with lower masses than those already discovered by direct imaging. However, spectroscopic observations of a few young giant planets have {only} shown weak absorption by methane in the H band, which could be explained by the low surface gravity of these objects \citep{Barman2011a, Barman2011b, Oppenheimer2013, Konopacky2013}.

In practice, SDI also produces a significant attenuation of the planet itself, because the latter is present in the reference image used for the speckle subtraction. {This attenuation has} to be quantified to derive its photometry accurately. To date, only two independent surveys {have been made} using SDI with the Very Large Telescope (VLT){, as well as with} the Multiple Mirror Telescope (MMT) \citep{Biller2007} {and} the Gemini South telescope \citep[][]{Biller2013, Nielsen2013, Wahhaj2013a}, but {they} have not reported {any detections of planetary-mass objects yet}. The non-detection results were exploited to assess the frequency of giant planets at long periods. \citet{Nielsen2008} did not consider the biases introduced by SDI for their analysis, {unlike} \citet{Biller2013}, \citet{Nielsen2013}, and \citet{Wahhaj2013a} for the statistical analysis of the NICI Campaign.

{The} observing strategy of the NICI Campaign is based on the complementarity of two observing modes in order to optimize the survey sensitivity: ADI \citep{Marois2006a} {and} the combination of SDI and ADI (ASDI). These two observing modes are not performed simultaneously on the same target, because the spectral filters used are different (large-band and narrow-band{, respectively}). The ADI and ASDI contrast curves presented in \citet{Biller2013}, \citet{Nielsen2013}, and \citet{Wahhaj2013a} are corrected from the attenuation and the artifacts produced by the reduction pipeline except for the SDI part (for the ASDI curves), because the attenuation {depends} on the spectral properties of a planet. Nevertheless, this point is taken into account for the {planet frequency study}. Both ADI and ASDI contrast curves are considered in this analysis, but only the best detection limit is finally used. The results are essentially consistent with the previous surveys.

{We} note that \citet{Biller2013} and \citet{Nielsen2013} do not report {any} ASDI mass detection limits for individual targets, because of the particularities of the SDI attenuation. \citet{Wahhaj2013a} present individual mass detection limits combining ADI and ASDI, using the contrast-mass conversion based on evolutionary models. We argue in this work that this method is inadequate for interpreting {the} dual-band imaging data analyzed with SDI-based algorithms. Our arguments are also relevant to IFS data processed with similar techniques \citep{Sparks2002, Crepp2011, Pueyo2012}.

In this paper, we present the outcome of a small survey of 16 stars performed with NaCo (Nasmyth Adaptive Optics System and Near-Infrared Imager and Spectrograph), the near-IR AO-assisted camera of the VLT \citep{Rousset2003, Lenzen2003}. Our prime objective is to observe a selected sample of young ($\lesssim$200~Myr) and nearby ($\lesssim$25~pc) stars to search for massive but cool gas giant planets at separations as small as 5--10~AU. For this purpose, we combine state-of-the-art high-contrast imaging techniques similar to those implemented in SPHERE \citep{Beuzit2008}. Our second objective is to address the {problem} of assessing detection limits in SDI data, which is important when it falls to very short angular {separations} ($<$0.5--1$''$), and to determine the condition(s) for which SDI gives the optimal performance. This last topic has not been addressed so far in the literature. {Unlike} the NICI Campaign, we {carried} out the observations of the survey with only one observing mode, ASDI. We consider ASDI and ADI for the reduction and analysis of the same data set, thus {allowing comparison} of the performances of these differential imaging techniques. This paper is designed to focus on the astrophysical exploitation of the survey, based on a simple, straightforward, and robust method {of accounting} for the photometric bias induced by SDI. A subsequent paper will analyze {the details of} the biases of SDI data reduction and {will correctly estimate} the detection performance (Rameau et al., in prep.). The methods and results presented in these papers may serve as a basis for {interpreting} future large surveys to be performed with SPHERE and GPI.

We describe the sample selection in Sect.~\ref{sec:targetselection}{, then explain} the observing strategy, the data acquisition, and the reduction pipeline {in Sect.~\ref{sec:obs}}. In Sect.~\ref{sec:sdianalysis}, we explain the {problem} of assessing detection limits in SDI, which requires {a different analysis} from the method usually considered for direct imaging surveys. {In this section, we} also introduce the method we {used for interpreting} our survey. We present {ADI and ASDI detection limits} and carry out a detailed study of the SDI performance in Sect.~\ref{sec:results}. Finally, we discuss the {broad trends} of the survey in Sect.~\ref{sec:discussion}.

% ------------------------------------------------------------------------------------
\section{Target sample}
\label{sec:targetselection}
% ------------------------------------------------------------------------------------

\begin{figure}[t]
	\centering
	\includegraphics[trim = 19mm 5mm 3mm 9mm, clip, width=.4\textwidth]{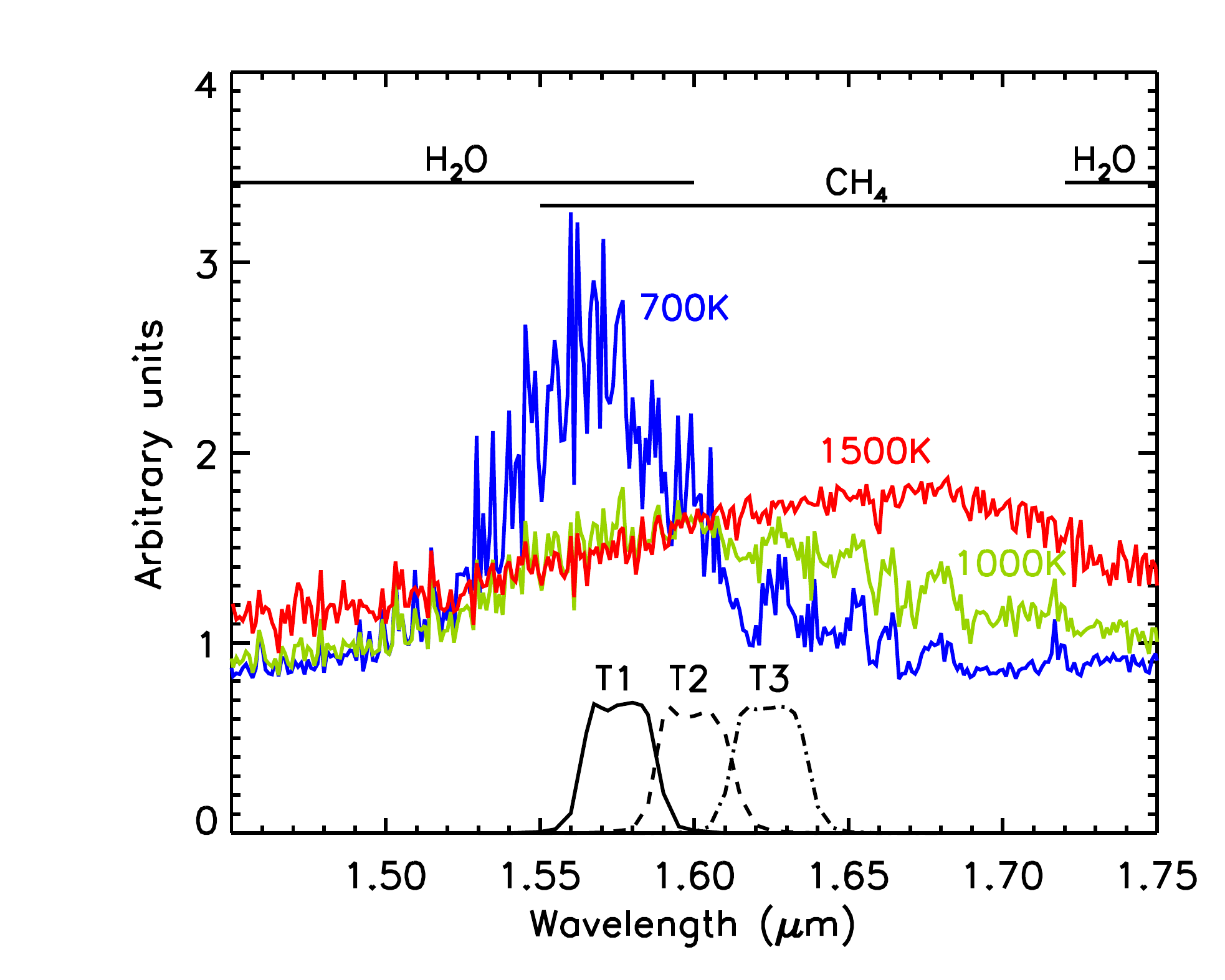}
	\caption{Spectra {($F_{\lambda}$)} of model atmospheres of giant planets for different effective temperatures \citep[colored solid lines, BT-Settl models from][]{Allard2012}. Each spectrum is normalized to its value at 1.6~$\muup$m and is vertically shifted by a constant. The transmission of the three SDI filters of NaCo ($T_1$, $T_2$, and $T_3$) are shown in black lines with different styles. Theoretical absorption bands of water and methane are also indicated. {The vertical scale is linear.}}
	\label{fig:spectracoldegp}
\end{figure}

\begin{figure}[t]
	\centering
	\includegraphics[trim = 19mm 5mm 3mm 9mm, clip, width=.4\textwidth]{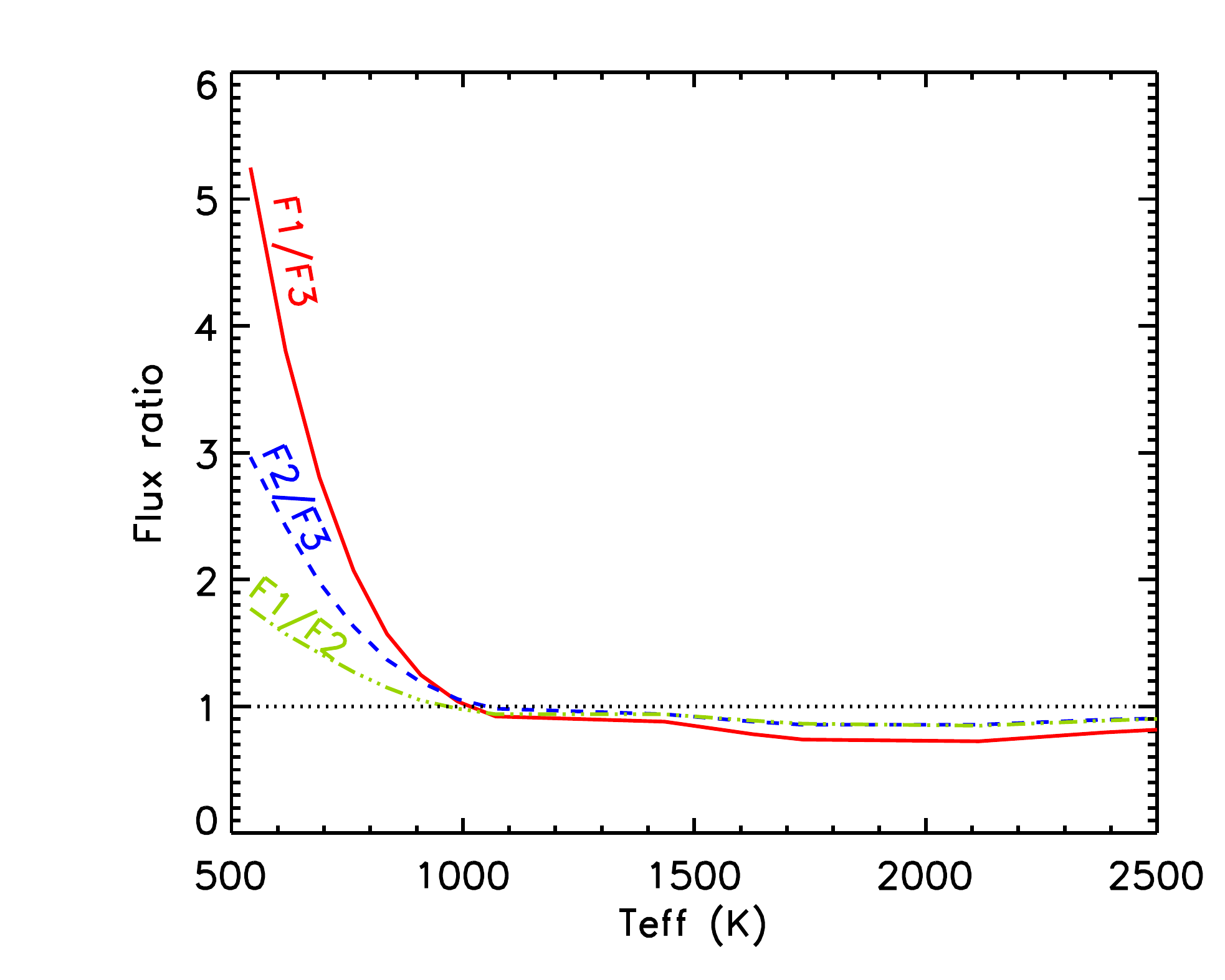}
	\caption{Flux ratio as a function of the effective temperature derived for the NaCo SDI filters (Fig.~\ref{fig:spectracoldegp}). $F_1$, $F_2$, and $F_3$ refer to the fluxes in the filters at 1.575, 1.600, and 1.625~$\muup$m, respectively. {The theoretical relations are derived for a given age of 70~Myr, so the surface gravity of the object {log(g)} is not constant. It increases from 3.75 to 4.75 for the plot range. We also consider the same photometric {zeropoint} for all the SDI filters.}}
	\label{fig:rfluxteff}
\end{figure}

The survey presented in this paper has been optimized to search for cool ($<$1300~K, the condensation temperature of methane) giant planets around the closest young stars. Our approach is to use NaCo with an observing strategy similar to one of the main observing modes foreseen in SPHERE, namely dual-band imaging coupled to coronagraphy and angular differential imaging, to improve the detection performance at small angular separations in the 0.2--0.5$''$ region (i.e., 5--12~AU for a star at 25~pc). We observed a defined sample of stars optimized in age and distance so as to explore the closest physical separations in the stellar environment and to fully exploit the NaCo differential imaging capabilities for {detecting} planets with cool atmospheres (i.e., with methane features at $\sim$1.6~$\muup$m). Figure~\ref{fig:spectracoldegp} shows theoretical spectra of giant planets for effective temperatures of 1\,500, 1\,000, and 700~K, as well as the transmission of the SDI filters of NaCo. We note three distinct regimes for the differential fluxes between the SDI filters (see also Fig.~\ref{fig:rfluxteff}). For effective temperatures {over} $\sim$1\,500~K, there is no methane absorption and the spectrum shows a positive slope ($F_3$\,$>$\,$F_2$\,$>$\,$F_1$, with $F_1$, $F_2$, and $F_3$ the fluxes in the filters at 1.575, 1.600, and 1.625~$\muup$m, respectively). When effective temperatures range from $\sim$1\,500 down to 1\,000~K, methane begins to condense in the atmosphere and to {partially absorb} the emergent flux longwards 1.55~$\muup$m. The fluxes in the SDI filters are nearly identical. This regime is the worst case for SDI as the self-subtraction results in little to no flux left in the final image for small separations. Finally, for temperatures lower than $\sim$1\,000~K, methane absorbs strongly the emergent flux for wavelengths beyond 1.55~$\muup$m and the SDI flux ratios are the largest. This regime is the optimal case for SDI.

Based on a complete compilation of young and nearby stars recently identified in young co-moving groups and from systematic spectroscopic surveys, a {subsample} of stars, mostly AFGK spectral types, {was} selected according to their declination ($\delta$~$\lesssim$~25$\degb$), age ($\lesssim$200~Myr), distance ($d$~$\lesssim$~25~pc), and R-band brightness ($R$~$\lesssim$~9.5) to ensure deep detection performances. The age cut-off ensures that the cool companions detected will have masses within the planetary mass regime. Most targets are members of the nearest young stellar associations, including the AB~Doradus (AB~Dor) and Hercules-Lyra (Her-Lyr) groups \citep{Lopez-santiago2006}. The distance cut-off ensures that {(i)} the probed projected separations are $>$5--10~AU, and {(ii)} these targets are the most favorable for {detecting} cool companions, necessary to fully exploit SDI. The targets are brighter than $H=6.5$ to ensure good sensitivity with the narrow ($\Delta \lambda$\,=\,0.025~$\muup$m, $\Delta \lambda/\lambda$\,$\sim$\,1.6\%) SDI filters of NaCo, while they are bright enough in the visible to allow good AO efficiency.

\begin{figure}[t]
	\centering
	\includegraphics[trim = 5mm 4mm 7mm 8mm, clip, width=.4\textwidth]{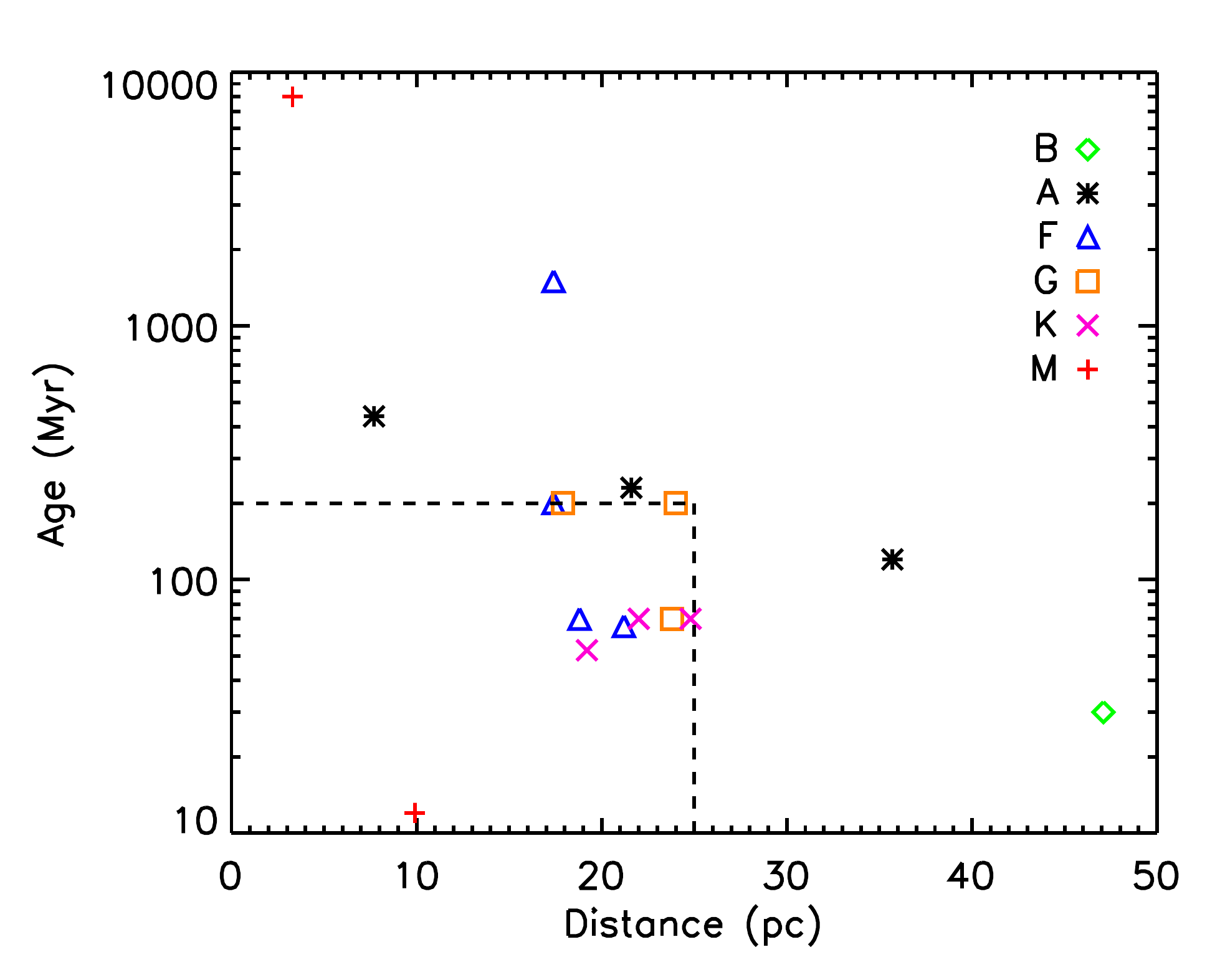}
	\caption{Age-distance diagram of the star sample observed in the NaCo SDI survey. The dashed lines indicate the criteria used for the sample selection (age $\lesssim$ 200~Myr and distance $\lesssim$ 25~pc).}
	\label{fig:sample}
\end{figure}

The properties of the 16 observed targets are summarized in Table~\ref{tab:sample} and Fig.~\ref{fig:sample}. The spectral types and the distances are taken from the SIMBAD database\footnote{\url{http://simbad.u-strasbg.fr/}.}, and the H-band magnitude from the 2MASS catalog \citep{Cutri2003}. The ages are derived from individual sources listed in Table~\ref{tab:sample}. The targets that do not comply with the selection criteria given above are either backup targets or targets for which the age was underestimated at the time of the observing proposal.

\begin{table*}[!ht]
 \caption{Observational and physical properties of the observed targets.}
 \label{tab:sample}
 \begin{center}
 \begin{tabular}{l c c c c c c c c}
 \hline\hline
 Name & $\alpha$ & $\delta $& SpT & $d$ & Moving & Age & Ref. & $H$ \\
 & [J2000] & [J2000] & & (pc) & group & (Myr) & & (mag) \\
 \hline
 HIP~98470 & 20h00m20s & $-$33$^{\circ}$42$'$12$''$ & F7V & 21.2 &  & 70\tablefootmark{a} & 1 & 4.64 \\
 HIP~107350 & 21h44m31s & $+$14$^{\circ}$46$'$19$''$ & G0V & 17.9 & Her-Lyr & 200 & 2,11 & 4.60 \\
 HIP~118008 & 23h56m11s & $-$39$^{\circ}$03$'$08$''$ & K2V & 22.0 & AB Dor & 70 & 3,9,11 & 6.00 \\
 HD~10647 & 01h42m32s & $-$53$^{\circ}$44$'$27$''$ & F9V & 17.4 & & 1\,500 & 11 & 4.40 \\
 HIP~76829 & 15h41m11s & $-$44$^{\circ}$39$'$40$''$ & F5V & 17.4 & Her-Lyr? & 200 & 1,2 & 3.73 \\
 HIP~102409 & 20h45m10s & $-$31$^{\circ}$20$'$27$''$ & M1V & 9.9 & $\beta$ Pic & 12 & 3,9 & 4.83 \\
 HIP~106231 & 21h31m02s & $+$23$^{\circ}$20$'$07$''$ & K5-7V & 24.8 & AB Dor & 70 & 3,9,11 & 6.52 \\
 HIP~114046 & 23h05m52s & $-$35$^{\circ}$51$'$11$''$ & M2V & 3.3 & & 8\,000 & 1 & 3.61 \\
 HIP~7576 & 01h37m35s & $-$06$^{\circ}$45$'$38$''$ & G5V & 24.0 & Her-Lyr & 200 & 2,11 & 5.90 \\
 HIP~14555 & 03h07m56s & $-$28$^{\circ}$13$'$11$''$ & K8V & 19.2 & & 50\tablefootmark{b} & 8 & 6.58 \\
 HIP~10602 & 02h16m31s & $-$51$^{\circ}$30$'$44$''$ & B8IV-V & 47.1 & Tuc-Hor & 30 & 1,3,9 & 3.95 \\
 HIP~18859 & 04h02m37s & $-$00$^{\circ}$16$'$08$''$ & F6V & 18.8 & AB Dor & 70 & 3,9,11 & 4.34 \\
 HD~31295 & 04h54m54s & $+$10$^{\circ}$09$'$03$''$ & A0V & 35.7 & & 120 & 5 & 4.52 \\
 HD~38678 & 05h46m57s & $-$14$^{\circ}$49$'$19$''$ & A2IV-V & 21.6 & Cas? & 230 & 6,7,10 & 3.31 \\
 HIP~30314 & 06h22m31s & $-$60$^{\circ}$13$'$07$''$ & G1V & 23.8 & AB Dor & 70 & 3,9,11 & 5.16 \\
 Fomalhaut & 22h57m39s & $-$29$^{\circ}$37$'$20$''$ & A4V & 7.7 & Cas? & 440 & 4 & 0.94 \\
 \hline
 \end{tabular}
 \end{center}
 \tablefoot{{Columns give} the name, right ascension, declination, spectral type, distance, co-moving group, mean estimated age, and magnitude in H band. The {comoving} groups indicated in the table are Hercules-Lyra (Her-Lyr), AB~Doradus (AB~Dor), $\beta$~Pictoris ($\beta$~Pic), Tucana-Horologium (Tuc-Hor), and Castor (Cas).} \\
 \tablefoottext{a}{{We adopt a prior age estimation with respect to the work of Desidera et al.{(submitted)}. Our age estimate is slightly younger {than} the mean value derived by the latter, but within the range of plausible values.}}
 \tablefoottext{b}{HIP~14555 is a nearby and very active late K dwarf for which the small amount of lithium in the spectrum \citep{Torres2006} indicates an age of $\sim$50~Myr. It is also reported to be a double-lined spectroscopic binary without further details \citep{Gizis2002}. Therefore, as discussed in \citet{Makarov2008}, we {cannot} exclude that it is a tidally locked binary with significantly older age. }
 \tablebib{(1)~Desidera et al., submitted; (2)~\citet{Lopez-santiago2006}; (3)~\citet{Malo2013}; (4)~\citet{Mamajek2012}; (5)~\citet{Rhee2007}; (6)~\citet{Su2001}; (7)~\citet{Su2006}; (8)~\citet{Torres2006}; (9)~\citet{Torres2008}; (10)~\citet{Vican2012}; (11)~Vigan et al., in prep..}
 \end{table*}

% ------------------------------------------------------------------------------------
\section{Observations and data reduction}
\label{sec:obs}
% ------------------------------------------------------------------------------------
\subsection{Observing strategy}
\label{sec:strategy}

NaCo offers several high-contrast imaging modes, and the objective of our program was to take advantage of those that are relevant to test the interpretation of the SPHERE data. We combined the four-quadrant phase mask \citep[FQPM,][]{Rouan2000}, the pupil-tracking mode {that} allows {ADI observations}, and the SDI mode, which is based on the concept of the TRIDENT instrument on the Canada-France-Hawaii Telescope \citep{Marois2003b}. In this section, we refer to the combination of these techniques as ASDI-4, or {as} ASDI when the FQPM is not used {(for reasons related to the observing conditions, see Sect.~\ref{sec:acquisition}). In the latter case, the images were saturated to compensate {for} the loss of dynamic at the cost of a higher photon noise in the inner part of the PSF.} Phase masks were installed in NaCo as soon as 2003 \citep{Boccaletti2004} and have produced astrophysical results \citep{Gratadour2005, Riaud2006, Boccaletti2009a, Boccaletti2012b}.

{Although} the star attenuation provided by the FQPM is chromatic, {\citet{Boccaletti2004} show that the contrast achieved for a given spectral band is not limited by chromaticity effects {induced by the low} spectral resolution and/or {by} small differential aberrations, because the coherent energies measured by the wavefront sensor are modest ($\lesssim$50\% at 2.17 microns). This will also be the case in H band, for which the AO correction is worse.} The SDI mode is based on a double Wollaston prism \citep{Lenzen2004, Close2005}, which produces four {subimages} on the detector in front of which is set a custom assembly of three narrow-band filters (Fig.~\ref{fig:sdiimages}, left). The central wavelengths of these filters are $\lambda_1$\,=\,1.575~$\muup$m, $\lambda_2$\,=\,1.600~$\muup$m, and $\lambda_{\rm{3a}}$\,=\,$\lambda_{\rm{3b}}$\,=\,1.625~$\muup$m\footnote{{The beams corresponding to the $I_{\rm{3a}}$ and $I_{\rm{3b}}$ images are formed by the double Wollaston prism but pass through the same SDI filter.}}. The platescale of the SDI camera is $\sim$17~mas/pix. SDI was upgraded in 2007 to provide a larger field of view (8$''\times$\,8$''$, limited by a field mask to avoid contamination between the {subimages}) and a lower chromatic dispersion of each point spread function (PSF). The differential aberrations between the four images {were measured to be lower than 10~nm rms per mode for the first Zernike modes using phase diversity \citep{Lenzen2004}}. The combination of the FQPM and SDI modes was commissioned by some of us using AB~Dor {($H$\,=\,4.845)} as a test bench \citep{Boccaletti2008a}. {The seeing conditions estimated by the Differential Image Motion Monitor (DIMM) were good (0.78\,$\pm$\,0.12$''$, $\lambda$\,=\,0.5~$\muup$m), as {was} the coherent energy measured by the visible (0.45--1~$\muup$m) AO wavefront sensor (52\,$\pm$\,4\%, $\lambda$\,=\,2.17~$\muup$m).} We measured a noise level, {expressed as the contrast to the star, of} $10^{-4}$ at only 0.2$''$ after applying SDI and ADI on ten images covering 50$^{\circ}$ of parallactic rotation (exposure time of 936~s for each image). However, this noise level does not {consider} the self-subtraction of off-axis {objects} occurring in SDI. Consequently, it cannot be used to derive mass constraints on detectable companions.

 \begin{table*}
 \caption{Log of the observed targets.}
 \label{tab:obs}
 \begin{center}
 \begin{tabular}{l c c c c c c c c c c c c}
 \hline\hline
 Name & Observing & Mode & $\Delta$PA & DIT & NDIT & $N_{\rm{exp}}$ & $\epsilon$ & $\tau_0$ & CE & Image & \# frames & Exp. \\
 & date & & ($^{\circ}$) & (s) & & & ($''$) & (ms) & (\%) & bin & mastercube & time (s) \\
 \hline
 HIP~98470 & 2010.08.20 & ASDI-4 & 102 & 8 & 8/2 & 24+39 & 1.03 & 8.3 & 9.5 & 2 & 118 & 1\,888 \\
 HIP~107350 & 2010.08.20 & ASDI-4 & 35 & 8 & 4 & 81 & 0.70 & 7.8 & 44.3 & 2 & 153 & 2\,448 \\
 HIP~118008 & 2010.08.20 & ASDI & 72 & 20 & 2/11 & 5+13 & 1.05 & 7.2 & 37.8 & 1 & 144 & 2\,880 \\
 HD~10647 & 2010.08.20 & ASDI-4 & 48 & 5 & 12/6 & 8+66 & 0.92 & 6.7 & 29.8 & 3 & 151 & 2\,265 \\
 HIP~76829 & 2010.08.21 & ASDI-4 & 15 & 3 & 10 & 41 & 0.83 & 5.5 & 44.3 & 5 & 73 & 1\,095 \\
 HIP~102409 & 2010.08.21 & ASDI-4 & 26 & 10 & 3 & 78 & 1.30 & 3.6 & 44.0 & 2 & 77 & 1\,540 \\
 HIP~106231 & 2010.08.21 & ASDI-4 & 10 & 30 & 1 & 31 & 1.83 & 2.3 & 2.9 & 1 & 30 & 900 \\
 HIP~114046 & 2010.08.21 & ASDI-4 & 94 & 3 & 10 & 74 & 1.26 & 3.6 & 40.2 & 4 & 135 & 1\,620 \\
 HIP~7576 & 2010.08.21 & ASDI-4 & 31 & 30 & 1 & 68 & 1.02 & 3.1 & 45.7 & 1 & 65 & 1\,950 \\
 HIP~14555 & 2010.08.21 & ASDI-4 & 66 & 30 & 1 & 76 & 1.16 & 3.3 & 36.8 & 1 & 74 & 2\,220 \\
 HIP~10602 & 2010.12.18 & ASDI-4 & 41 & 4 & 8 & 70 & 0.92 & 5.1 & 42.9 & 4 & 136& 2\,176 \\
 HIP~18859 & 2010.12.18 & ASDI-4 & 41 & 5 & 6 & 69 & 0.85 & 5.7 & 35.5 & 2 & 201 & 2\,010 \\
 HD~31295 & 2010.12.19 & ASDI-4 & 32 & 8 & 4 & 62 & 1.40 & 4.0 & 25.0 & 2 & 115 & 1\,840 \\
 HD~38678 & 2010.12.19 & ASDI & 62 & 2.5 & 100 & 11 & 1.39 & 3.3 & 31.2 & 8 & 127 & 2\,540 \\
 HIP~30314 & 2010.12.19 & ASDI-4 & 15 & 15 & 2 & 19 & 1.70 & 2.9 & 20.1 & 1 & 30 & 450 \\
 Fomalhaut & 2011.10.09 & ASDI & 155 & 0.5 & 100 & 128 & 0.85 & 4.0 & 38.1 & 50 & 244 & 6\,100 \\
 \hline
 \end{tabular}
 \end{center}
 \tablefoot{$\Delta$PA refers to the amplitude of the parallactic rotation, DIT {(Detector Integration Time)} to the single exposure time, NDIT {(Number of Detector InTegrations)} to the frame number in a single datacube, $N_{\rm{exp}}$ to the number of datacubes of the observing sequence, $\epsilon$ to the seeing, $\tau_0$ to the mean correlation time of the atmospheric turbulence, and CE to the coherent energy during the observations. {The seeing is measured by the DIMM at 0.5~$\muup$m, hence includes high-frequency terms unseen by the AO wavefront sensor. The two last quantities are estimated by the AO system at 0.55~$\muup$m and 2.17~$\muup$m, respectively.} The next three columns {indicate} the image binning used for {constructing} the science mastercube (Sect.~\ref{sec:mastercubes}), the number of frames in the mastercube, and the corresponding total exposure time on the target. For the observing mode, ASDI-4 refers to FQPM+SDI+ADI and ASDI to SDI+ADI.}
 \end{table*}

Assuming that atmospheric speckles average out over time, both SDI and ADI are intended to attenuate the quasi-static stellar speckles. In ADI, the quasi-static speckles located in pupil planes are kept at a relatively stable position with respect to the detector, while the field of view rotates in a deterministic way. The ability of ADI to suppress speckles {directly depends} on the stability of the telescope, the AO correction, and the instrument. In SDI, images at different wavelengths are obtained simultaneously, and the quasi-static speckles move radially while an off-axis object (a planet for instance) remains at the same position. The performance of SDI is limited by the differential aberrations and the chromatic dependence of the speckles, which is considered linear in first approximation. If this is true for the geometrical aspect (position and size of the speckles), the {wavelength dependence of the phase} can be nonlinear due to chromatism \citep{Marois2006b}. The two techniques {benefit each} other if combined. In this work, they are combined sequentially with SDI first to attenuate the temporal dependence of the speckles and then {with} ADI to reduce the impact of chromaticity.

% ------------------------------------------------------------------------------------
\subsection{Data acquisition}
\label{sec:acquisition}

The data result from three programs carried out in {visitor mode in August 2010 (085.C-0257A) and December 2010 (086.C-0164A), and in service mode} in October 2011 (088.C-0893A). The observing log is given in Table~\ref{tab:obs}. The observing conditions vary from {medium} (mean seeing $\epsilon$\,=\,0.7--1$''$) to very bad ($\epsilon$\,$>$1.4$''$). The durations of the observing sequences are 30--180~min. The amplitude of the parallactic rotation $\Delta\theta$ varies from 10$^{\circ}$ to 155$^{\circ}$ according to the observing time and the star's declination. For the ASDI-4 observations, the single exposure time (DIT) is chosen to fill $\sim$80\% of the full well capacity of the detector (15\,000~ADU). For the ASDI observations, the star's image is saturated to a radius {about five to seven} pixels to increase the dynamic range in the regions of interest. Because the field of view is relatively small, the azimuthal smearing of off-axis point sources remains negligible even for the {highest} DIT value (30~s). A data set is composed of $N_{\rm{exp}}$ datacubes containing DIT$\times$NDIT frames.

{A} defect in the setting of the {pupil-tracking} mode {leads} the star image {to drift} across the detector\footnote{This problem was corrected in October 2011 \citep{Girard2012}.}. This is a very unfavorable situation for coronagraphy, especially for high zenithal angles. For this reason, each datacube is limited to a duration of 30--60~s in order to adjust the star image behind the coronagraphic mask online. For HIP~118008 and HD~38678, we did not use the coronagraph because of a large drift and very unstable observing conditions, respectively. For Fomalhaut, the coronagraphic mask was not allowed in service mode at the time of the observation.

At the end of each observing sequence, a few sky frames are acquired around the observed star, essentially for the purpose {of subtracting} the detector bias (the sky itself being fainter than the bias for the narrow SDI filters\footnote{{The exposure time beyond which the background noise regime is reached for the SDI filters without the neutral density is $\sim$1\,400~s (\url{http://www.eso.org/observing/etc/}).}}. {Before each observing sequence}, we record out-of-mask images of the star, with or without a neutral density filter depending on the star magnitude, to determine the PSF and to serve as a photometric calibration.

% ------------------------------------------------------------------------------------
\subsection{Data reduction}

The data are reduced with the pipeline described in \citet{Boccaletti2012b}, which was modified to take {the specificities of SDI data into account}. This pipeline operates in two steps:
\begin{enumerate}
\item Step 1: {constructing} the ``reduced'' mastercubes for the science images and the PSF;
\item Step 2: {applicating} the differential imaging methods, SDI and/or ADI.
\end{enumerate}

\begin{figure*}[t]
	\centering
	\includegraphics[trim = 10mm 31.5mm 12.2mm 23.3mm, clip, height=.26\textheight]{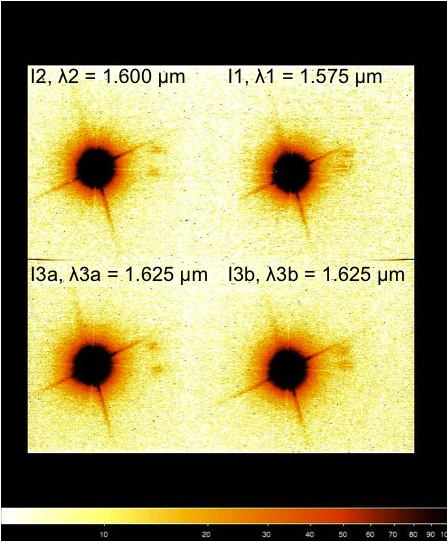}
	\includegraphics[trim = 13mm 25mm 16mm 7mm, clip, height=.26\textheight]{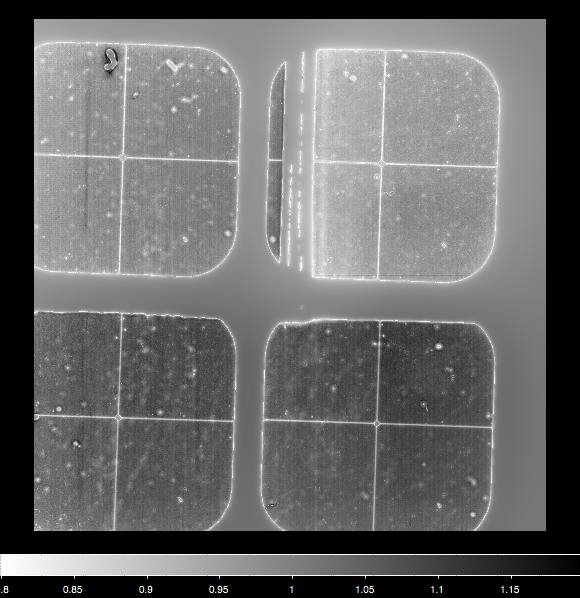}
	\caption{\textit{Left}: Single raw image of a target star provided by the SDI mode of NaCo. The wavelength of the four {subimages} is from the top right in the anticlockwise direction: 1.575, 1.600, 1.625, and 1.625~$\muup$m. The image is thresholded and the scale is logarithmic. \textit{Right}: Flat-field image of the SDI filters (the vertical bars of the filter support are visible in the top half of the image), the double Wollaston prism (which produces the four {subimages}) and the FQPM (the cross visible in each sub-image). The image scale is linear. Both image sizes are 1\,026~pixels~$\times$~1\,024~pixels, and the field of view of each {subimage} is 8$''\times$\,8$''$. The cut of the images seen on their left side is due to the position of the field selector on the light beam. Note the inverted tints of both images.}
	\label{fig:sdiimages}
\end{figure*} 

% ------------------------------------------------------------------------------------
\subsubsection{Science and PSF ``reduced'' mastercubes}
\label{sec:mastercubes}

An example of a single frame recorded at the detector is shown in Fig.~\ref{fig:sdiimages} (left). In the first step, all single frames are corrected for bad pixels and flat field. However, when using the FQPM combined to SDI, the flat field (FF) correction requires particular attention. Several optical elements are involved in this configuration (the FQPM, the Wollaston prism, the SDI filters), each being located on different filter wheels. Some of them are close to a focus (the FQPM and the SDI filters), so are very sensitive to dust features. The {dust features} (Fig.~\ref{fig:sdiimages}, right) generate variations {in} the pixel transmission as large as 20\%. Given the accuracies of the wheel positioning, the relative positions of the optical elements are not the same from one observation to another.

The detector FF is measured on twilights with the H-band filter. Standard calibrations obtained with the internal lamp provide the FF of some combinations of elements (SDI filters, SDI filters + Wollaston prism, SDI filters + Wollaston prism + FQPM). A FF of all the optical elements (SDI filters + Wollaston prism + FQPM) is shown in Fig.~\ref{fig:sdiimages} (right) for illustration. We can see for each {subimage} the square-rounded-corner field stop and the trace of the FQPM {transitions. We also note} the edges of the filter assembly (one edge appears in the upper right subimage). From these calibrations, we extract the FF of each individual element. Then, for most targets, an image of the configuration SDI filters + Wollaston prism (+ FQPM) is recorded at the beginning of the observation. This image cannot be used as a FF because of a poor signal-to-noise ratio (integration time of 2~s instead of 9~s for a regular FF). Instead, it is used as a reference for the realignment of the individual FFs to produce a combined FF. For two stars (HIP~10602 and HIP~18859), the preimaging of the configuration is not recorded. For these cases, in addition to the detector FF, we only correct for the averaged transmission in each subimage. Therefore, the small-scale patterns like dust features on the FQPM are not removed. Finally, the FF is normalized to the median value over the pixels containing signals, hence excluding the field stop and the edges of the filter assembly. {The averaged transmission factors in the SDI quadrants are $\sim$0.93, $\sim$0.95, and $\sim$1.07 for the {subimages} $I_1$, $I_2$, and $I_{\rm{3a}}$/$I_{\rm{3b}}$ respectively. The precision of the FF correction is {10\% to 20\%} for the two stars without the pre-imaging of the instrumental configuration. It is $\sim$4\% for the other targets and set by the centering precision of the regular FF (0.2~pixels).}

For the PSF images, several datacubes are recorded each containing several frames. For a datacube, {we subtract a mean sky background image of the same duration than the individual images, extract the SDI quadrants, recenter the star image on the central pixel using Gaussian fitting, and average these frames}. All datacubes are processed similarly and then averaged, producing a ($x$,\,$y$,\,$\lambda$) PSF mastercube normalized in ADU/s and corrected for the neutral density transmission {\citep[1.23\%$\pm$0.05\% as reported in][we actually use 1/80 in the data reduction]{Boccaletti2008a, Bonnefoy2013a}}.

For the science images, the first important step is {to select frames, in particular, to reject open AO loops}. The selection is based on the number of pixels $\nu_f$ ($f$, the index of a frame) in a given {subimage} (Fig.~\ref{fig:sdiimages}, left) for which the flux is superior to an intensity threshold. This quantifies the width of the star image, a close-loop image having more pixels above this threshold than an open-loop image. After trials, we set this threshold to 60\% of the maximum flux measured on a single datacube. Then, we retain the images for which {the number of pixels} $\nu_f$ is greater than 40\% (70\%) of the maximum value of $\nu_f$ for coronagraphic (saturated) data. The threshold is higher for saturated images, since the flux can be high even in open-loop images. 

{Our} procedure performs an efficient rejection of the open-loop images. The selected frames are averaged per groups of a few units in order to reduce the frame number in the final mastercube to about a hundred or so (issue related to computing time {when applying the differential imaging, Sect.~\ref{sec:sdiadiprocessing}}). The binning values ensure that the smearing of off-axis point sources at the edges of the images remains negligible. If the frame bin is not an integer multiple of the number of frames in a datacube after selection, the remaining frames are averaged and accounted {for} in the mastercube. For the purpose of ADI, a vector of parallactic angles (averaged in a frame bin) is saved. Then, an averaged sky background image is subtracted out frame by frame to the mastercube. {We do not apply any linearity correction on the raw frames. However, the nonlinear regions are not taken into account for the image normalization (the PSF are not saturated) or the estimation of the flux rescaling factor for the SDI processing (Sect.~\ref{sec:sdiadiprocessing}).} {No astrometric calibration is applied, {so that} the orientation of the True North is known with low accuracy (0.5--1$\degb$).}

In addition to filter open-loop images during the construction of the mastercube, a second level of selection is applied to the mastercube in order to reject the frames of lower quality (poor AO correction, large seeing, large offset from the coronagraphic mask, etc.). For these data, we base our statistical analysis on the total flux in one subimage, but other criteria are available in our pipeline. The frames departing from the median flux by $X$ times the standard deviation are eliminated from the mastercube. For most stars in our sample, we use $X$=2 (5\% of images rejected), except for Fomalhaut for which we select $X$=3 (1\% of images rejected). These values are chosen to reject the worst frames, while reducing the discontinuities in the temporal sequence. Indeed, large temporal discontinuities introduce biases in the ADI construction of the star reference image. In particular, the observing sequence on Fomalhaut suffers from large discontinuities. After selection, the vector of parallactic angles is updated, and the temporal mastercube ($x$,~$y$,~$t$) is separated to form a four-dimensional ($x$,~$y$,~$t$,~$\lambda$) mastercube.

Before the registration of the ``reduced'' mastercube, the frames for each {subimage} are recentered using function-fitting. For saturated images, the most robust results are obtained with Moffat function-fitting\footnote{The fit is performed using the IDL library \texttt{mpfit} developed by C. Markwardt and available at \url{http://www.physics.wisc.edu/~craigm/idl/fitting.html}.} \citep{Moffat1969}, in which the central saturated region is assigned a null weight. For coronagraphic images, centroiding algorithms are not efficient, since the coronagraph alters the intensity and the shape of the star image as a nonlinear function of the pointing. Therefore, we follow the same process as described in \citet{Boccaletti2012b}, where the fit of a Moffat function is applied on images thresholded at 1\% of their maximum. It has the advantage of putting the same weight in all pixels within a given radius to the star ($\sim$0.75$''$) and to damp the impact of nonlinearities in coronagraphic images.

In Table~\ref{tab:obs} (three last columns), we {provide the} frame binning (number of co-added frames), the frame number in the science mastercube, and the total exposure time {for all targets}. The value of the latter is {approximate} as long as the number of frames per datacube is not always an integer multiple of the frame bin.

% ------------------------------------------------------------------------------------
\subsubsection{SDI and ADI algorithms}
\label{sec:sdiadiprocessing}
Since SDI is obtained from simultaneous images, it has the capability {of reducing} the impact of evolving stellar speckles, which usually limit the performance of ADI. Therefore, SDI should be applied in the first stage. The principle of SDI is already described in \citet{Racine1999}, \citet{Marois2000}, and \citet{Biller2007}, but we briefly recall it for the purpose of this paper. Two images of a star ({science image \iscience and reference image \irefnospace}) at different wavelengths but spectrally adjacent are aligned, set to the same spatial scale, scaled in intensity to correct for differential filter transmission and stellar flux variations, and finally subtracted. If we {assume $\lambda_{\rm{r}}$~$>$~$\lambda_{\rm{s}}$}, which corresponds to our data, this implies that {the reference image \iref} is reduced in size by a {factor $\lambda_{\rm{r}}/\lambda_{\rm{s}}$}. {In our pipeline, the intensity scaling factor is derived from the ratio of the total fluxes measured in {annuli} of inner and outer radii 0.2$''$ and 0.5$''$ respectively. {We did not optimize these values, but we {have} checked that we are in the intensity linear regime.} After the subtraction, the star contribution is strongly attenuated, while the signature of a putative planet is composed of a positive component at the separation of the planet and a negative component closer in {separated by}
\begin{equation}
\Delta r = r_0 \left(1-\frac{\lambda_{\rm{s}}}{\lambda_{\rm{r}}}\right)
\label{eq:sdispacing}
\end{equation}
with $r_0$ the planet separation in {the science image \isciencenospace}. As $\Delta r$ increases with $r_0$, the overlapping between the positive and negative components of the planet decreases with the angular separation. The bifurcation point ($r_{\rm{b}}$), as defined by \citet{Thatte2007}, is the angular separation beyond which the spacing between the positive and negative companion components is greater than the PSF width {in the science image $\lambda_{\rm{s}}/D$} (with $D$ the telescope diameter). Using this definition in Eq.~(\ref{eq:sdispacing}) and rearranging the terms for expressing {the bifurcation point}, {we obtain}
\begin{equation}
r_{\rm{b}}\,=\,\frac{\lambda_{\rm{s}}}{D}\times\frac{\lambda_{\rm{r}}}{(\lambda_{\rm{r}}-\lambda_{\rm{s}})}.
\label{eq:rb}
\end{equation}

In our data reduction, where four {subimages} are available, we consider all subtractions. We note the {subtraction $I_{\rm{sr}}$\,=\,$I_{\rm{s}}$$-$$I^{\prime}_{\rm{r}}$, with $I^{\prime}_{\rm{r}}$ the reference image} rescaled spatially and photometrically. Thus, the subtractions are \iacnospace, \iadnospace, \ibcnospace, \ibdnospace, and \iab (Fig.~\ref{fig:sdiimages}, left). For the VLT ($D$\,=\,8.2~m), {the bifurcation point} is 1.3$''$ {for the two first subtractions and and 2.6$''$  for three last ones.} However, this is valid for diffraction-limited images (full width at half maximum FWHM~=~40~mas), which is not the case for our data (FWHM~=~57--120~mas). {This implies that the point-source overlapping will be more {significant} at a given separation, so the SDI performance will be degraded (Sect.~\ref{sec:sdisignature}).}

We apply SDI to each couple of frames in the temporal sequence. An output four-dimensional datacube is produced, with the number of subtractions as {the} fourth dimension. ADI is applied at a second stage. We consider two algorithms, classical ADI \citep[cADI,][]{Marois2006a} and the Karhunen-Lo\`eve Image Processing \citep[KLIP,][]{Soummer2012}. The processed image is smoothed on boxes of 3$\times$3~pixels, approximately the PSF FWHM, in order to filter the high-frequency noise. Finally, we derive the contrast levels of the residual noise using the standard deviation of the image residuals on rings {with one-pixel width} for all separations. Figure~\ref{fig:limdetection} shows the median 1-$\sigma$ noise level of the survey with ADI and ASDI, together with the worst and best cases based on the performance reached at 0.5$''$. We consider the image $I_1$ for ADI and the {subtraction} \iad for ASDI. We account for the attenuation of ADI for both methods by injecting {seven} synthetic planets in the raw data at separations of {0.3, 0.5, 1, 1.5, 2, 3, and 4$''$} in two directions separated by 180$^{\circ}$ to avoid overlapping{. The} largest parallactic rotation in the survey is 155$^{\circ}$ {(Table~\ref{tab:obs}).} Nevertheless, we neglect the SDI attenuation, {since} it depends on the spectral properties of companions {that} could have been detected (Sect.~\ref{sec:sdianalysis}). This is equivalent to {assuming} that the companion contains methane and does not emit flux in {the $I_{\rm{3b}}$ image}. We note that the worse the observation quality the smaller the difference between the ADI and ASDI noise levels. For the median curves, we see that ASDI improves by a factor $\sim$2--3.5 the noise level at separations closer than 1.5$''$, where speckle noise {dominates background} noise. Nonetheless, ASDI slightly degrades the detection performance for larger separations, where background noise is the dominant source of noise. In particular, ASDI potentially offers a gain in angular resolution of 0.2--0.4$''$ to search for faint companions of a given contrast at separations of 0.3--1$''$. We explain in Sect.~\ref{sec:sdianalysis} that while the residual noise can be used to determine the sensitivity limits in planet mass in ADI imaging, this method is no longer valid in SDI or ASDI imaging.

\begin{figure}[t]
	\centering
	\includegraphics[trim = 9mm 4mm 7mm 8mm, clip, width=.4\textwidth]{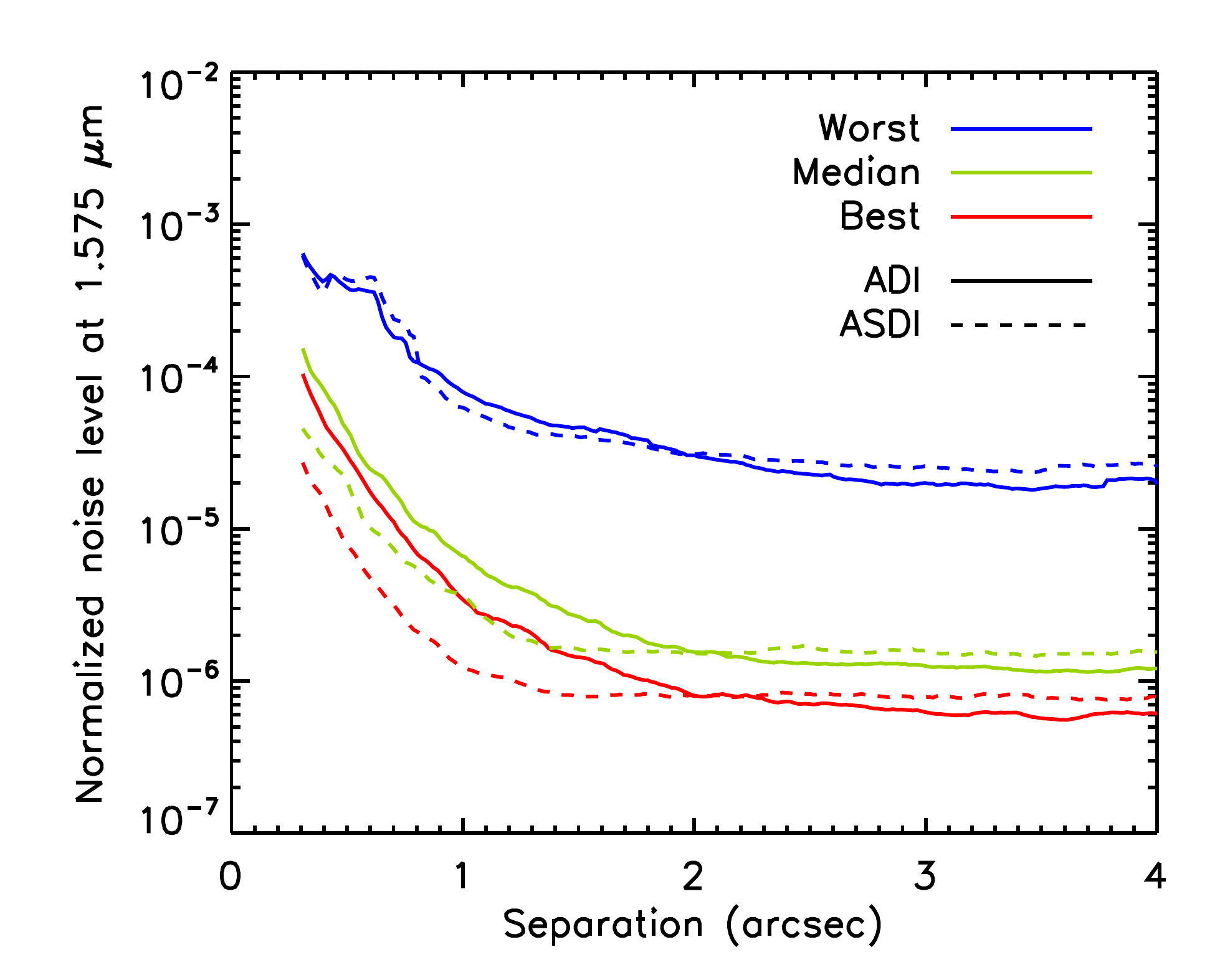}
	\caption{Residual noise in the $I_1$ image expressed in contrast with respect to the star after ADI alone (solid lines) and after ASDI (dashed lines). We consider for ASDI the {subtraction} \iadnospace. {We represent the median, best, and worst curves in the survey (colored curves, see text)}. We account for self-subtraction from ADI for all curves {using synthetic planets, but not from SDI for the ASDI curves (see text)}.}
	\label{fig:limdetection}
\end{figure}

\subsection{{Photometric errors}}
\label{sec:errorphot}
{We summarize here the different sources of photometric errors in the data reduction and analysis {that} affect the noise levels shown in Fig.~\ref{fig:limdetection}:
\begin{itemize}
\item the correction of the neutral density transmission. This transmission is measured with a precision of 4\%;
\item the correction of the pupil Lyot stop associated with the FQPM. This diaphragm is undersized by 10\% in diameter with respect to the full aperture, so the geometrical throughput is 0.808. However, \citet{Boccaletti2008a} report an uncertainty of 4\% on this value, probably due to optical misalignments of the entrance pupil of NaCo. For the coronagraphic observations, we kept the Lyot stop for the measurement of the PSF to minimize this error source, except when the neutral density is used;
\item the photometric stability due to variations {in} the AO correction and, for the coronagraphic observations, variations in the FQPM centering. The PSF is measured once at the beginning of the observing sequence. Consequently, we cannot estimate the temporal variability of the PSF. We use the science images and measure the variability of total intensity in an annulus between 0.2$''$ and 0.5$''$ (region in the linear regime) for all targets. The median value at 1~$\sigma$ is 12\% (range 6--17\%).
\end{itemize}}

% ------------------------------------------------------------------------------------
\section{SDI data analysis}
\label{sec:sdianalysis}
% ------------------------------------------------------------------------------------

{In this section, we first introduce the theoretical formalism of the flux measurement in SDI-processed images (Sect.~\ref{sec:sdisignature}) and argue that, contrary to single-band imaging surveys, the residual noise cannot be converted into mass limits through evolutionary models, because it represents a differential flux. This differential flux can be accounted for by several planet masses, resulting in degeneracies for the mass limits (Sect.~\ref{sec:degendifffluxmass}). The degeneracies can be broken with the use of single-band detection constraints. Finally, we describe the method used for the analysis of this survey, which is based on synthetic planets and model fluxes to directly determine the mass limits (Sect.~\ref{sec:detectionlimits}).}

\subsection{SDI signature}
\label{sec:sdisignature}

In the general context of high-contrast imaging with a single spectral filter, the detection limits are measured on contrast maps. In most cases, the azimuthal standard deviation is used to derive one-dimensional contrast curves. These curves are then corrected from various attenuations (ADI and/or coronagraph) and converted into mass limits according to a given evolutionary model. The {problem} of SDI is different because we measure a differential intensity. The sign and modulus of this residual intensity at the location of an object depend on its spectral shape (at the first order determined from its temperature), as shown in Fig.~\ref{fig:spectracoldegp}, and on its separation in the image. It can be expressed as follows (no coronagraph and no ADI):

{\begin{equation}
F_{\rm{sr}}=F_{\mathrm{s}}-F_{\mathrm{r}} \times \alpha \times \phi(\vec{r})
\label{eq:fsdi}
\end{equation}}
with {\fscience and \fref the object fluxes in the science and reference images}, $\alpha$ the intensity rescaling factor, $\phi(\vec{r})$ the attenuation due to the spatial rescaling, and $\vec{r}$\,=\,$(r,\theta)$. The dependency of $\phi(\vec{r})$ on $\theta$ is related to the PSF structure (see below){, where} $\phi(\vec{r})$\,=\,1 means that the planetary companion is close to the star so that the positive and negative components of its SDI signature mostly overlap and {cancel each other out}. This is the worst case for SDI in terms of performance. On the contrary, $\phi(\vec{r})$\,$\simeq$\,0 means that the processing does not attenuate the planet intensity, i.e. {$F_{\rm{sr}}$\,$\simeq$\,$F_{\rm{s}}$}, which is the optimal case. The factor $\phi(\vec{r})$ can be determined for each radial separation and azimuthal direction using measured {or modeled} PSF in each couple of filters. Knowing the characteristics of $\phi(\vec{r})$ is important for data reduction and interpretation of the detection limits (Sect.~\ref{sec:results}). We briefly discuss here key points relevant to the survey analysis, while a companion paper will address the SDI biases in detail (Rameau et al., in prep.).

{First, {the SDI geometrical attenuation} $\phi(\vec{r})$ depends on the wavelengths of the images used for the SDI subtraction. For a given separation in the image, the larger the spacing between the wavelengths, the smaller $\phi(\vec{r})$ (Eq.~(\ref{eq:sdispacing})). In particular, we saw that the bifurcation point is smaller for the \iad subtraction than for the \ibc subtraction (Sect.~\ref{sec:sdiadiprocessing}). We also note that the flux ratio $F_1/F_3$ is greater than the flux ratio $F_2/F_3$ for cool giant planets (Fig.~\ref{fig:rfluxteff}). As a result, the self-subtraction will be less for the \iad subtraction (Eq.~(\ref{eq:fsdi})). Thus, we consider {only this subtraction for the data analysis}.}

\begin{figure}[t]
	\centering
	\includegraphics[trim = 14mm 5mm 5mm 10mm, clip, width=.4\textwidth]{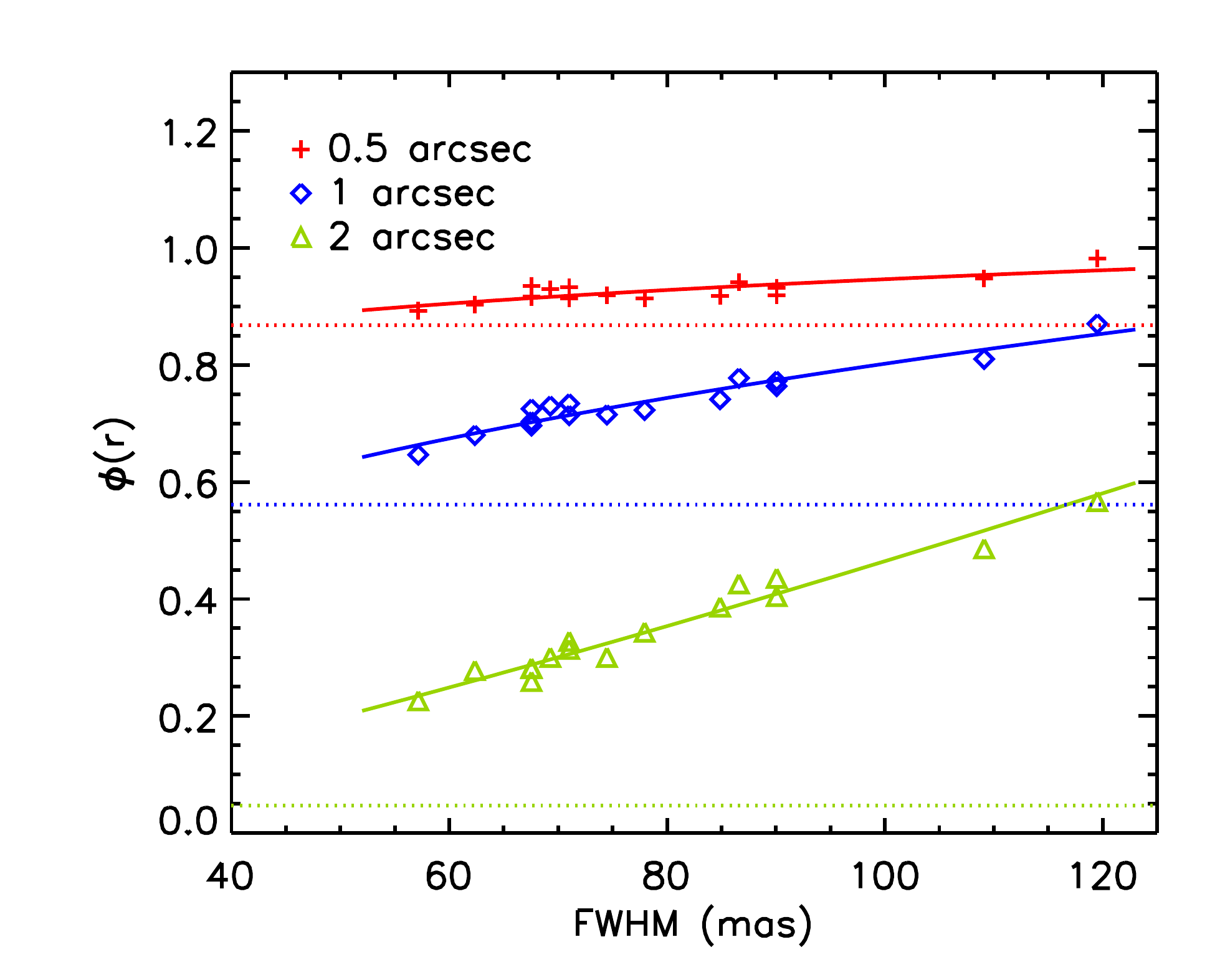}
	\caption{{Azimuthal mean of the geometrical SDI attenuation factor $\phi(\vec{r})$} derived for all the observed stars and the \iad subtraction at separations of 0.5, 1, and 2 arcseconds (colors) as a function of the FWHM (see text). The dotted lines indicate {the values} for a diffraction-limited PSF. The solid lines are power-law fitting laws.}
	\label{fig:phirseeingecmeant0mean}
\end{figure}

{For a given wavelength {couple ($\lambda_{\rm{s}}$,~$\lambda_{\rm{r}}$)}, {the SDI geometrical attenuation} $\phi(\vec{r})$ is a function of the position in the image field, both radially and azimuthally. These dependencies are intimately related to the PSF properties. We use real data to {derive} $\phi(\vec{r})$ to take {PSF structures that cannot be {modeled} into account}. We determine $\phi(\vec{r})$ from the flux ratios of a PSF measured in the $I_1$ and \iad images. Both fluxes are summed in apertures of 3$\times$3~pixels centered on the PSF location in {the $I_1$ image}. Figure~\ref{fig:phirseeingecmeant0mean} represents the azimuthal mean {of the SDI geometrical attenuation (noted $\phi(r)$ in the remainder of this paragraph)} measured at three separations for all the stars in the survey as a function of the FWHM. The FWHM is estimated using the mean value from several methods (Gaussian fitting, radial profile). The typical standard deviations are 0.3--0.4~pixels, but can be as much as $\sim$1~pixel for elongated PSF. {The {values of $\phi(r)$} for a diffraction-limited PSF are also indicated as dotted lines}. For a given star (same FWHM), we observe that $\phi(r)$ decreases with the separation. The measured values are greater than the theoretical values, since the images are not in the diffraction-limited regime (Sect.~\ref{sec:sdiadiprocessing}). We also note that the larger the separation{, the greater} the discrepancy between the measured and theoretical values. For a given separation, we observe that the range of the measured $\phi(r)$ increases. At small separations, $\phi(r)$ is {close to 1} (Eq.~(\ref{eq:sdispacing})), and the self-subtraction is large and {depends little} on the PSF properties. As the separation increases, $\phi(r)$ diminishes, and the dependency of the self-subtraction on the PSF quality is stronger. We determine that the best fitting law for this behavior is a power law. We note that the fitting laws {agree} with the diffraction-limited values of $\phi(r)$ at the theoretical value of the FWHM for the smallest separations. The discrepancy seen for 2$''$ could be accounted for by a different behavior of $\phi(r)$ with this parameter in (nearly) diffraction-limited regimes, as expected for SPHERE and GPI. We tested possible correlations of $\phi(r)$ with other observational factors, the Strehl ratio, coherent energy, correlation time of the turbulence, and seeing {(results not shown, see the detail of the estimation methods in Appendix~\ref{sec:corrnoiseatten}). We find correlations with the Strehl ratio alone.} We conclude that the PSF FWHM is an important parameter for {interpreting the} SDI performance at separations beyond 0.5$''$. This is also true for ASDI {(see Sect.~\ref{sec:results})}.

{Finally, {the SDI geometrical attenuation} $\phi(\vec{r})$ (so the SDI performance) is a function of the direction in the field of view $\theta$, owing to PSF asymmetries. For a few observations, the PSF is elongated because of an astigmatism not corrected by the active optics system of the telescope. This must be accounted for when deriving SDI (so ASDI) sensitivity limits. The method that we use for {assessing the} detection limits of the survey accounts for this point (Sect.~\ref{sec:detectionlimits}).}

\subsection{Degeneracy of the differential flux with the planet mass}
\label{sec:degendifffluxmass}

When a low-mass object is detected in an SDI-processed image and providing $\phi(\vec{r})$ is calibrated, one can test the individual intensities {\fscience and \fref} for all planet masses in an evolutionary model that reproduce the measured differential flux {$F_{\rm{sr}}$} (Eq.~(\ref{eq:fsdi})). Although no practical case has been published in the literature yet, we expect that several values of {\fscience and \fref} can match the observations, which results in degeneracies for the mass estimation. The number and the values of the mass solutions depend on $\phi(\vec{r})$, so on the position in the image.

\begin{figure*}[t]
	\centering
	\includegraphics[trim = 11.5mm 5mm 5mm 10mm, clip, width=.4\textwidth]{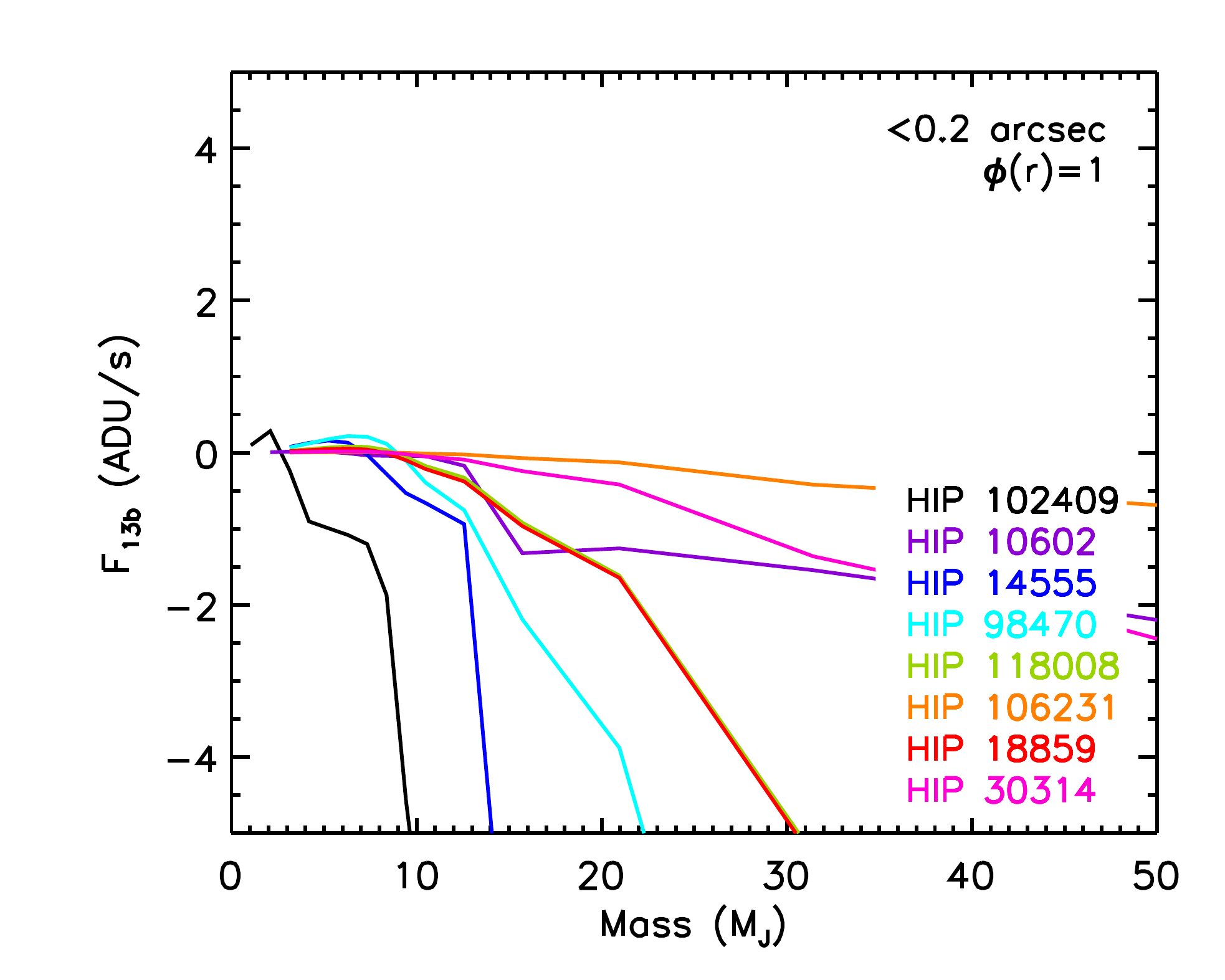}
	\includegraphics[trim = 11.5mm 5mm 5mm 10mm, clip, width=.4\textwidth]{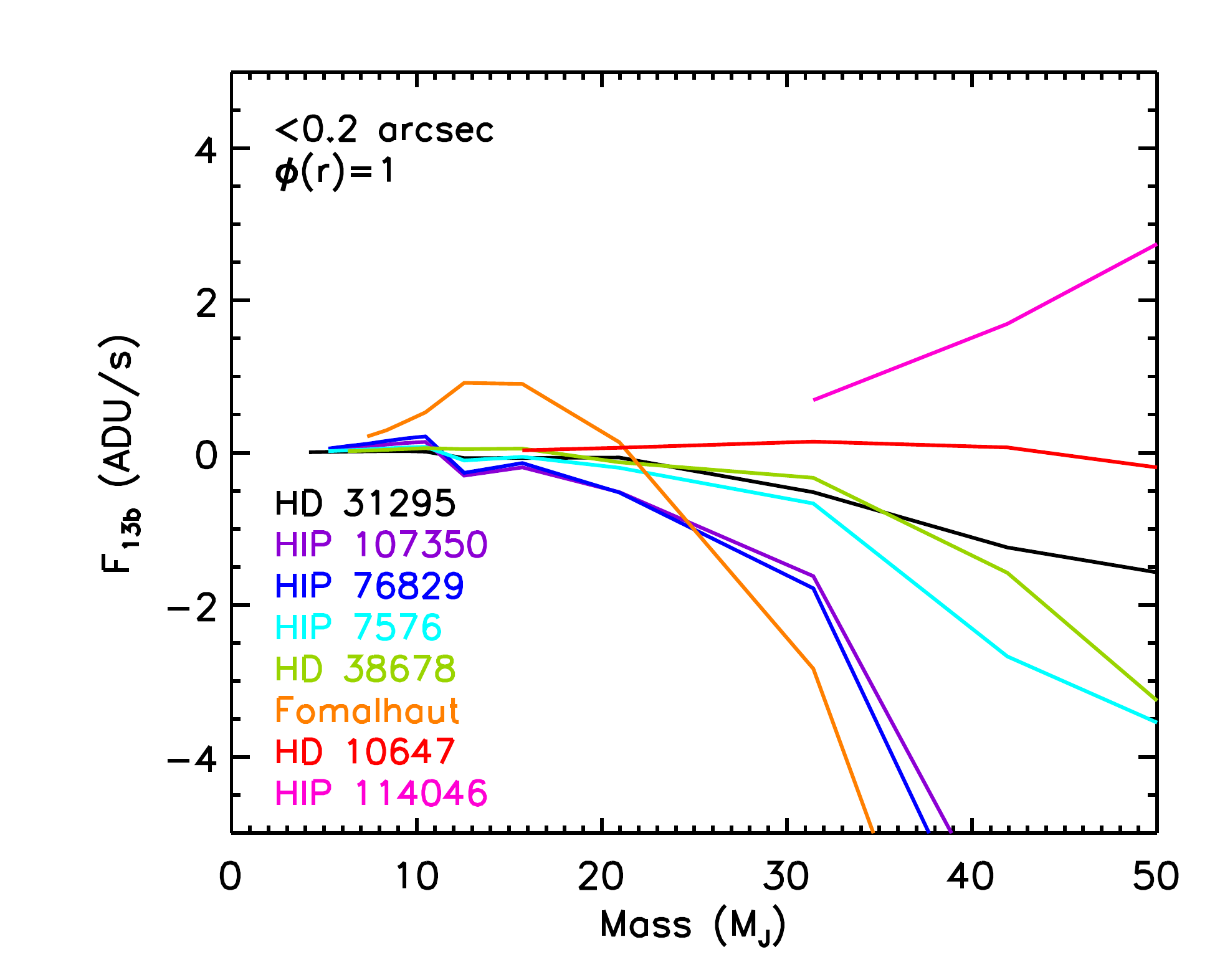}
	\includegraphics[trim = 11.5mm 5mm 5mm 10mm, clip, width=.4\textwidth]{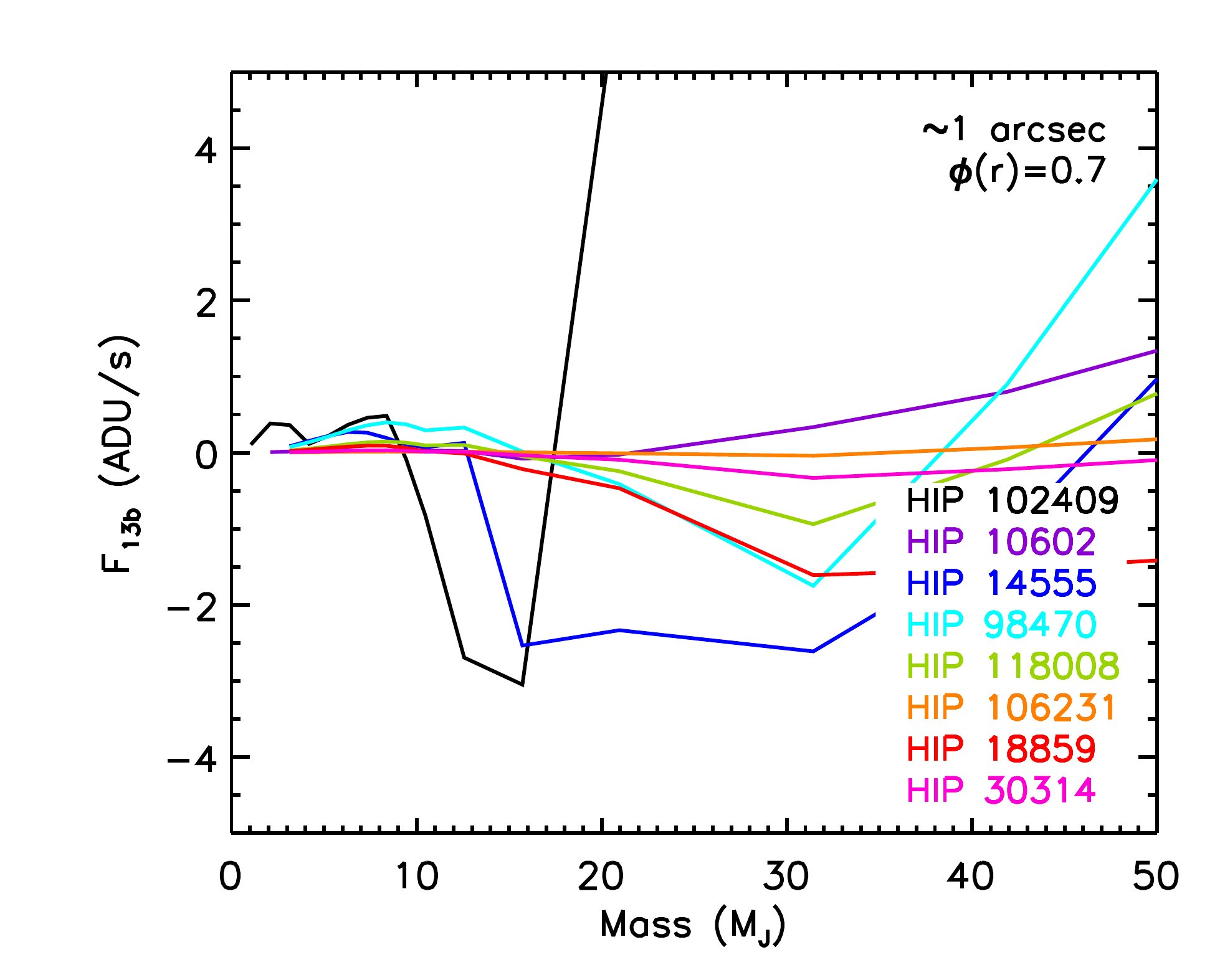}
	\includegraphics[trim = 11.5mm 5mm 5mm 10mm, clip, width=.4\textwidth]{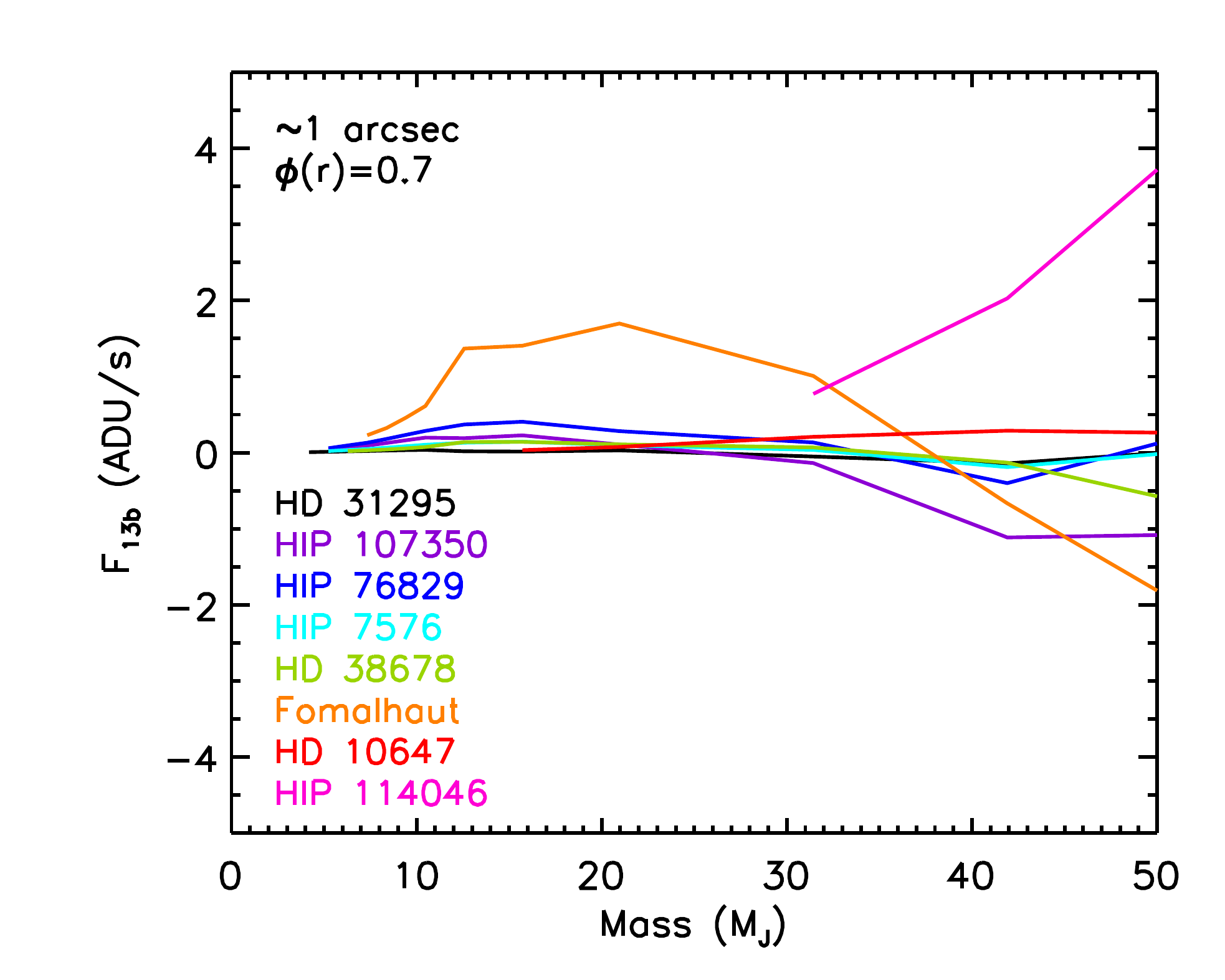}
	\includegraphics[trim = 11.5mm 5mm 5mm 10mm, clip, width=.4\textwidth]{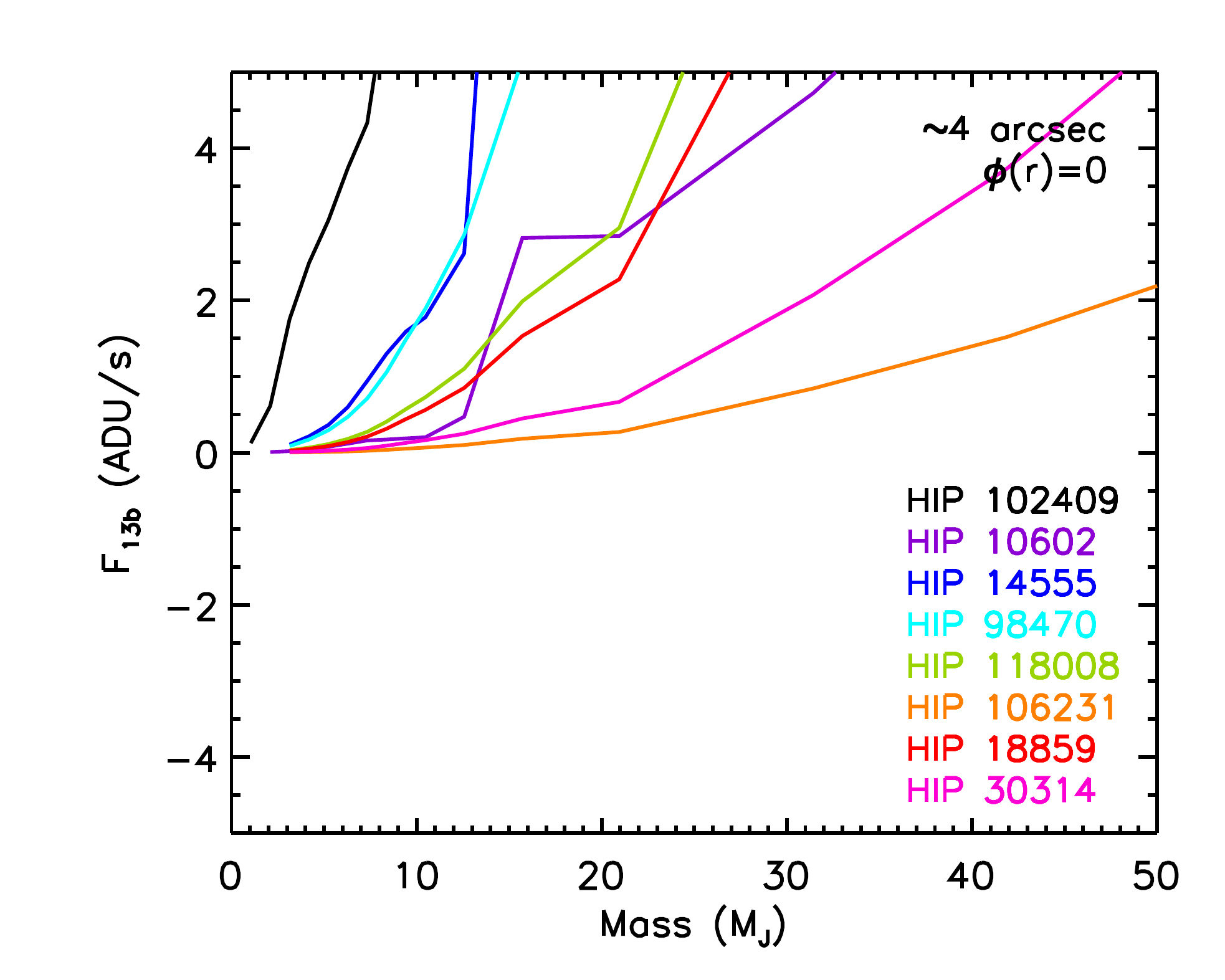}
	\includegraphics[trim = 11.5mm 5mm 5mm 10mm, clip, width=.4\textwidth]{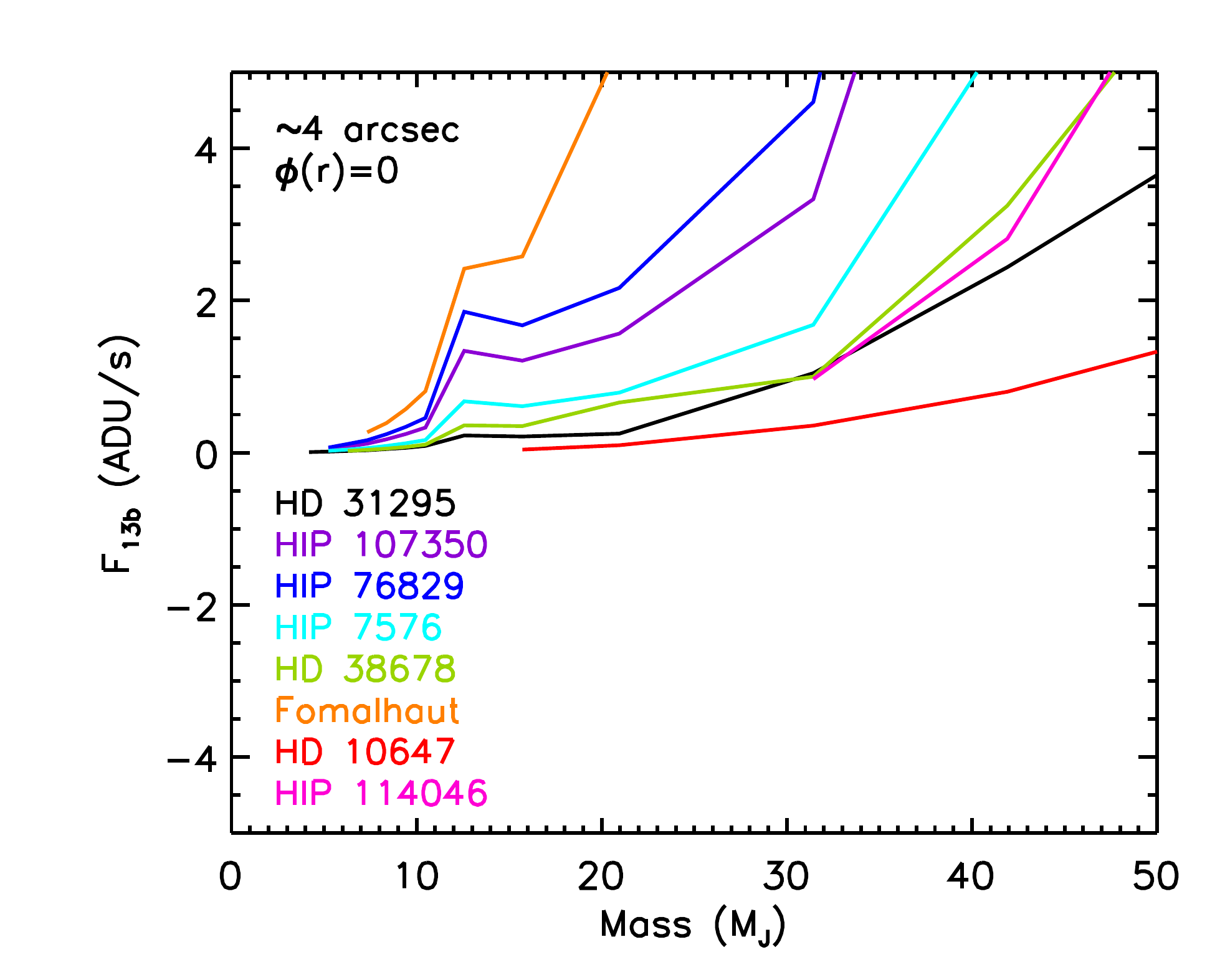}
	\caption{Differential flux \fad of a planet predicted by the evolutionary model BT-Settl \citep{Allard2012} as a function of mass for all targets and three separations (rows). The stars are sorted by age, $<$100~Myr (\textit{left column}) and $\geq$100~Myr (\textit{right column}). {The} top right part of the right panels {shows} the curve for HIP~114046, which is a very old star (8~Gyr, Table~\ref{tab:sample}).}
	\label{fig:difffluxmass}
\end{figure*}

Figure~\ref{fig:difffluxmass} shows the differential flux \fad predicted by the BT-Settl models \citep{Allard2012} as a function of mass for three values of $\phi(\vec{r})$. To obtain these curves, we derive the fluxes in Eq.~(\ref{eq:fsdi}) from the model absolute magnitudes and assume $\phi(\vec{r})$\,=\,$\phi(r)$ and $\alpha$\,=\,1; i.e., {the flux measured in the reference image \fluxd} is corrected for the differences in flux and filter transmission. For the conversion of absolute magnitudes into fluxes, we consider the star distance, the star magnitude in H band, and the maximum of the measured PSF in the corresponding filters {for the photometric zeropoints}. Using the H-band magnitude for the magnitude in the SDI filters is equivalent to assuming that the stellar spectrum in this spectral region does not exhibit features (Rameau et al., in prep.). {The curves shown in Fig.~\ref{fig:difffluxmass} are specific to the choice of stellar parameters considered in this paper. Using different parameters (in particular the star age) will produce different curves. We chose to express the differential flux \fad in ADU/s in order to represent the data for all stars. We must take {the actual integration time of the observation into account} to test the ability to distinguish potential degeneracies of \fad with the mass.}

For separations closer than 0.2$''$ (top row), the differential flux \fad is negative for masses greater than 3--25~\mj according to the stellar age, while it is positive for {lower masses.} This is consistent with Fig.~\ref{fig:spectracoldegp}. For temperatures higher than $\sim$1\,000~K ($\gtrsim$4~\mj at 10~Myr and $\gtrsim$10~\mj at 100~Myr), there is no methane absorption at 1.625~$\muup$m, while {methane absorbs the flux for lower temperatures}. In some cases, we find that \fad is accounted for by two or three values of the companion mass. The number and the values of the degeneracies depend on the stellar properties (and on the position in the image, see below). The degeneracies correspond to low values of \fadnospace, typically inferior to 0.5~ADU/s for stars younger than 200~Myr. Consequently, long integration times will be required to reach the photometric accuracy needed to detect the degeneracies.

For separations around 1$''$ (middle row), we note three regimes for \fadnospace. It is positive for low masses, then negative for intermediate masses, and finally positive for massive objects. The reason for the existence of this last regime is that after the flux ratio \fluxratio diminishes below 1 when there is no methane absorption anymore, it increases towards one for temperatures higher than $\sim$2\,100~K (Fig.~\ref{fig:rfluxteff}). We retrieve {degeneracies of \fad with the mass for some cases}, but different due to a different $\phi(r)$ value. We also find new degeneracies for negative \fadnospace.

For separations of $\sim$4$''$ (bottom row), \fad is positive {regardless of} the mass because \fad=~$F_1$, and increases with the mass. For a few targets (HIP~10602, and the stars younger than Fomalhaut in the right panel), degeneracies appear for $\sim$10--20~\mjnospace. These degeneracies are inherent to the evolutionary model.

{One possible solution {for removing} the mass degeneracies is to measure the flux of an object in a single filter if it is detected, after data processing with ADI, for instance. We show in the next section that in the case where no object is detected, the mass degeneracies inherent to SDI also affect the assessment of the detection limits, implying the need to compare the constraints of ASDI to those of single-band differential imaging.}

\subsection{Detection limit}
\label{sec:detectionlimits}
When no object is detected in an SDI-processed image, the situation is more complex. The residual image exhibits positive and negative intensities with strong pixel-to-pixel variations resulting from the image subtraction. As long as we should expect a low-mass object to also have either positive or negative flux, it becomes difficult to disentangle the noise from a signal. Thus, the method used for single-band surveys (ADI, coronagraphy){, which consists of building} a contrast map from the standard deviation and {converting it into mass,} is no longer valid. A variation of this method will be analyzed thoroughly in a companion paper (Rameau et al., in prep.). Such a method requires good determination of both ADI and SDI attenuations ($\phi(\vec{r})$). Here, we apply a more straightforward and robust technique based on {injecting synthetic planets into} the data set at the cost of a longer computing time and sparsity in the detection map. The use of synthetic planets is common in high-contrast imaging{. For this work, we use the measured PSF for each data set for the whole corresponding temporal sequence. Thus, we assume} that the PSF is the same in the science and the PSF images. This hypothesis strongly depends on the AO-loop and photometric stabilities, and we should expect variations from one data set to another.

We consider a given evolutionary model where the absolute magnitudes of low-mass objects are tabulated for different ages and masses in the SDI filters. Assuming the age and the distance for a given star, we obtain the expected flux for a given planet mass. {Twelve synthetic planets} are simultaneously injected in empty datacubes at the positions {0.3, 0.5, 0.7, 0.9, 1.1, 1.4, 1.7, 2, 2.5, 3, 3.5, and 4$''$}. They are also introduced at several position angles. For each case, we take care that the planets do not overlap, because of the SDI spatial rescaling and the ADI field rotation. The resulting datacube is processed with ASDI and ADI as described in Sect.~\ref{sec:sdiadiprocessing}. We measure the residual intensities of the planets using aperture photometry (Sect.~\ref{sec:sdisignature}). For each separation, we azimuthally average the fluxes.

{The} noise is measured on the data processed without the synthetic planets (Sect.~\ref{sec:sdiadiprocessing}) and scaled to the same aperture size assuming white noise. The process is repeated for each model mass in order to obtain a three-dimensional array of signal-to-noise ratio versus mass and separation. For each separation, we interpolate the signal-to-noise ratio versus mass relation to determine the mass and effective temperature achieved at 5~$\sigma$. We checked that the {detection limits (Sect.~\ref{sec:results})} are consistent with the reduced datacubes containing the data and the synthetic planets. Figure~\ref{fig:snrmass} shows examples of curves of the signal-to-noise ratio as a function of planet mass for HIP~118008 and several separations for ADI and ASDI. For a given mass, it increases with the separation, because the noise level diminishes. For a given separation, it increases monotonously with the mass for ADI but not for ASDI. For the latter case and a signal-to-noise ratio of 5, a degeneracy is observed around 0.6$''$ (6, 8.5, and 10.5~\mjnospace). However, it is broken when comparing it to the ADI constraint, which is deeper {(objects $\geq$5.5~\mj excluded)}. We encountered several degeneracies in the data analysis, but we excluded all of them using ADI.

{The errors on the signal-to-noise ratios (and the sensitivity limits, Sects.~\ref{sec:results} and \ref{sec:discussion}) are induced by the errors on the noise level (Fig.~\ref{fig:limdetection}) and on the measured flux of the synthetic planets. We summarize the contributors to the latter below:
\begin{itemize}
\item the photometric stability of the PSF ($\sim$12\%, Sect.~\ref{sec:errorphot});
\item the stability of the PSF FWHM. We evaluate it using the coherent energy estimated by the wavefront sensor: the median variability of the coherent energy is 14\% (range 6--30\%). This translates into FWHM variability of 12\% (median value, range 8--20\%);
\item the uncertainties on the star magnitude and distance (median combined error for the flux 9\%). We assume that the star magnitude is the same in the H band and in the SDI filters;
\item the photometric extraction. We derive the fluxes in apertures of 3$\times$3~pixels (width 52~mas), while the measured FWHM are 57--120~mas (Fig.~\ref{fig:phirseeingecmeant0mean}).
\end{itemize}
We evaluate the impact on the signal-to-noise ratio for an error budget of 20\% on the PSF photometry to $\sim$30\%. This translates in errors on the planet mass of 5--1~\mj ($\sim$200--100~K) for a range of 50--7~\mj (Sects.~\ref{sec:results} and \ref{sec:discussion}).}

\begin{figure}[t]
 	\centering
 	\includegraphics[trim = 9mm 5mm 5mm 10mm, clip, width=.4\textwidth]{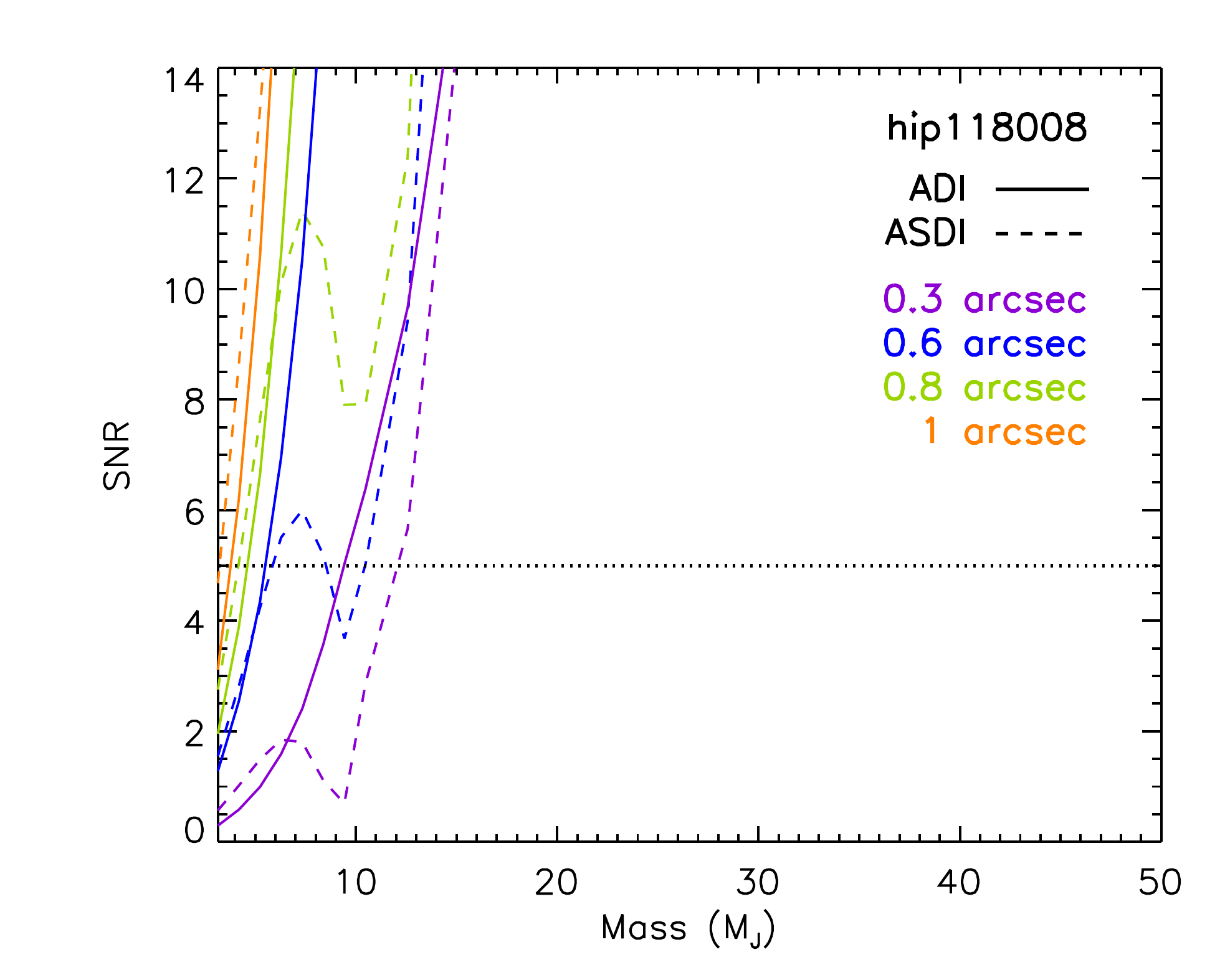}
\caption{Signal-to-noise ratio of synthetic planets measured in the images of HIP~118008 after ADI (solid lines) and ASDI (dashed lines) as a function of mass for several separations. The cut-off at low masses is due to the evolutionary model BT-Settl (see text). The black dotted horizontal line indicates the signal-to-noise ratio used to derive the mass sensitivity limits (see text).}
 	\label{fig:snrmass}
 \end{figure}

For this work, we use the 2012 release of the evolutionary model BT-Settl \citep{Allard2012}. This model simulates the emergent spectrum from {the} atmospheres of giant planets, brown dwarfs, and very low-mass stars for given values of effective temperature, surface gravity, and metallicity\footnote{The synthetic spectra are available at \url{http://phoenix.ens-lyon.fr/Grids/BT-Settl/CIFIST2011/SPECTRA/}.}. It then uses the evolutionary tracks of the COND model \citep{Baraffe2003} to associate the spectrum parameters to {the age and the mass of the object}. With respect to previously published models \citep{Burrows1997, Chabrier2000, Baraffe2003}, BT-Settl accounts for the cloud opacity. We select a subgrid in the model, spanning effective temperatures from $\sim$500 to 4\,000~K. The low cut-off in effective temperature implies that the minimum planet mass available at a given age increases with the latter (from 1~\mj at 10~Myr to 5~\mj at 200~Myr). Although BT-Settl {models atmospheres with} effective temperatures well below the condensation temperature of methane, we reach {the model boundary for half of the targets} (Sect.~\ref{sec:results}). For these cases, we cut the sensitivity limits at this value. The limitation of {the BT-Settl predictions} to temperatures above $\sim$500~K is due to poor knowledge of infrared molecular lines of methane and ammonia at lower temperatures \citep{Allard2012}. Modeling colder atmospheres could be necessary for the data analysis of SPHERE and GPI, since we expect higher contrast performance than for NaCo. We discuss the use of {the BT-Settl evolutionary model and the dependence of the results on this model} in Sect.~\ref{sec:criticalanalysis}.

{We summarize the key conclusions of the whole section.}
\begin{enumerate}
\item {The residual noise measured in SDI-processed images cannot be converted into mass limits in the same way as used for single-band imaging surveys. This {led} us to develop a method based on synthetic planets and flux predictions to directly estimate the mass limits.}
\item {The assessment of the detection limits, as well as the photometric characterization {based on SDI data}, face {degeneracies in} the planet properties. This implies that {interpreting SDI data requires an analysis that couples} ASDI and single-band imaging.}
\item {The spectral overlapping of an off-axis source $\phi(\vec{r})$ is strongly {affected} by the PSF FWHM for large separations ($>$0.5$''$). We thus expect that the PSF FWHM is a parameter to {consider when studying} the ASDI detection limits.}
\end{enumerate}

\section{Results}
\label{sec:results}

The data analysis does not yield any detections. {We discuss below the mass sensitivity limits (Sect.~\ref{sec:detlimmass}). Then, we analyze the limits in effective temperatures (Sect.~\ref{sec:detlimteff})}. We conclude with a detailed study of the ASDI performance (Sect.~\ref{sec:analysesdi}).

\begin{figure*}[t]
	\centering
	\includegraphics[trim = 11.5mm 5mm 5mm 10mm, clip, width=.4\textwidth]{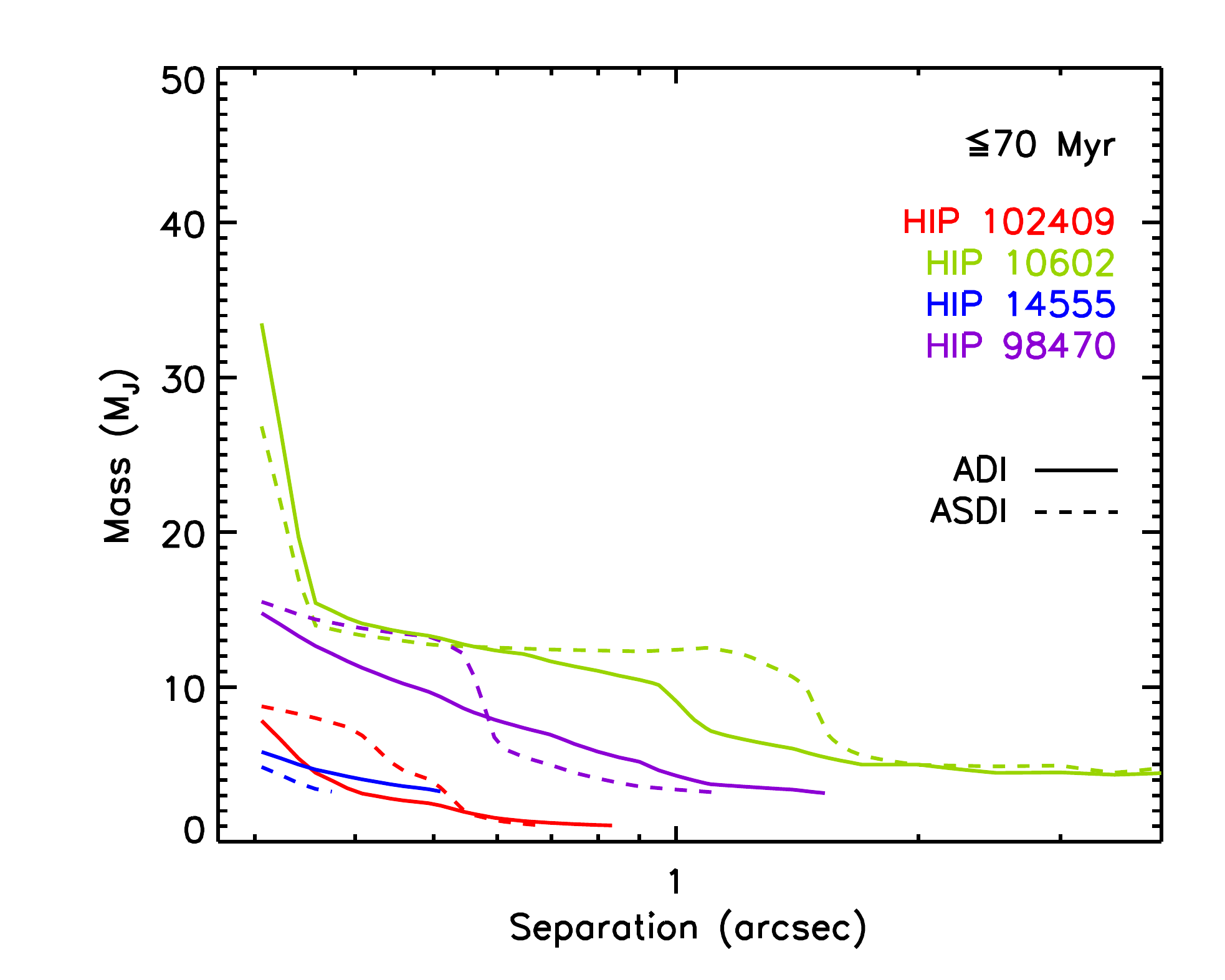}
	\includegraphics[trim = 11.5mm 5mm 5mm 10mm, clip, width=.4\textwidth]{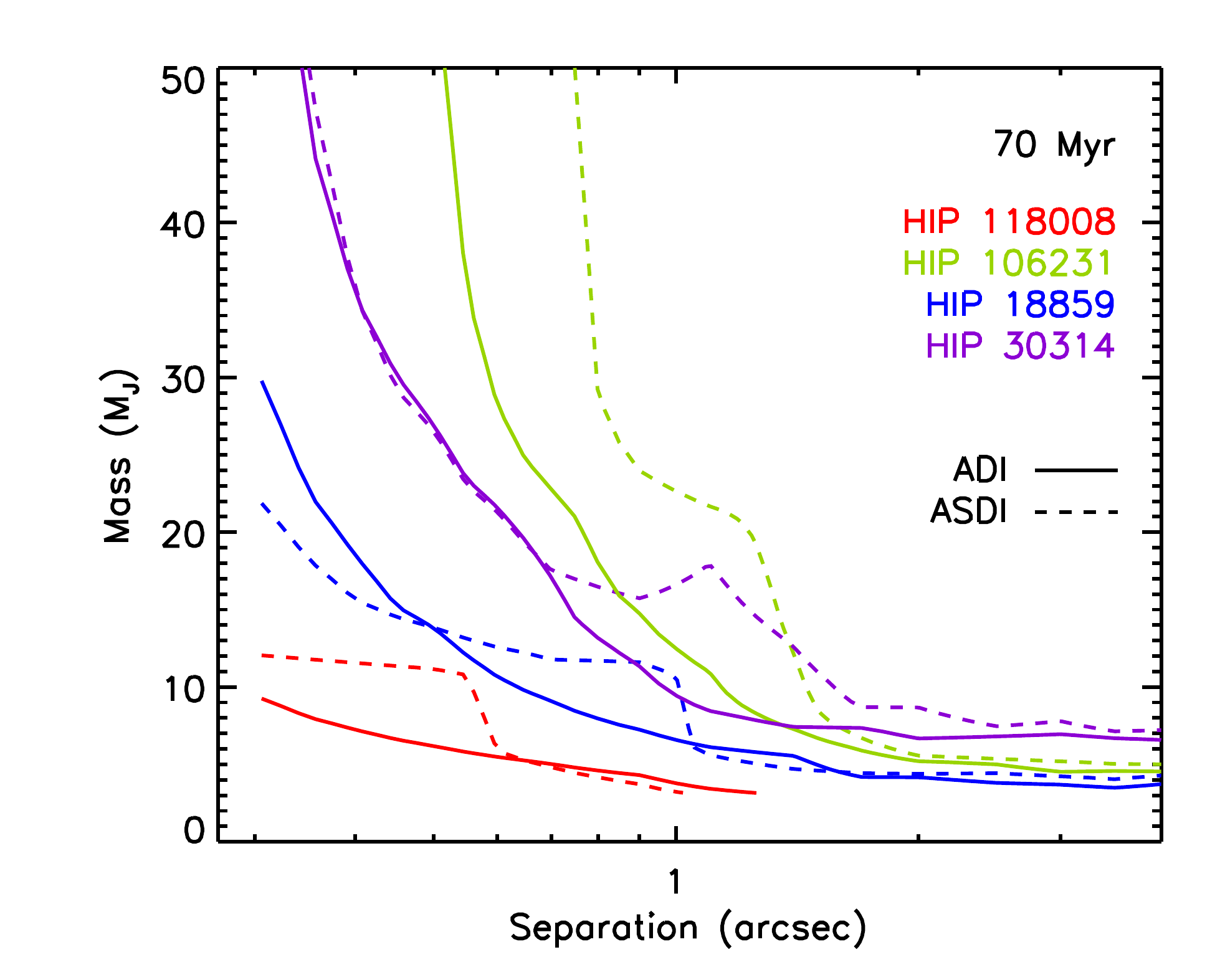}
	\includegraphics[trim = 11.5mm 5mm 5mm 10mm, clip, width=.4\textwidth]{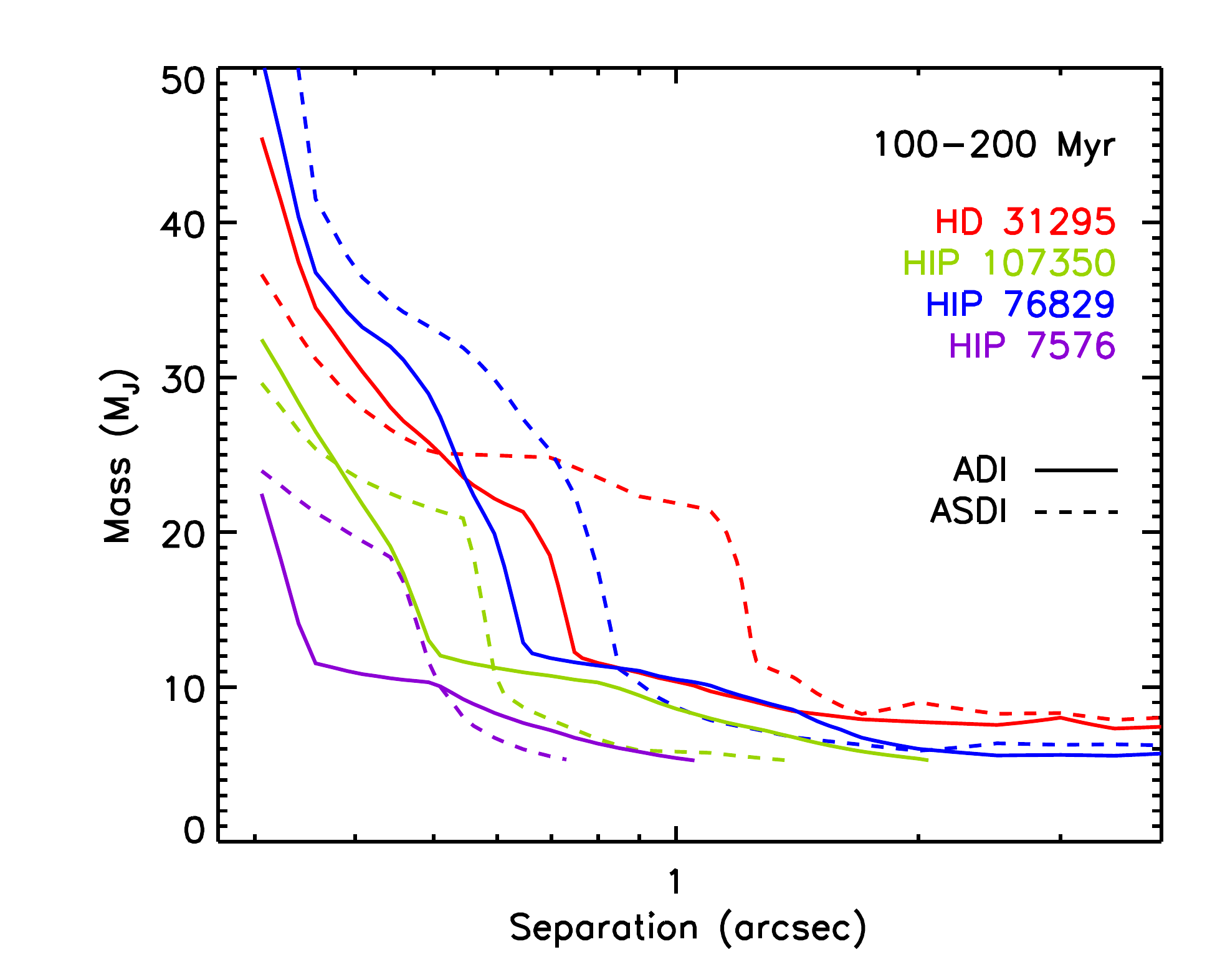}
	\includegraphics[trim = 11.5mm 5mm 5mm 10mm, clip, width=.4\textwidth]{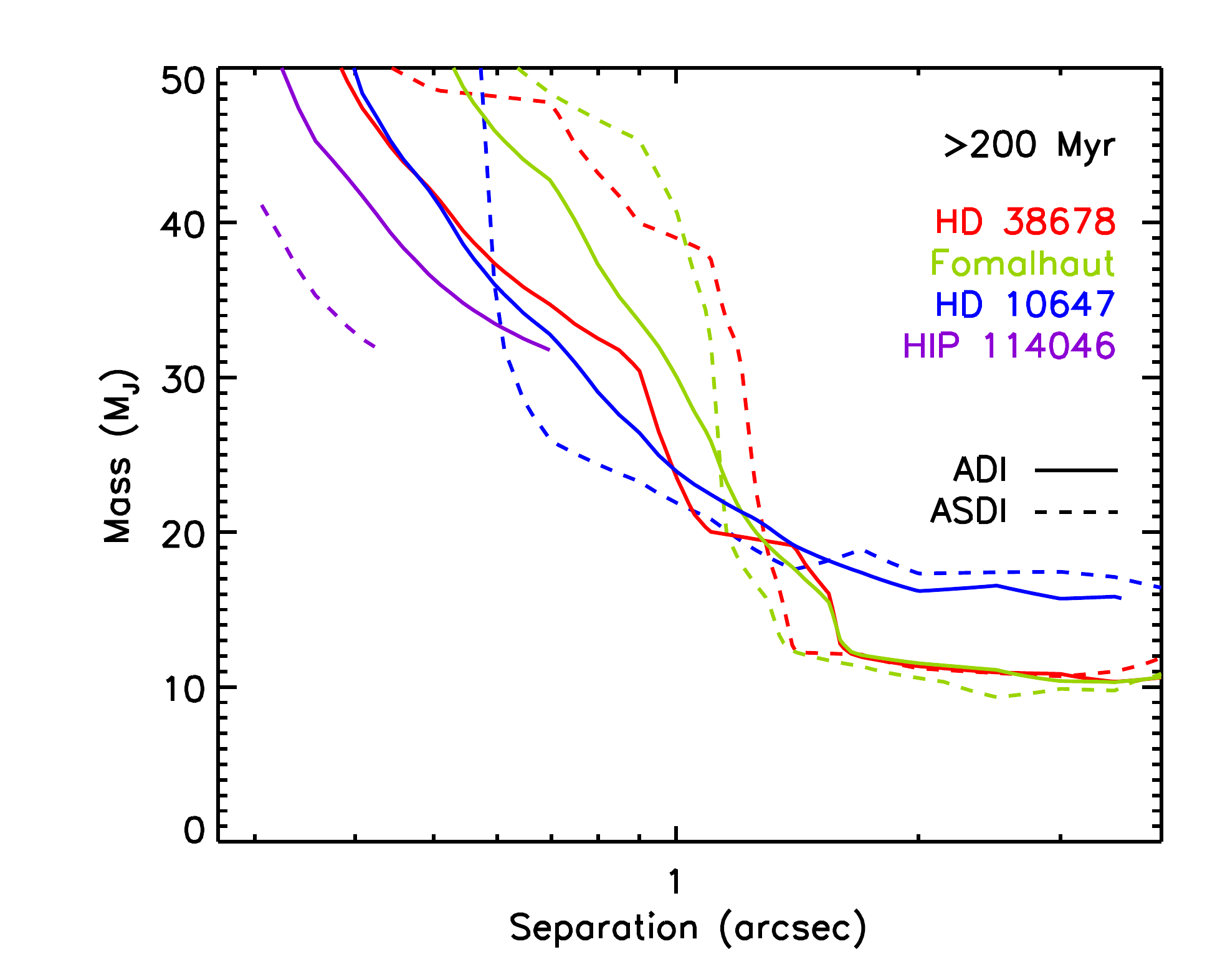}
	\caption{Detection limits in mass of the survey according to the BT-Settl evolutionary model \citep{Allard2012} after ADI applied to the $I_1$ image (solid curves) and after ASDI performed for the \iad subtraction (dashed curves). The stars are sorted by increasing age from left to right and from top to bottom. The curves are cut when the minimum mass covered by the evolutionary model is reached (see text).}
	\label{fig:detlimmass}
\end{figure*}

\subsection{{Mass detection limits}}
\label{sec:detlimmass}

{We represent the mass detection limits derived for ADI as a function of the angular separation in Fig.~\ref{fig:detlimmass}. The stars are sorted by age categories in the panels. As expected, the lowest mass achievable increases with the star age, from 1--4~\mj for stars younger than 70~Myr (top left), to 10--31~\mj for stars older than 200~Myr (bottom right). As mentioned in Sect.~\ref{sec:detectionlimits}, the sensitivity limits are cut when the lower mass available in the evolutionary model is reached. Nevertheless, the performance at small separations ($\lesssim$1$''$) {clearly depends on} other parameters, as we can see in the {top right-hand panel}, which shows stars with the same age\footnote{We note that age uncertainties also {affect} the detection limits, but we do not consider them in this work.}. Good observing conditions (HIP~18859), large parallactic rotations (HIP~118008), and good image dynamic (indicated by the star magnitude and the total exposure time, Tables~\ref{tab:sample} and \ref{tab:obs}) ultimately account for the performance. The median sensitivity limits are {47, 19, and 9~\mj} at 0.3, 0.6, and 1$''$.}

{The ASDI mass limits are shown in Fig.~\ref{fig:detlimmass}. We give some broad tendencies of the ASDI performance based on our data below as well as in Sects.~\ref{sec:detlimteff} and \ref{sec:analysesdi}. They may not be extrapolated to other published SDI data. We do not see any correlation of the gain provided by ASDI with the star age. There is no particular separation range for which the processing improves or degrades the detection limits. The mass gains can be as large as 10\% to 35\%. Beyond 1.5$''$, ASDI degrades the sensitivity since background noise dominates. The exception is Fomalhaut, because of a higher image dynamic (Table~\ref{tab:obs}). For HIP~102409, HIP~30314, and HIP~106231, ASDI gives worse performance with respect to ADI for all separations. The combined ADI-ASDI median limits are {37, 19, and 9~\mj} at 0.3, 0.6, and 1$''$.}

\subsection{Effective temperatures}
\label{sec:detlimteff}

\begin{figure*}[t]
	\centering
	\includegraphics[trim = 9mm 5mm 5mm 10mm, clip, width=.4\textwidth]{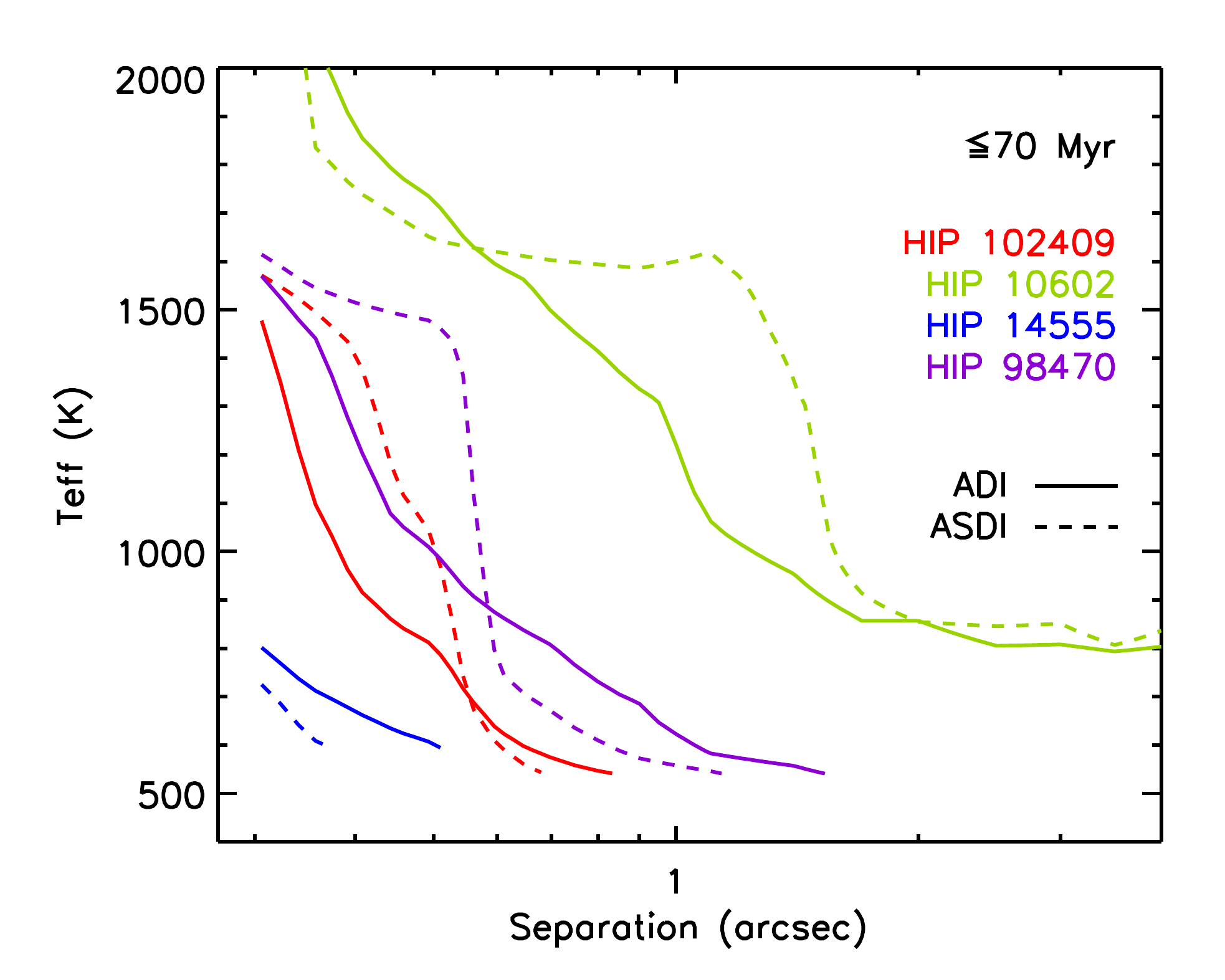}
	\includegraphics[trim = 9mm 5mm 5mm 10mm, clip, width=.4\textwidth]{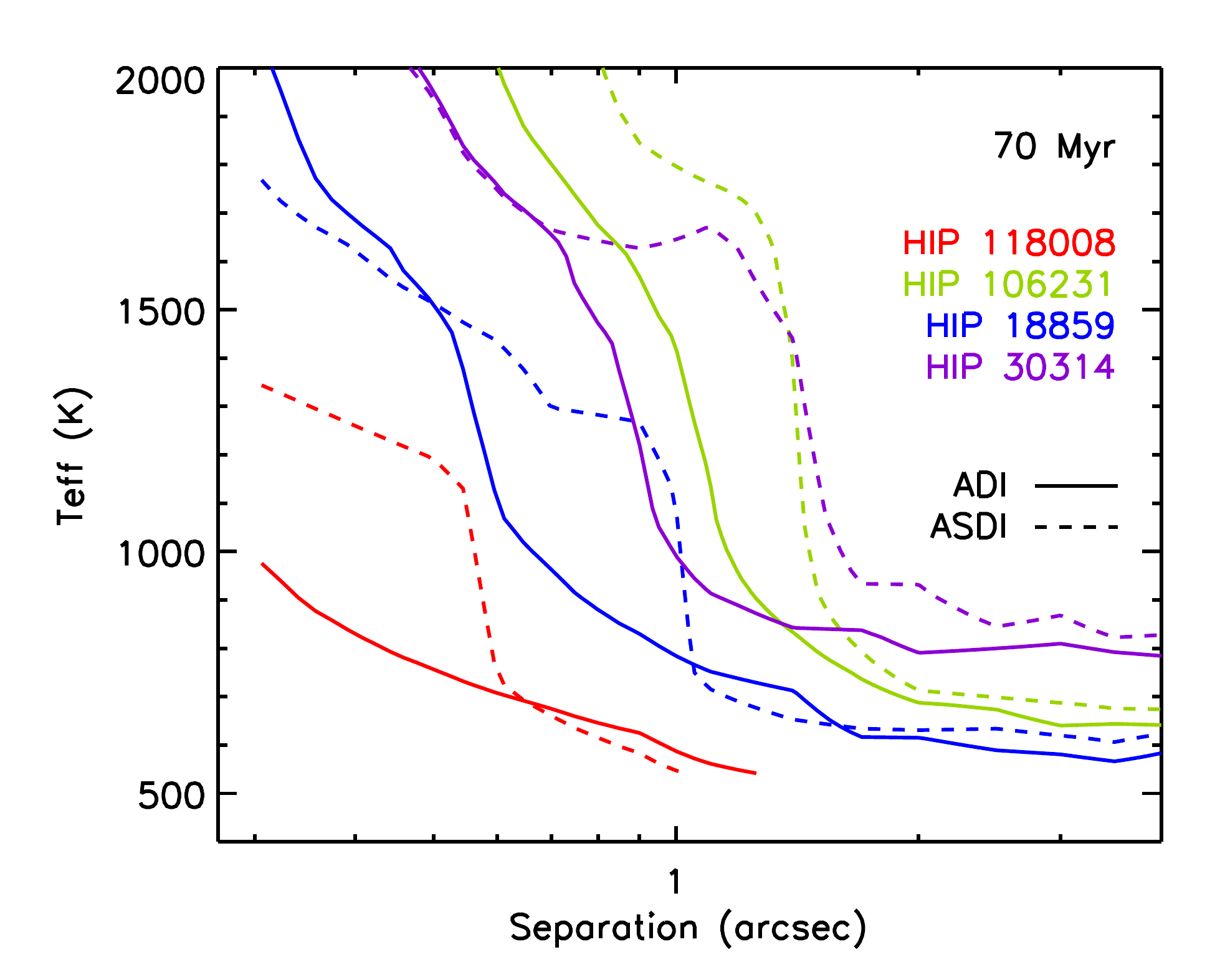}
	\includegraphics[trim = 9mm 5mm 5mm 10mm, clip, width=.4\textwidth]{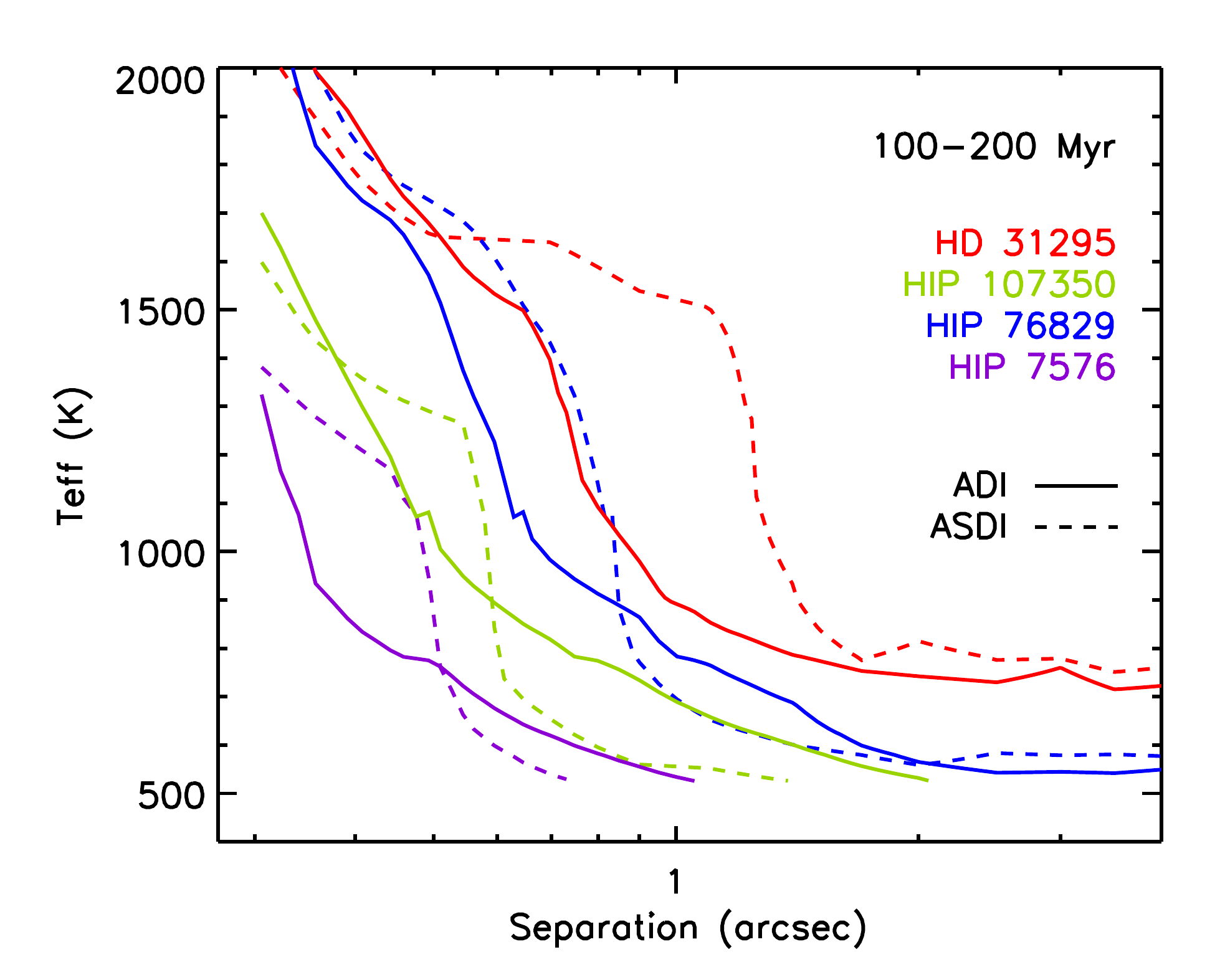}
	\includegraphics[trim = 9mm 5mm 5mm 10mm, clip, width=.4\textwidth]{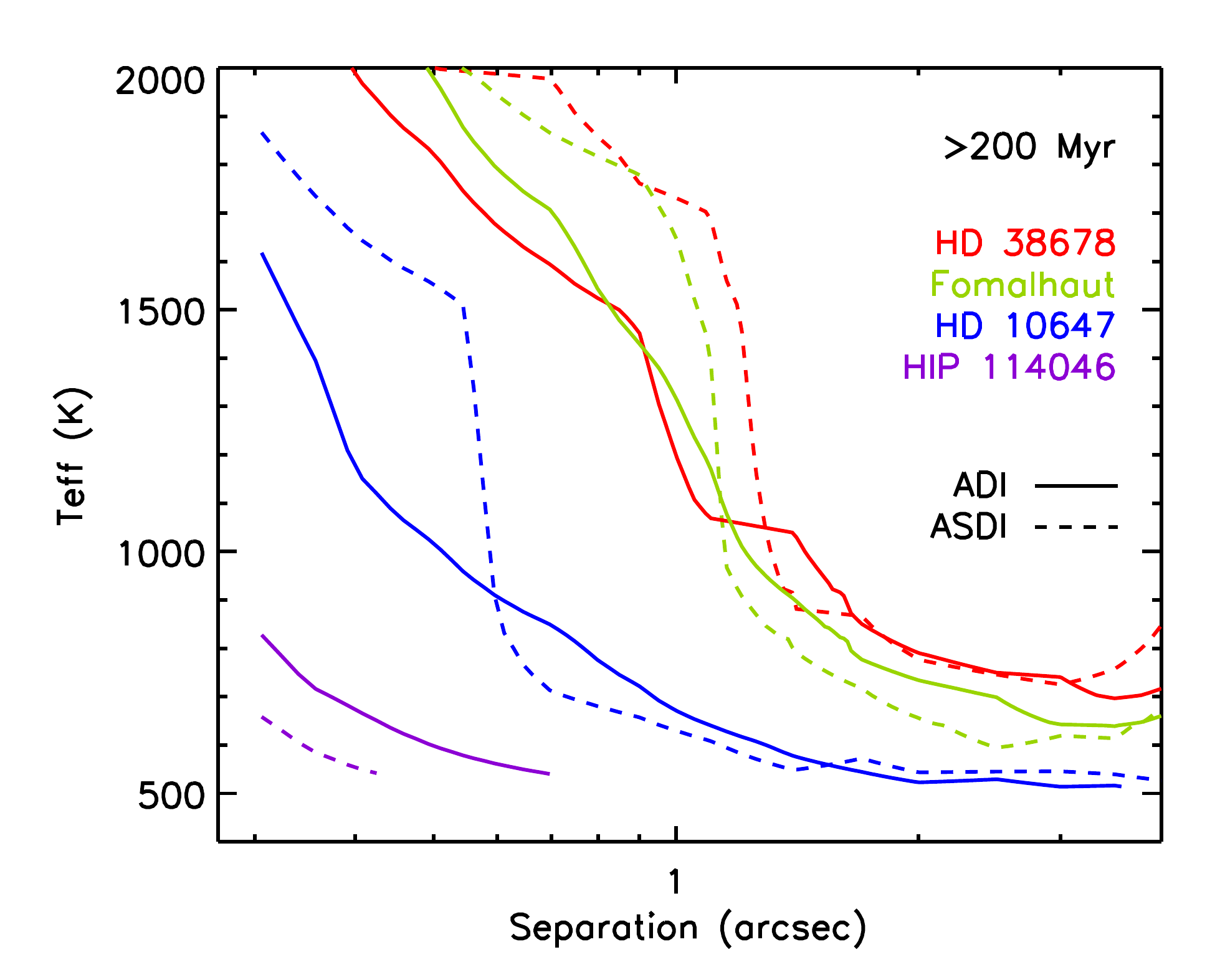}
	\caption{Same as Fig.~\ref{fig:detlimmass} but for the effective temperature. Note the different vertical scale with respect to Fig.~\ref{fig:detlimmass}.}
	\label{fig:detlimteff}
\end{figure*}

A critical parameter {that affects} the self-subtraction of point sources in ASDI is the flux ratio \fluxratio (Fig.~\ref{fig:rfluxteff}). The value of \fluxratio depends strongly on the effective temperature. It is {high} for values below $\sim$1\,000~K, close to 1 for the range 1\,000--1\,500~K, and it decreases for values between 1\,500~K and 2\,100~K. Therefore, we expect a bad performance for ASDI if the {raw noise level (before differential imaging) is not sensitive to objects} colder than 1\,000~K, especially in the range 1\,000--1\,500~K. We show the sensitivity limits for both ADI and ASDI in Fig.~\ref{fig:detlimteff}. We clearly see the degradation in sensitivity of ASDI with respect to ADI as breaks {in} the curves at {separations of {0.3--1.4$''$}}. Considering all the ADI curves, we reach median temperatures of {1\,510--784~K} at 0.5--1$''$. When combining ADI and ASDI, we reach median values of {1\,507--696~K} {in} the same separation range. We note that the upper limit of the range is {nearly} the same without or with including ASDI{, because it does not usually improve the sensitivity below $\sim$0.5$''$.}

\subsection{Analysis of the ASDI performance}
\label{sec:analysesdi}

The purposes of this section are to determine, for our data, the condition(s) for which ASDI improves or degrades the sensitivity with respect to ADI and to understand the behaviors seen in Figs.~\ref{fig:detlimmass} and \ref{fig:detlimteff}. The detection limits are determined by comparing the residual intensity of synthetic planets to the residual quasi static speckles (Sect.~\ref{sec:detectionlimits}). A planet flux is attenuated by ASDI by a quantity determined in part by its spectral properties (presence/absence of methane absorption bands, Fig.~\ref{fig:spectracoldegp}) and $\phi(r)$, which quantifies the geometric self-subtraction and diminishes with the separation (Sect.~\ref{sec:sdisignature}). The quasi static speckles are partially calibrated by ASDI at short separations ($\lesssim$1--2$''$), except for HIP~106231 (Fig.~\ref{fig:limdetection}). {ASDI improves (respectively degrades) the sensitivity if the speckle attenuation is larger (resp. smaller) than the planet self-subtraction.}

We first determine the ASDI speckle attenuation as a function of the ASDI self-subtraction of point sources for three angular separations in Fig.~\ref{fig:analysegainsdi}. The speckle attenuation gains are derived as the ratios of the ADI and ASDI noise levels (Fig.~\ref{fig:limdetection}). The self-subtraction factors are measured from the ratios of the fluxes of synthetic planets measured before and after ASDI (Sect.~\ref{sec:detectionlimits}), thus including both effects of $\phi(r)$ and of the planet spectrum. For information purposes, the points in Fig.~\ref{fig:analysegainsdi} are indicated with different symbols, {according to the mass gain} (Fig.~\ref{fig:detlimmass}, but the curves are not cut to the minimum model mass). While this distinction is consistent with the ratio speckle attenuation/self-subtraction for the two largest separations, it is not always the case for 0.5$''$. {At this separation, the steepness of the detection limits (Fig.~\ref{fig:detlimmass}) and/or the azimuthal structure of the residual speckles may bias the determination of the mass at 5~$\sigma$.} As expected, we observe that both the speckle attenuation and the point-source self-subtraction diminish with the separation. We see in particular that ASDI {affects} the sensitivity limits up to a separation of $\sim$2.5$''$ (Fig.~\ref{fig:detlimmass}).

\begin{figure}[t]
	\centering
	\includegraphics[trim = 7mm 2mm 2mm 8mm, clip, width=.4\textwidth]{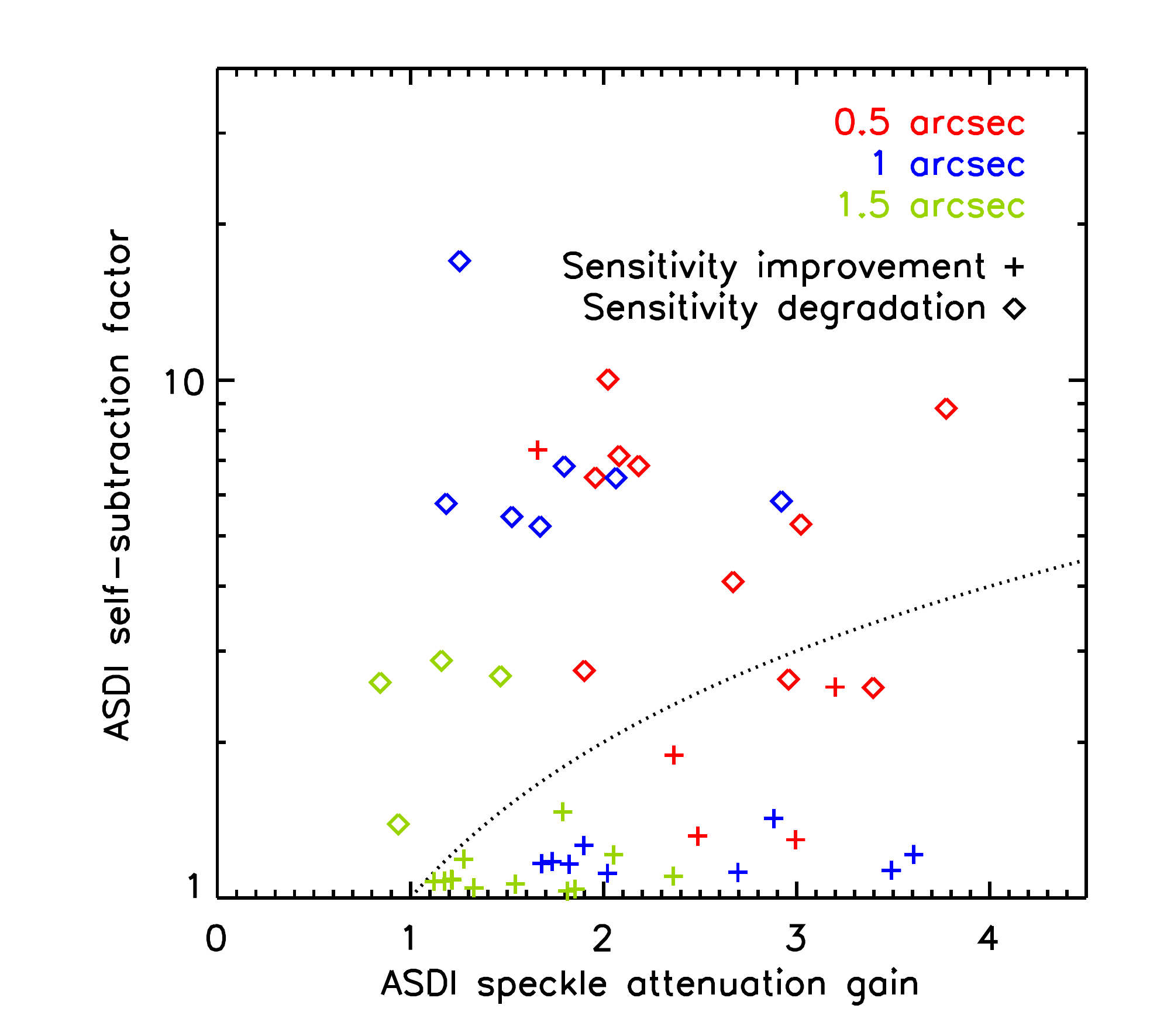}
	\caption{ASDI self-subtraction of point sources as a function of the ASDI speckle attenuation for three angular separations (colors). The points are indicated with different symbols according to the ASDI performance in Fig.~\ref{fig:detlimmass}. {The dotted line indicates where the ASDI speckle attenuation equals the ASDI self-subtraction.} For the sake of clarity, we do not include the point pertaining to HIP~106231 measured at 0.5$''$ (speckle attenuation = 0.9, self-subtraction = 100).}
	\label{fig:analysegainsdi}
\end{figure}

\begin{figure*}[t]
	\centering
	\includegraphics[trim = 11.5mm 5mm 5mm 10mm, clip, width=.4\textwidth]{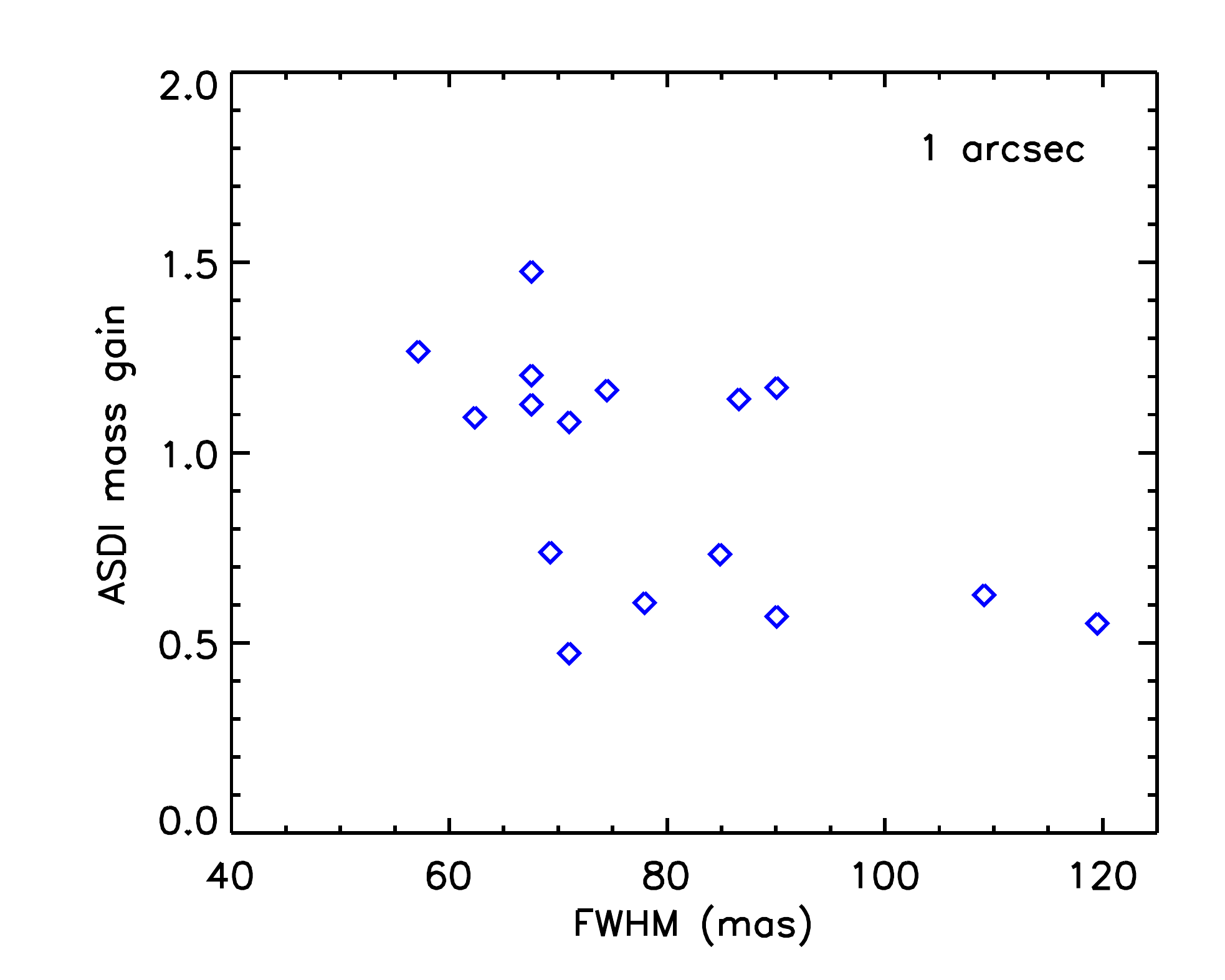}
	\includegraphics[trim = 11.5mm 5mm 5mm 10mm, clip, width=.4\textwidth]{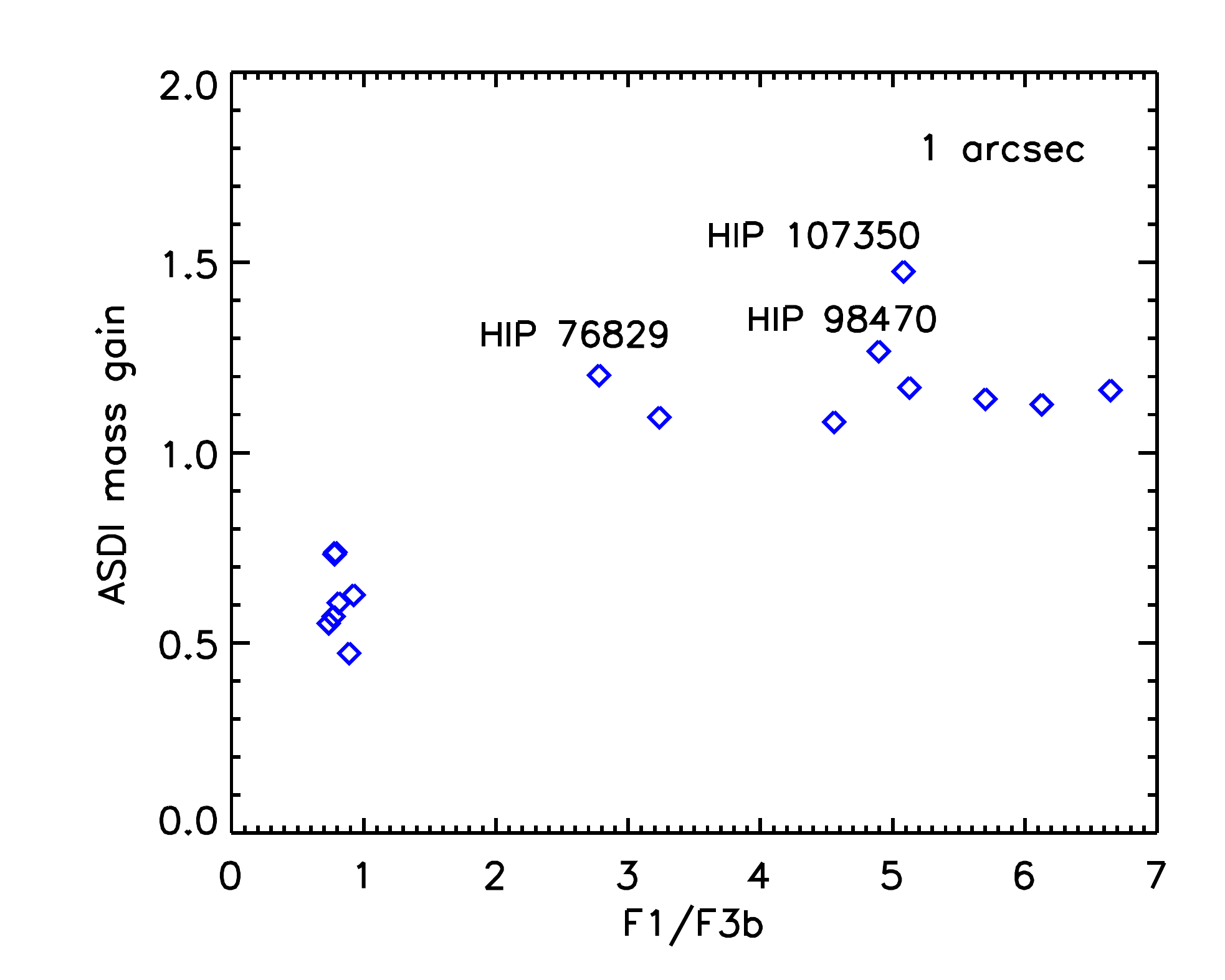}
	\caption{ASDI mass gain as a function of the PSF FWHM (\textit{left}) and the SDI flux ratio of the companion (\textit{right}) for an angular separation of 1$''$. The labels refer to the points with the best mass gains.}
	\label{fig:corrmassgain}
\end{figure*}

{Another} tendency as the separation increases is that the number of stars for which ASDI improves the sensitivity increases ({5, 9, and 12 at 0.5$''$, 1$''$, and 1.5$''$, respectively}). For a given separation closer than $\sim$1$''$, the companion self-subtractions can reach higher values than the speckle attenuations (up to $\sim$10 against $\sim$4), implying that this is the parameter that ultimately drives the ASDI performance.} We confirm this point below. {We briefly discuss the sources of errors on the quantities represented in the plots described in this section. There is no error bar on the speckle attenuation because normalization errors will affect the images $I_1$ and $I_{\rm{3b}}$ in the same way and differential errors will not affect the SDI subtraction because the $I_{\rm{3b}}$ image is rescaled relatively to the $I_1$ image. The ASDI self-subtraction is affected by the PSF errors. The median error is 1. The error on the FWHM due to the estimation methods is 5--7~mas (17~mas for elongated PSF, Sect.~\ref{sec:sdisignature}). The error on $F_1/F_{\rm{3b}}$ is induced by the error on the planet mass and is $\sim$0.5 typically. Finally, the error on the ASDI mass gain stems from the errors on the sensitivity limits, and its median value is $<$0.1.}

{We {conducted} a correlation study of the ASDI speckle attenuation and the ASDI self-subtraction in Appendix~\ref{sec:appendixanalysesdi}. We summarize the results. The ASDI speckle attenuation decreases with the PSF FWHM and the raw noise level (without differential imaging). As for the ASDI self-subtraction, we show that it is mainly correlated to the flux ratio \fluxratio and $\phi(\vec{r})$, but it also depends on the PSF FWHM for high values at large angular separations.}

Finally, we analyze possible tendencies of the ASDI mass sensitivity gain with the FWHM and \fluxratio in Fig.~\ref{fig:corrmassgain} for a separation of 1$''$. We {computed} the mass gains as the ratio of the ASDI and ADI limits (Fig.~\ref{fig:detlimmass}), but without thresholding the curves to the minimum model mass. The mass gains range from 0.5 to 1.5. They no longer exhibit correlations with the FWHM, while they do with \fluxrationospace. The mass gains are smaller than 1 for \fluxrationospace\,$\lesssim$\,2 (T$_{\rm{eff}}$\,$\gtrsim$\,800~K) and greater than 1 above. ASDI provides an improvement {in} the sensitivity for nine stars, for which we derived the lowest point-source self-subtractions (Appendix~\ref{sec:corrselfsub}). The point-source self-subtraction is thus the critical parameter that determines the performance of ASDI, at least at first order. This confirms the conclusion deduced from Fig.~\ref{fig:analysegainsdi}. We note that for the mass gains greater than 1, the values are quite constant ($\sim$1.1--1.2) with \fluxrationospace. Nevertheless, {three points stand out and present the best mass gains {($\geq$1.2)}. The corresponding stars exhibit {some of the} best speckle attenuations (Fig.~\ref{fig:analysegainsdi}, blue points above $\sim$2.7) and have rather good PSF (FWHM\,$\lesssim$\,67~mas) and/or low raw noise level ($\lesssim$4.6$\times$10$^{-5}$). Age also affects the mass gains. HIP~114046 presents the best speckle attenuation (blue point at 3.6), but the smallest mass gain.} As for the data points with mass gains below 1, we see an important vertical dispersion. This feature is accounted for by differences in speckle attenuation, the stars with smaller mass gains also having smaller speckle attenuations. The latter are related to the poorer quality of the PSF and of the raw noise level (Appendix~\ref{sec:corrnoiseatten}).

{To summarize the conclusions of this study:
\begin{enumerate}
\item the planet flux ratio \fluxratio determines whether ASDI improves or degrades the sensitivity at first order for separations closer than $\sim$1$''$;
\item when \fluxratio is favorable, good PSF and noise properties improve the mass gain further.
\end{enumerate}}

\section{Discussion}
\label{sec:discussion}

We first focus on the results for two targets of particular interest (Sect.~\ref{sec:indivstars}). Then, we analyze the {broad characteristics} of the detection limits. Since the observed sample is quite modest and inhomogeneous, we do not intend to carry out a study of the giant planet frequency, as for published large imaging surveys \citep[e.g.,][]{Lafreniere2007b, Chauvin2010, Vigan2012, Rameau2013a, Wahhaj2013a, Nielsen2013, Biller2013}. We summarize {the properties} of the limits in mass (Sect.~\ref{sec:masssep}) and in effective temperature (Sect.~\ref{sec:teffsep}). Then, we discuss the validity of our hypotheses and the dependency of the results on them (Sect.~\ref{sec:criticalanalysis}). Finally, we discuss some implications of this work for the data analysis and interpretation of the upcoming imaging instrument SPHERE (Sect.~\ref{sec:implicationssphere}).

\begin{figure*}[t]
	\centering
	\includegraphics[trim = 11.5mm 5mm 5mm 10mm, clip, width=.4\textwidth]{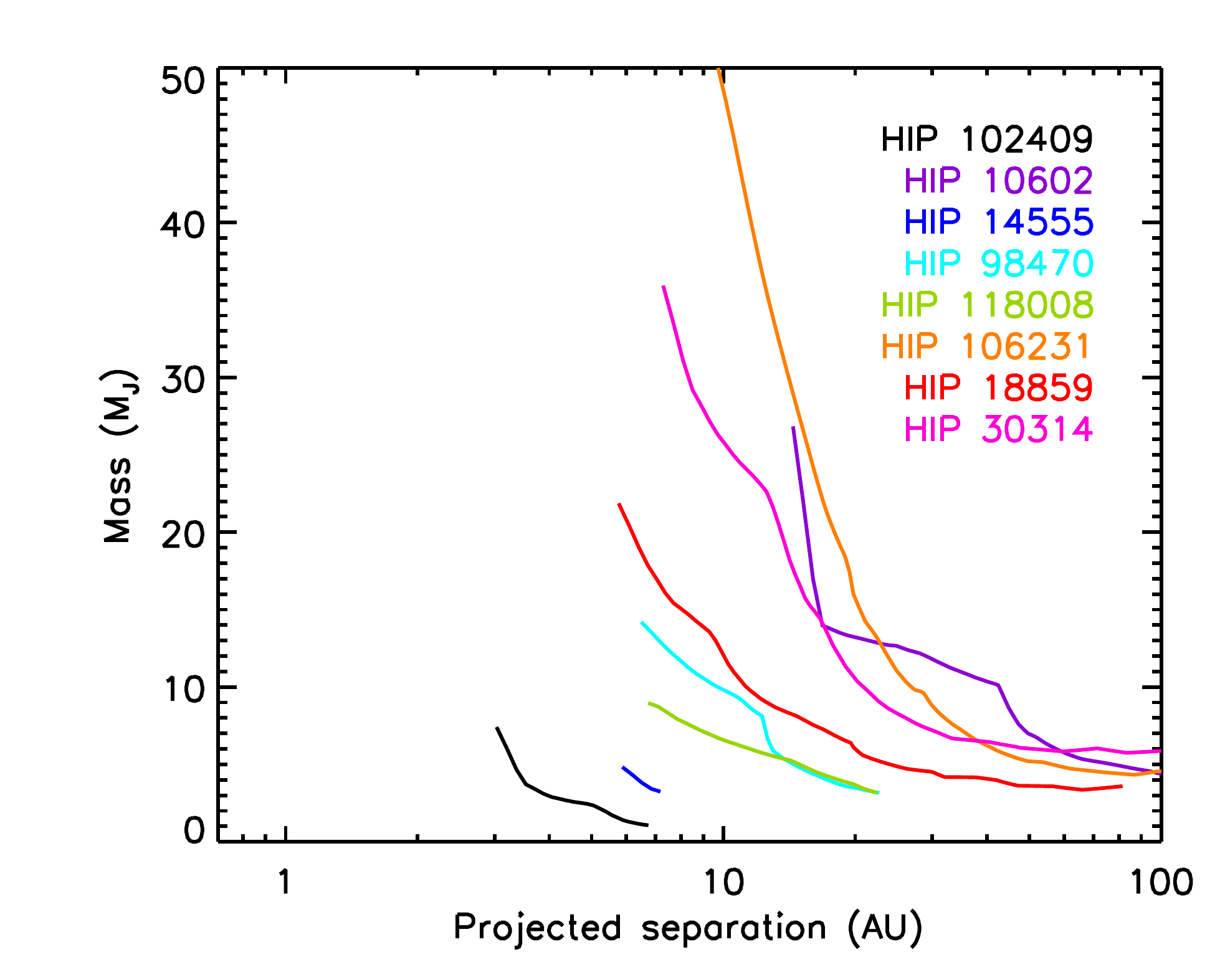}
	\includegraphics[trim = 11.5mm 5mm 5mm 10mm, clip, width=.4\textwidth]{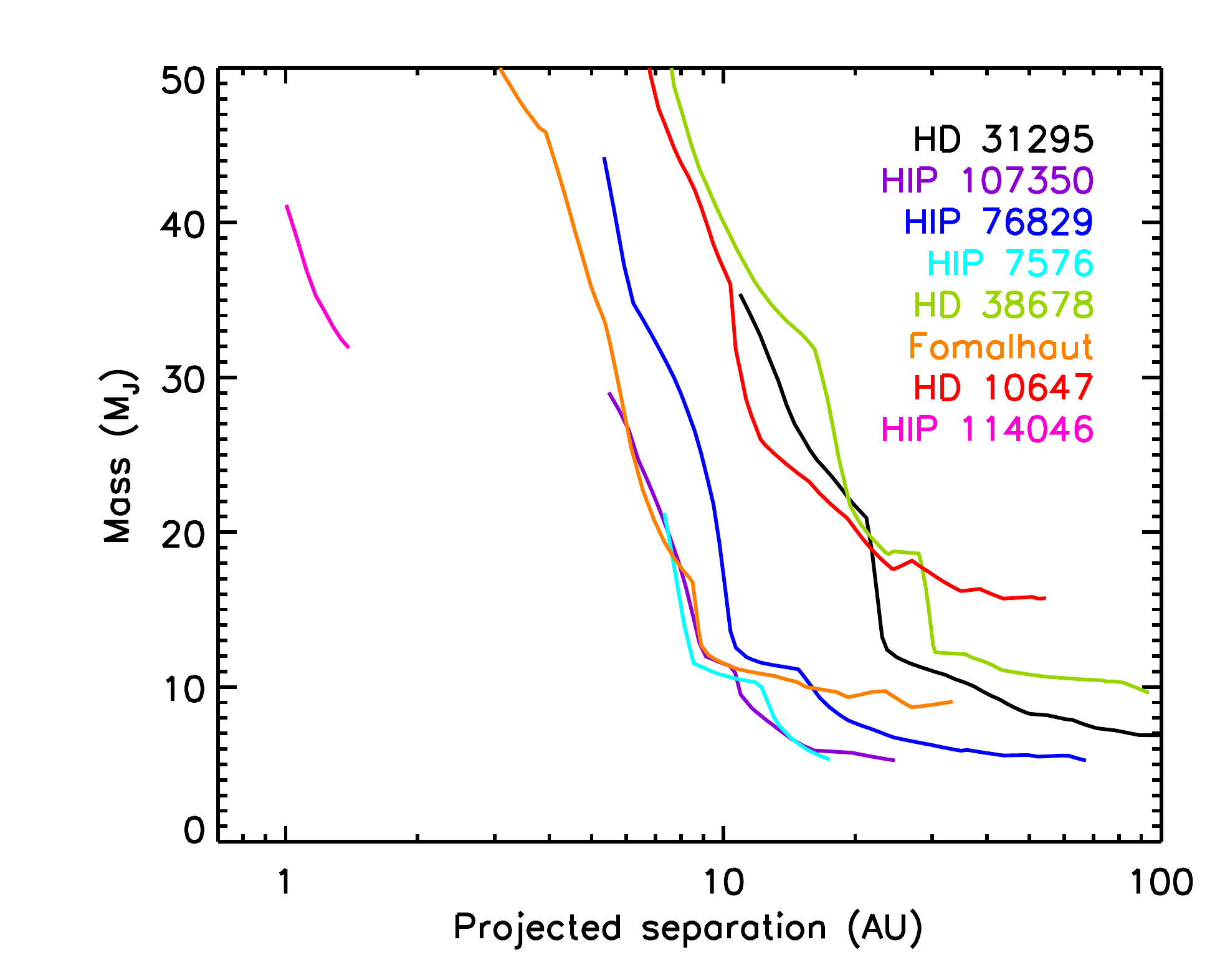}
	\caption{Detection limits in mass using BT-Settl \citep{Allard2012} and combining several differential imaging methods (see text) as a function of the projected separation. The left panel represents the results for the stars younger than 100~Myr and the right panel those for older targets. {The curves are cut according to the maximum projected separation accessible in the SDI field (see text).}}
	\label{fig:detlimmassau}
\end{figure*}

\subsection{Results for individual stars}
\label{sec:indivstars}

\subsubsection{Fomalhaut}
Fomalhaut is known to harbor a complex system, composed in particular of an outer debris disk with a sharp inner edge at 133~AU resolved by \citet{Kalas2005} and a planetary-mass object closer-in on a highly-eccentric orbit \citep{Kalas2008, Kalas2013}. {Because the SDI field is} limited to 4$''$, our images probe the inner part ($\lesssim$30~AU) of the system. Several teams have observed these regions in order to set constraints on the mass of putative planetary companions. {A radial velocity study} excludes companions with masses higher than 6~\mj for separations below 2~AU {\citep{Lagrange2013}}. {The analysis of VLT images at 4.05~$\muup$m does not yield the detection of objects more massive than 30 and 11~\mj at $\sim$2 and 8~AU \citep[COND model,][]{Baraffe2003}, respectively \citep{Kenworthy2013}}. Imaging at 4.7~$\muup$m constrains the mass of putative companions below 4~\mj (2.6~\mjnospace) at 9~AU (30~AU) \citep[][COND model]{Kenworthy2009}. The imaging sensitivity limits are derived assuming the star age from \citet{Mamajek2012}. Our narrow-band observations probe the same separation range as in \citet{Kenworthy2009}, but with a lower sensitivity of 9--12~\mj {based on the same age estimate} (Fig.~\ref{fig:detlimmassau}, right).

\subsubsection{HIP~102409/AU~Mic}
AU Mic is surrounded by an edge-on debris disk discovered by \citet{Kalas2004}. The disk has a parallactic angle of 127$\degb$ and extends to projected separations of 290.7~AU\footnote{\url{www.circumstellardisks.org}.}. Our SDI images are sensitive to regions closer than $\sim$40~AU. Based on HST and/or ground-based observations and an age estimate of 12~Myr\footnote{We note that \citet{Binks2014} reassess the age of the $\beta$~Pic moving group to 21~Myr. This produces an increase of {1--3~\mj in the mass detection limits quoted in this paragraph}}, {several studies} set constraints on the mass of planetary companions. According to the predictions of \citet{Burrows1997}, {HST images presented in \citet{Fitzgerald2007} do not reveal} companions of $>$3~\mj and $>$1~\mj embedded in the disk at separations larger than 10~AU and 30~AU, respectively. Based on the COND model \citep{Baraffe2003}, {VLT/NaCo imaging by \citet{Delorme2012a} excludes} giant planets more massive than 0.6~\mj beyond 20~AU, while {the analysis of Gemini/NICI data discussed in \citet{Wahhaj2013a} rejects} objects of $>$5~\mj and $>$2~\mj at separations $>$3.6~AU and $>$10~AU, respectively. Although the SDI mode of NaCo is not designed for the observation of extended sources, we detect in the SDI images collapsed in wavelength ($I_1$\,+\,$I_2$\,+\,$I_{\rm{3b}}$) and processed with KLIP for several input parameters \citep[][and Sect.~\ref{sec:masssep}]{Soummer2012} the AU Mic debris disk with a poor signal-to-noise ratio per resolution element \citep[method described in][]{Boccaletti2012b} of $\geq$3 for separations of $\sim$1.5--3.7$''$ for the southeast part and $\sim$2.1--3.5$''$ for the northwest part. As for the search for planetary companions, our detection limits extend to projected separations down to {$\sim$3~AU (>7~\mjnospace)} and exclude Jupiter-mass companions beyond 6~AU (Fig.~\ref{fig:detlimmassau}, left) according to the BT-Settl model \citep{Allard2012}. These detection limits are azimuthally averaged on the whole field of view. The limits measured along the disk midplane are not significantly different, {because} the signal-to-noise ratio measured on the disk {is weak}.

\begin{figure*}[t]
	\centering
	\includegraphics[trim = 12mm 5mm 3mm 10mm, clip, width=.4\textwidth]{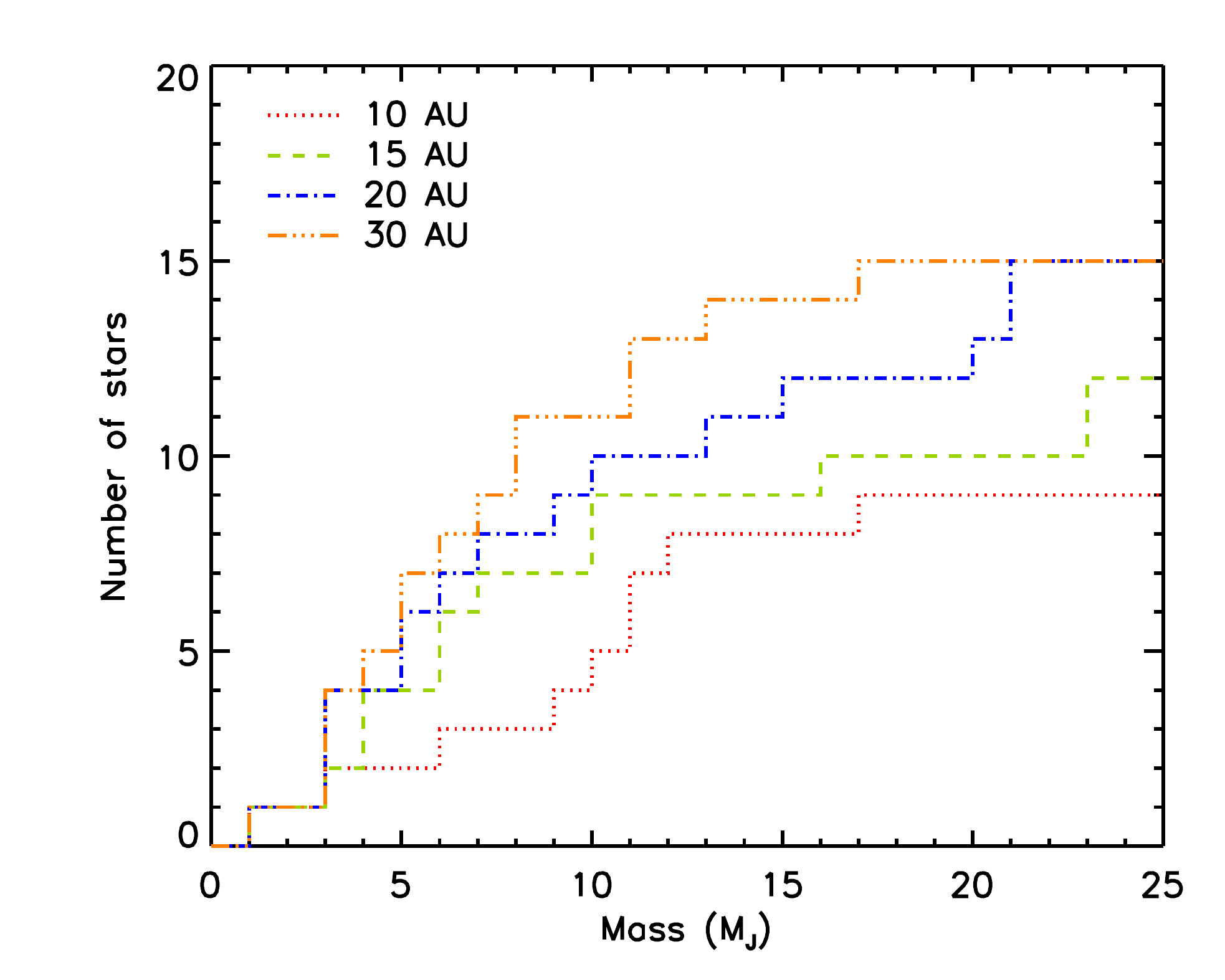}
	\includegraphics[trim = 12mm 5mm 3mm 10mm, clip, width=.4\textwidth]{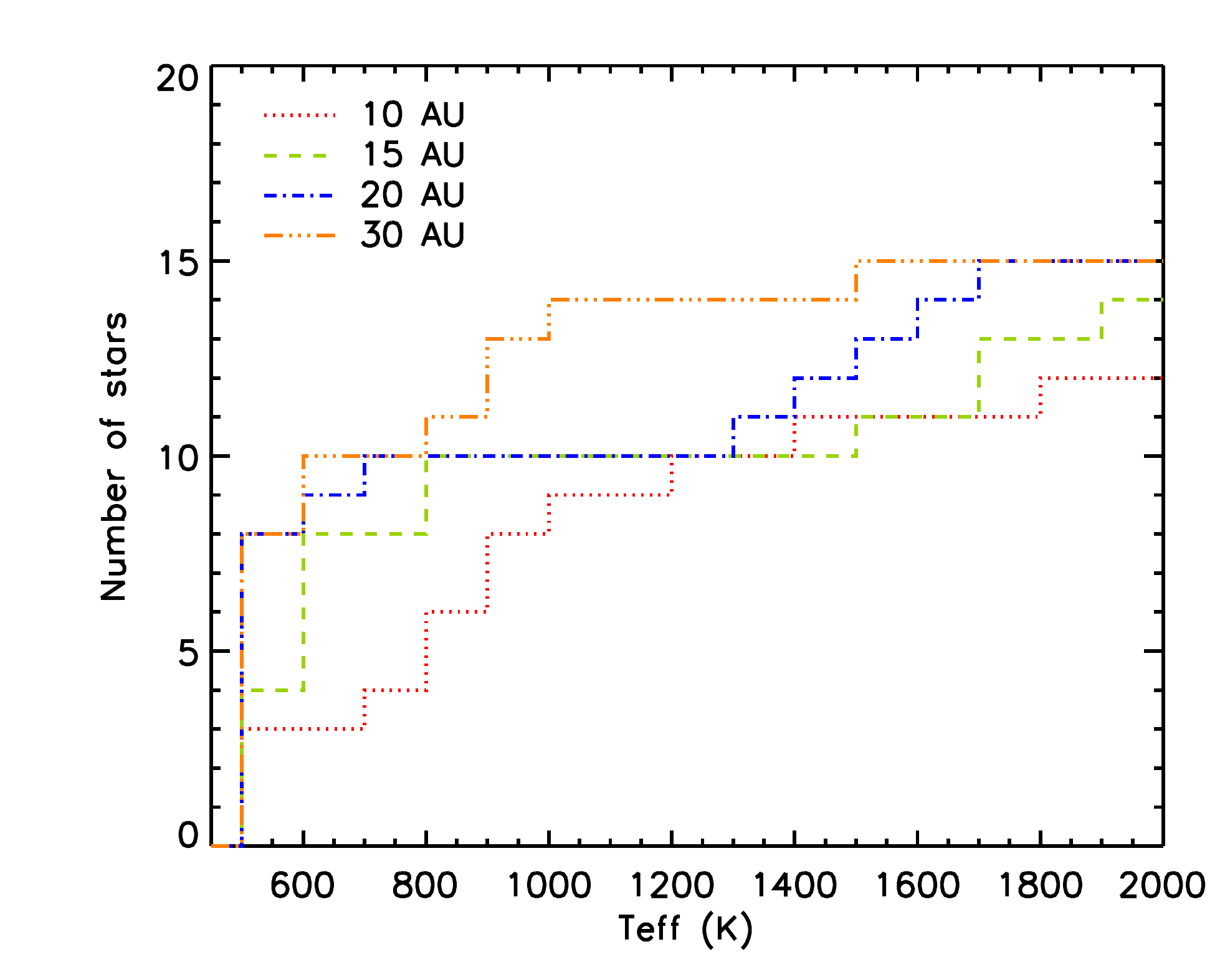}
	\caption{Histograms of the mass (\textit{left}) and effective temperature (\textit{right}) detectable at 5~$\sigma$ (Fig.~\ref{fig:detlimmassau}) for several physical separations. The maximum number of stars is 16 {for 10~AU} and 15 beyond (see text).}
	\label{fig:histograms}
\end{figure*}

\subsection{Characteristics of the mass limits}
\label{sec:masssep}

Figure~\ref{fig:detlimmassau} shows the ultimate mass limits combining several algorithms as a function of the projected separation. We {considered} cADI and KLIP \citep{Soummer2012} applied to the $I_1$ image, cASDI and the combination of KLIP and SDI on the \iad subtraction, as well as cADI on the sum of the images $I_1$, $I_2$, and $I_{\rm{3b}}$ (Fig.~\ref{fig:sdiimages}, left). For each separation, we {selected} the best limit between the algorithms. We applied KLIP to improve the sensitivity at close-in separations ($<$0.5$''$). The number of modes truncated for {constructing the} reference images of the stellar speckles is a free parameter of KLIP \citep{Soummer2012}. After some trials, we set the number of modes to one for all stars, except for HIP~98470, HD~38678, and Fomalhaut, for which we {chose} to retain two, three, and five modes, respectively. To estimate the flux of the synthetic planets after the processing, we first used KLIP on the datacubes containing the data and the planets to derive the principal components for a given number of modes{. Then we performed} KLIP with these components on the datacubes containing only the planets. We cut the curves at the minimum model mass. We note that certain curves are limited in projected separations due to the size of the SDI field ($\sim$4$''$ in radius). This especially concerns two stars, HIP~114046 ($\sim$13~AU, but the mass cut-off is reached before) and Fomalhaut ($\sim$30~AU).

We represent the histogram of the detectable masses for several projected separations in the {left-hand} panel of Fig.~\ref{fig:histograms}. The prime objective of our survey was to search for giant planets at separations as close as $\sim$5--10~AU. Considering an upper mass limit of 13~\mj \citep{Burrows1997}, we observe that eight targets (55\% of the sample) satisfy this objective (Fig.~\ref{fig:detlimmassau}). {We did not take HIP~114046 and HD~10647 into account} because of the cut-offs. The limit of 13~\mj is achieved for all targets beyond $\sim$30~AU. {The lowest mass achieved is 1~\mj beyond 6~AU around HIP~102409. This star combines several favorable properties, a young age, a late spectral type, and a close distance (Table~\ref{tab:sample}), although the observing conditions are not optimal (Table~\ref{tab:obs}).}

\subsection{Effective temperature vs. projected separation}
\label{sec:teffsep}

The histogram of the temperatures as a function of the physical separation is shown in the {right-hand} panel of Fig.~\ref{fig:histograms}. As for the detectable masses, we cut the detection limits (Fig.~\ref{fig:detlimteff}). The survey is sensitive to companions as cool as 1\,000~K around 65\% of the targets for projected separations larger than 10~AU. The optimal regime for ASDI (T$_{\rm{eff}}$\,$\lesssim$\,800~K, Fig.~\ref{fig:rfluxteff} and Sect.~\ref{sec:analysesdi}) is accessible for two-thirds of the sample at separations beyond $\sim$20~AU.

\subsection{{Dependency of the results on the hypotheses}}
\label{sec:criticalanalysis}

{We focus here on the relevance of the hypotheses used for this work and their influence on the results presented in Sect.~\ref{sec:results}. {We recall that the objective of this paper is not to address the unknowns inherent to the considered models, but to outline the difficulties for deriving mass limits from SDI images {regardless of} the models. We decided to pick one of the available models, so the mass limit presented here should not be considered as absolute values.} We based the study on BT-Settl \citep{Allard2012} because the model temperature range spans values below the methane CH$_4$ condensation temperature, suitable for the search for cool giant planets. This model has {the strong point of modeling} non-equilibrium chemistry, {which might be} a key ingredient for understanding the spectral properties of cool young giant planets \citep{Barman2011a, Barman2011b}. However, it also has caveats, in particular, it is unable to correctly reproduce the transition between M and L dwarfs \citep{Bonnefoy2013b, Manjavacas2013}, and it lacks of opacity in the CH$_4$ band at $\sim$1.6~$\muup$m \citep{King2010, Vigan2012}. {We note that the second point is a problem for all current models of planet spectra, because of the incompleteness of the molecular line lists \citep{Allard2012}.}}

{The hypothesis used for the relation between the SDI flux ratio \fluxratio and the effective temperature of substellar companions is critical. Indeed, we showed that \fluxratio is the parameter that determines {the ASDI performance at close-in separations to first order} (Sect.~\ref{sec:analysesdi}). We considered the BT-Settl relation (Fig.~\ref{fig:rfluxteff}), which could be pessimistic due to the lack of CH$_4$ absorption at $\sim$1.6~$\muup$m. Another possibility is to {model} the relation on the spectra of observed cool brown dwarfs, as {described in \citet{Wahhaj2013b} in the context of the data interpretation of the NICI Campaign}. For \teffnospace\,$\sim$\,750~K, they find \fluxrationospace\,$\sim$\,8 based on the spectrum of 2MASS J04151954-093506 \citep[][]{Knapp2004, Burgasser2006}. BT-Settl gives a value of $\sim$2.1\footnote{The NICI filters are at the wavelengths 1.578 and 1.652~$\muup$m and are larger than those of NaCo \citep[$\Delta \lambda/\lambda$\,$\sim$\,4\%,][]{Wahhaj2013b}. Nonetheless, the spectral differences are quite small (Fig.~\ref{fig:rfluxteff}).} for the same temperature (Fig.~\ref{fig:rfluxteff}). {Because $\phi(\vec{r})$ is} larger for NaCo than for NICI ($r_{\rm{b}}$\,$\sim$\,1.3$''$ vs. $\sim$0.9$''$, Eq.~(\ref{eq:rb})), this results in larger ASDI self-subtractions and worse sensitivity when reaching this temperature at close-in ($\lesssim$1$''$) separations for the NaCo data (Fig.~\ref{fig:detlimteff}). Nevertheless, and as noticed in \citet{Nielsen2013} and \citet{Biller2013}, we note that strong CH$_4$ absorptions are measured in the atmospheres of brown dwarfs at temperatures for which we detect little absorptions in those of cool young giant planets \citep{Barman2011b, Konopacky2013, Oppenheimer2013}. The only known young giant planet with measured strong CH$_4$ absorption to date is GJ~504~b \citep[][\teffnospace\,$\sim$\,510~K]{Kuzuhara2013, Janson2013}. Using the photometry reported in \citet{Janson2013}, we derive \fluxrationospace\,$>$\,2.6 (the object is not detected in the CH$_4$ band), whereas BT-Settl predicts \fluxrationospace\,$\gtrsim$\,5\footnote{The wavelengths of the HiCIAO filters are 1.557 and 1.716~$\muup$m \citep[$\Delta \lambda/\lambda$\,$\sim$\,9\%,][]{Janson2013}.} (Fig.~\ref{fig:rfluxteff}). We also compare the predictions to other cool (\teffnospace\,$\lesssim$\,1\,000~K) and young (or proposed as young) {objects}, HR~8799~b \citep{Barman2011a} and three T dwarfs \citep[HN~Peg~B, ROSS~458~C, CFBDSIR2149-0403,][]{Luhman2007, Burningham2011, Delorme2012b}. We find close agreement (factors $\lesssim$1.2) for the objects with \teffnospace\,$\gtrsim$\,1\,000~K. However, the two cool (\teffnospace\,$\sim$\,700~K) T dwarfs ROSS~458~C and CFBDSIR2149-0403 show stronger absorptions than predicted by BT-Settl (factors $\sim$2.4). We estimate the maximum values for the differences in magnitude and mass between the BT-Settl predictions and the observed cool brown dwarfs. This corresponds to close-in separations ($\lesssim$0.5$''$), where $\phi(r)$\,$\sim$\,1 (Fig.~\ref{fig:phirseeingecmeant0mean}). We find $\sim$0.85, $\sim$0.5, and $\sim$0.45~mag for the spectral types T4, T6, and T8, respectively (\teffnospace\,$\sim$1\,000--700~K). Using BT-Settl isochrones, these values translate into underestimations for our ASDI detection limits with respect to the use of observed brown dwarfs of $\sim$1--2~\mj for an age of 70--200~Myr (Figs.~\ref{fig:detlimmass} and \ref{fig:detlimteff}). For larger separations, these estimates are decreased because $\phi(r)$ is smaller.}

{To conclude this analysis, we outline that the interpretation of SDI data is model-dependent, as it is when assessing sensitivity limits in terms of planet parameters. In particular, the ASDI self-subtractions (Fig.~\ref{fig:analysegainsdi} and Appendix~\ref{sec:corrselfsub}), ASDI mass gains (Fig.~\ref{fig:corrmassgain}), and optimal \teff range for ASDI ($\lesssim$800~K) are derived for the BT-Settl model. However, the flux ratio \fluxratio optimal for ASDI ($\gtrsim$2) is not model-dependent when taken on its own.}

\subsection{Data reduction and analysis of SPHERE}
\label{sec:implicationssphere}

We outline in this paper a few points relevant to the analysis and interpretation of the SPHERE data. In particular, we advocate the need to analyze SDI data with both ASDI and ADI {to remove} the mass degeneracies with the measured flux inherent to ASDI (Sects.~\ref{sec:degendifffluxmass} and \ref{sec:detectionlimits}) and to improve the sensitivity when \teffnospace\,$\gtrsim$\,800~K (Sects.~\ref{sec:detlimteff} and \ref{sec:analysesdi}). Indeed, the main observing mode of SPHERE, NIRSUR, combines coronagraphy, ADI, and SDI \citep{Beuzit2008}. NIRSUR has been optimized for the search for young giant planets. It benefits from simultaneous observations with the dual-band imager IRDIS \citep{Dohlen2008a} and the integral field spectrometer IFS \citep{Claudi2008} to {distinguish} physical companions and uncorrected stellar speckles.

The methods described in this paper can be applied in a straightforward way to the data analysis of IRDIS{, which} offers five pairs of filters for SDI over the YJHKs spectral domain. \citet{Vigan2010} have studied the science performance of this instrument for the characterization of young giant planets. They determined that the H2H3 filter pair is the most suitable for this task when no a priori information on the object is available. The wavelengths of this filter pair are $\lambda_1$\,=\,1.587~$\muup$m and $\lambda_2$\,=\,1.667~$\muup$m. The separation between the filters is larger than for the SDI filters of NaCo (80~nm against 50~nm), resulting in a smaller bifurcation point (Eq.~(\ref{eq:rb})) of $\sim$0.8$''$ (vs. $\sim$1.3$''$). Considering this point and that IRDIS will deliver H-band images very close to the diffraction-limited regime, we thus expect that the self-subtraction of an off-axis companion for a given angular separation will be smaller for IRDIS than for NaCo. This would result in better performance of the ASDI mode of IRDIS at closer separations. This could imply for certain cases that the degeneracies in planet mass with the differential flux may not be {removed} when comparing ASDI limits to single-band sensitivity\footnote{For this kind of analysis, it could be useful to consider the narrow-band images outside the CH$_4$ band for ADI, especially if the dominant source of noise is speckle noise. Indeed, companions colder than $\sim$1\,000~K are brighter at these wavelengths than in H band, because of the CH$_4$ absorption longward $\sim$1.6~$\muup$m (Fig.~\ref{fig:spectracoldegp}). For the BT-Settl model, this translates into a difference of $\sim$2~\mj for ages younger than 200 Myr.}. While this will {affect} the detection and characterization of planetary-mass companions around individual targets, we note that it will not be a critical problem for statistical analyses.

{As for the IFS, we retrieve some aspects similar to the analysis of SDI data, despite different approaches to the speckle subtraction. SDI is based on assumptions on the spectral properties of planets (Sect.~\ref{sec:intro}), whereas IFS data analysis {takes advantage of the chromatic behavior of the speckles \citep{Sparks2002, Crepp2011, Pueyo2012}}. In both cases, the residual planet signature is affected by spectral overlapping and self-subtraction. However, in the IFS case, the self-subtraction depends on the planet intensity with respect to the speckles {\citep{Pueyo2012}}. The detection limits can be estimated using synthetic planets and model fluxes, as employed for this work.}

\section{Conclusions}

We presented the outcome of a high-contrast imaging survey of 16 stars combining the coronagraphic, the spectral differential imaging, and the angular differential {imaging modes} of VLT/NaCo. We did not detect any companion candidates in the reduced images. {We analyzed the sensitivity limits taking {the SDI photometric bias into account} and determined the optimal conditions in terms of SDI performance. The key results are
\begin{itemize}
\item {By} combining the best detection limits (ADI, ASDI), the survey is sensitive to cool giant planets (T$_{\rm{eff}}$\,$<$\,1\,000--1\,300~K) at projected separations $>$10~AU for 65\% of the sample and $>$30~AU for all targets. We are able to probe the range optimal for ASDI (\teffnospace\,$\lesssim$\,800~K according to the BT-Settl model) for two-thirds of the targets {beyond $\sim$}20~AU. Evolutionary models spanning temperatures $\lesssim$500~K would be needed for the data interpretation of SPHERE and GPI, since we reach this limit for half of the targets.
\item {Determination} of the detection limits in ASDI-processed images requires a different analysis than for images processed in single-band differential imaging. In particular, the residual noise level cannot be converted into planet mass through evolutionary models {({regarless of} the considered models), since} it represents a differential flux. This differential flux has to be corrected from the self-subtraction produced by SDI, which depends on the spectral properties assumed for the detectable companions. Thus, detection limits in SDI data should only be considered in terms of physical properties of companions. To derive the detection limits of our survey, we used the signal-to-noise ratio of synthetic planets introduced in the raw data and processed{, as well as the flux predictions of the BT-Settl model}.
\item {The} SDI differential flux may be reproduced by several flux couples, hence planet masses. Consequently, the data should also be processed with single-band differential imaging methods to {remove} the degeneracies.
\item ASDI can either improve or degrade the sensitivity {regardless of} the angular separation and the star age. For the favorable cases, the gains in detectable planet mass can be as {much as 10\% to 35\%}. The parameter that determines to first order the performance is the SDI flux ratio of the companion. This parameter reaches optimal values when it is $\gtrsim$2 for separations $\lesssim$1$''$. Using BT-Settl, this translates into \teffnospace\,$\sim$\,800~K, which is significantly lower than the methane condensation temperature (1\,300~K). The PSF quality is a second-order factor that modulates the ASDI gain.
\end{itemize}}

We finally discussed some implications of this work for the data analysis of SPHERE. We expect better performance in particular for ASDI with the dual-band imager IRDIS with respect to NaCo, thanks to more favorable filter characteristics and an extreme-AO system. We envision {a future project of applying} the methods that we developed for this paper {for analyzing} laboratory and commissioning data of IRDIS and IFS.

\begin{acknowledgements}
We thank the ESO Paranal staff for support during the visitor-mode observations and for performing the service-mode observations. {We also thank our anonymous
referee for constructive report that improved the readability of the paper.} Part of this work was carried out during A.-L.M.'s PhD thesis, supported through a fellowship from the Minist\`ere de l'\'Education Nationale, de la Recherche et de la Technologie. J.R., G.C., and A.-M.L. are supported by the French National Research Agency (ANR) through project grant ANR10-BLANC0504-01. S.D. acknowledges support from PRIN-INAF 2010 ``Planetary Systems at Young Ages''. We thank Ben Burningham and Philippe Delorme for sending us the spectra of ROSS~458C and CFBDSIR2149-0403. This research has benefitted from the SpeX Prism Spectral Libraries, maintained by Adam Burgasser at http://pono.ucsd.edu/~adam/browndwarfs/spexprism. This research has made use of the SIMBAD database, operated at the CDS, Strasbourg, France. This publication makes use of data products from the Two Micron All Sky Survey, which is a joint project of the University of Massachusetts and the Infrared Processing and Analysis Center/California Institute of Technology, funded by the National Aeronautics and Space Administration and the National Science Foundation. This research made use of the VizieR catalog access tool, the CDS, Strasbourg, France. The original description of the VizieR service was published in A\&AS 143, 23.
\end{acknowledgements}

\bibliographystyle{aa}
\bibliography{biblio}

\begin{appendix}

\section{Study of the speckle attenuation and point-source self-subtraction induced by ASDI}
\label{sec:appendixanalysesdi}

\begin{figure*}[!t]
	\centering
	\includegraphics[trim = 11.5mm 5mm 5mm 10mm, clip, width=.4\textwidth]{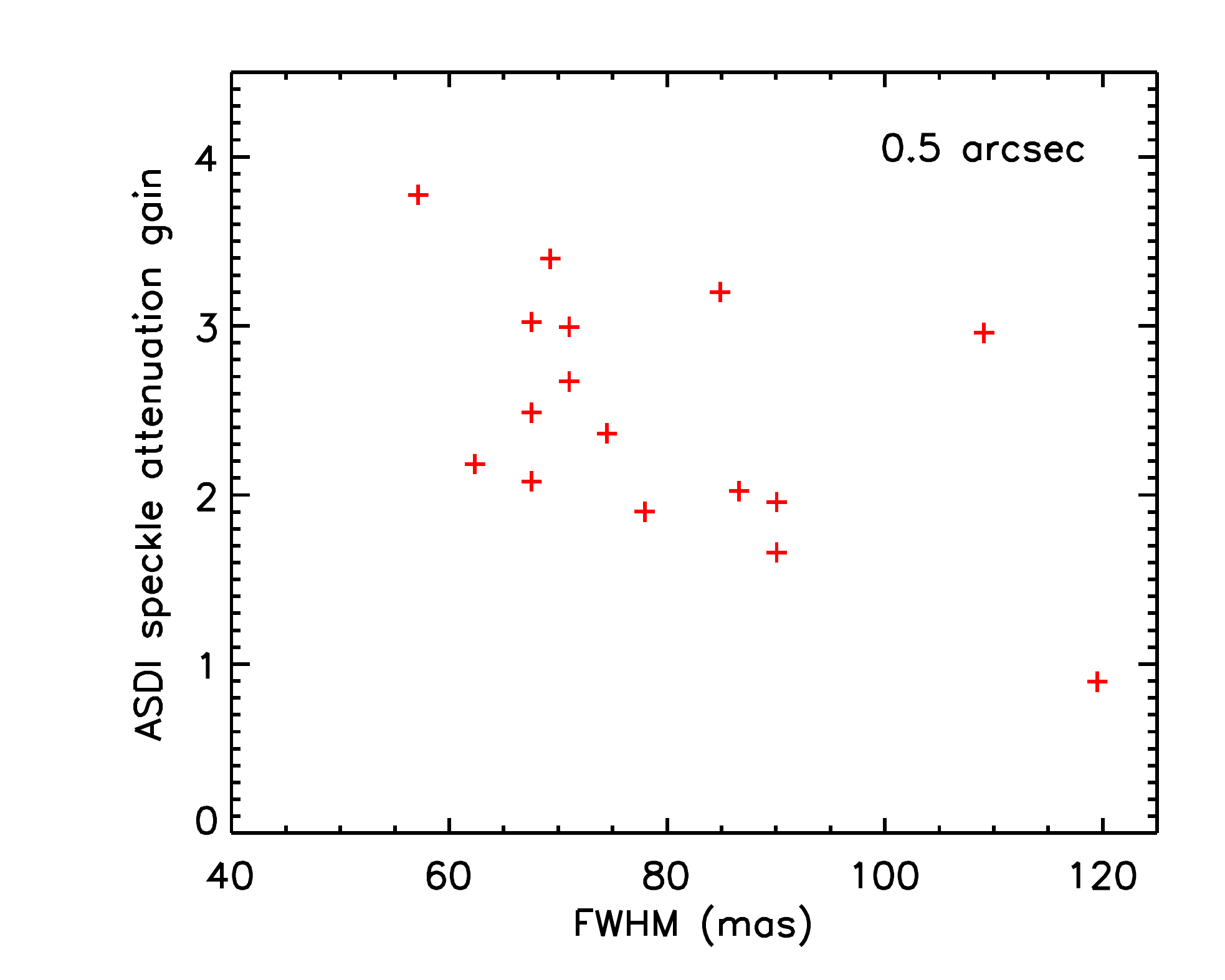}
	\includegraphics[trim = 11.5mm 5mm 5mm 10mm, clip, width=.4\textwidth]{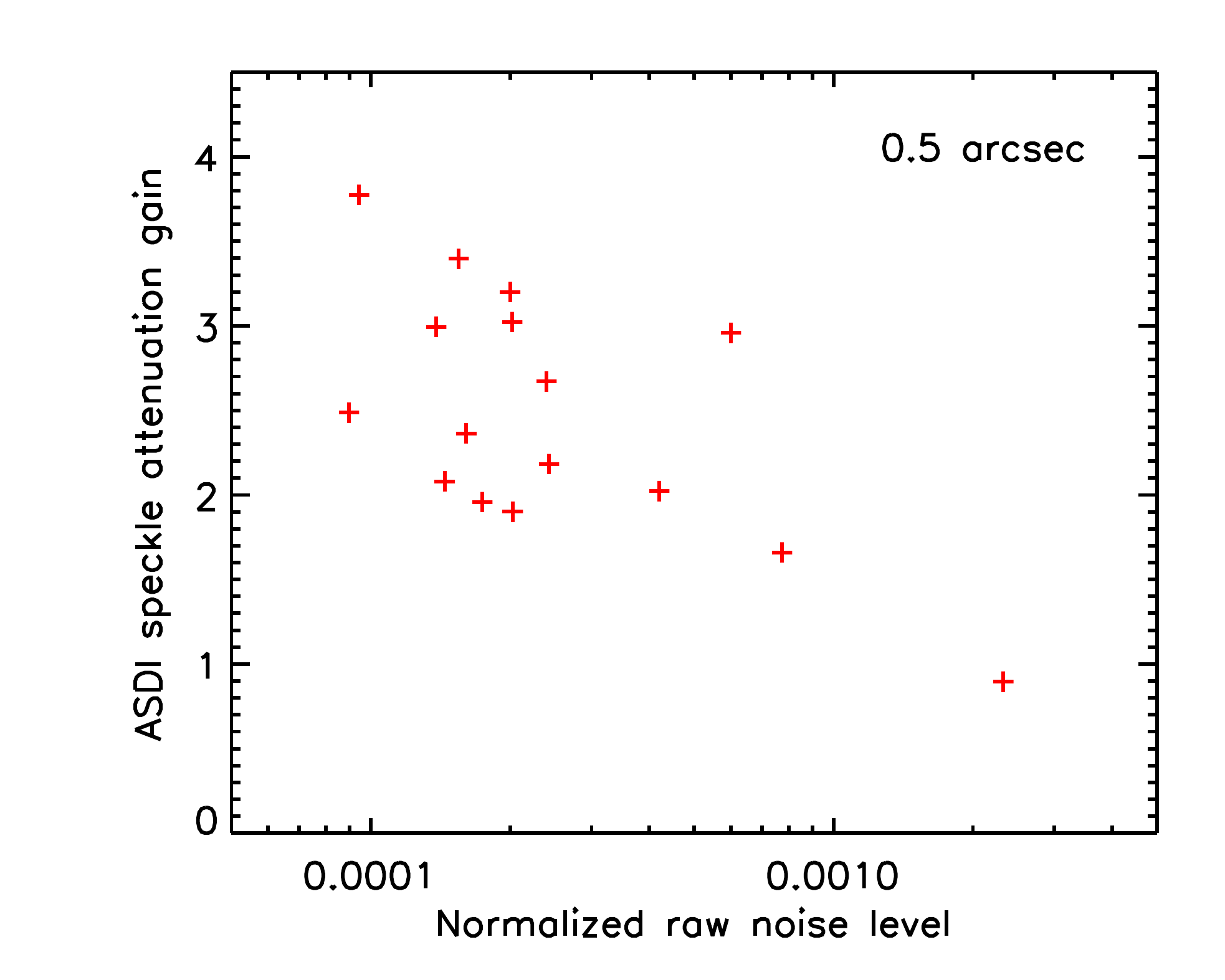}
	\caption{ASDI speckle attenuation factor {at 0.5$''$} as a function of the PSF FWHM (\textit{left}) and the noise level of the raw images (\textit{right}).}
	\label{fig:corrnoiseatten}
\end{figure*}

\begin{figure*}[!t]
	\centering
	\includegraphics[trim = 11.5mm 5mm 5mm 10mm, clip, width=.4\textwidth]{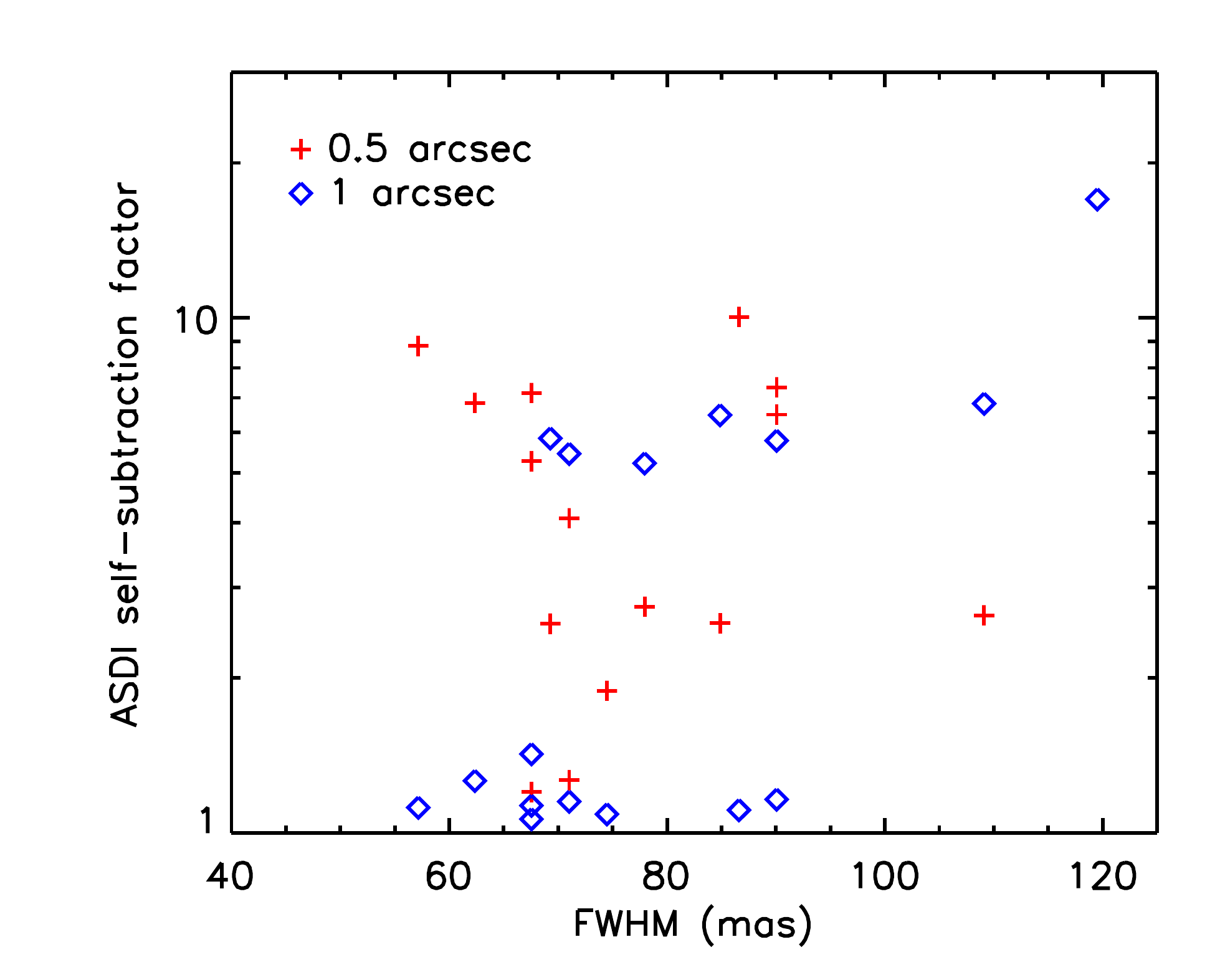}
	\includegraphics[trim = 11.5mm 5mm 5mm 10mm, clip, width=.4\textwidth]{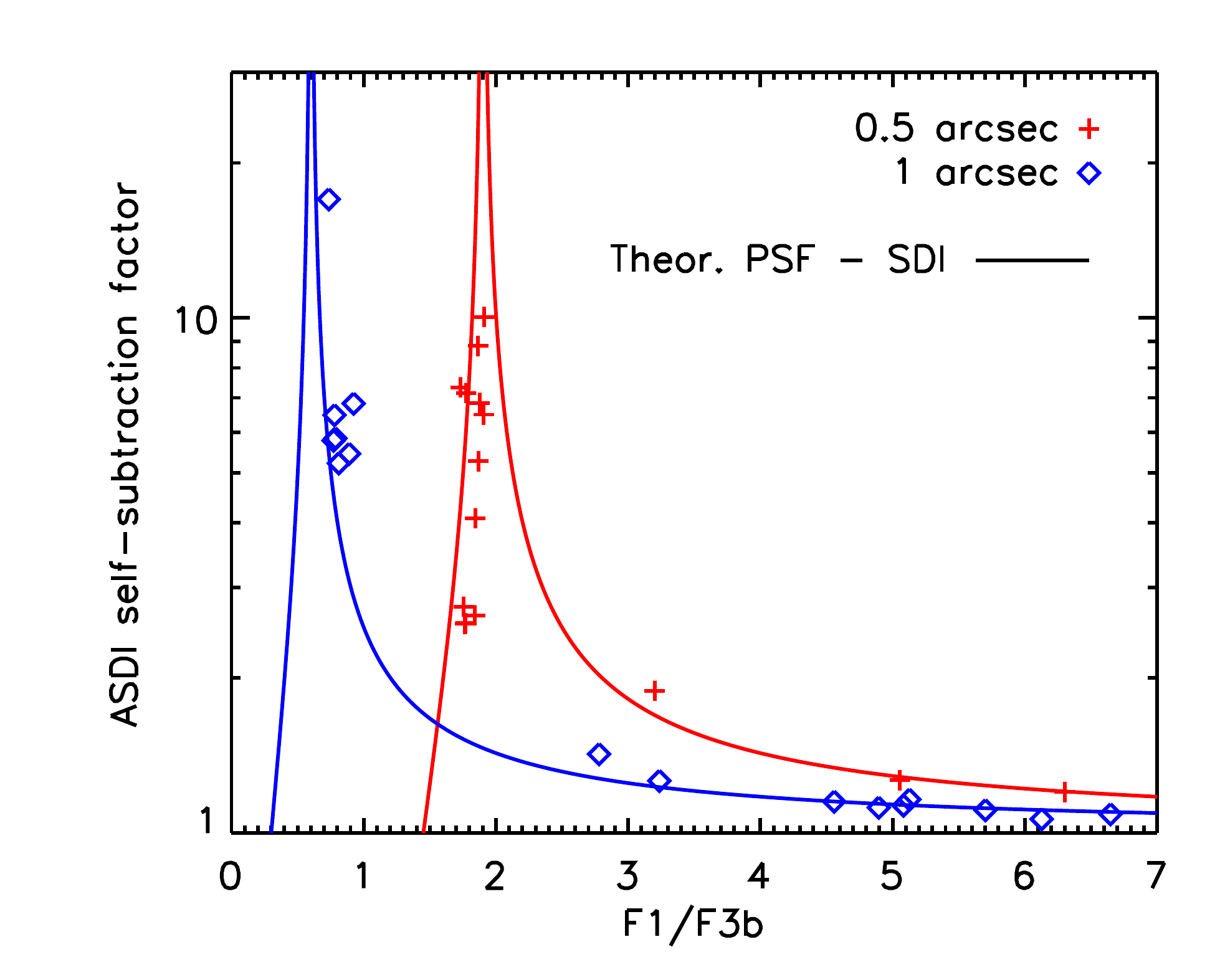}
	\caption{ASDI self-subtraction factor as a function of the PSF FWHM (\textit{left}) and the planet flux ratio (\textit{right}) derived for two separations to the image center. For the right panel, we also represent the theoretical curves for the self-subtraction measured on a diffraction-limited PSF processed in SDI alone (no ADI) as a function of the flux ratio (solid lines). The measurements and the theoretical curve determined at 0.5$''$ are shifted horizontally by 1 unit towards the right with respect to the data obtained at 1$''$ for the sake of clarity. Also note the point representing HIP~106231 at 0.5$''$, which is outside the plot ranges (FWHM\,=\,119~mas, \fluxrationospace\,=\,0.9, ASDI self-subtraction\,=\,100)}
	\label{fig:corrselfsub}
\end{figure*}

{This appendix completes Sect.~\ref{sec:analysesdi} by analyzing possible correlations of the speckle attenuation and self-subtraction with quantities representative of the observations and/or of the spectral properties of point sources.}

\subsection{Properties of the ASDI speckle attenuation}
\label{sec:corrnoiseatten}

{We focus here on possible trends of the ASDI speckle attenuation with observational factors. We test six quantities: the PSF FWHM, the Strehl ratio, the raw noise level (without differential imaging), the coherent energy, the turbulence correlation time, and the seeing. The first three factors are derived directly from the images. {The next two parameters are the median of the values estimated by the {visible} AO wavefront sensor (at a wavelength of 2.17~$\muup$m and 0.55~$\muup$m, respectively). The last quantity is derived by the DIMM at 0.5~$\muup$m. The coherent energy and the Strehl ratio are two methods for determining the quality of the AO correction \citep{Fusco2004}. The coherent energy of the NaCo-corrected images is assessed using the slope measurements of the Shack-Hartmann wavefront sensor. We derive the Strehl ratio as the ratio of the maxima of the measured and theoretical normalized PSF}. We find general tendencies with the FWHM and the raw noise level at 0.5$''$ {in Fig.~\ref{fig:corrnoiseatten}, where the attenuations are greater} for lower values of these parameters. We retrieve a general decreasing trend with the Strehl ratio. We observe the same trends (inverted for the Strehl ratio) with these three factors{. This is expected, because these factors} are related. We note in Fig.~\ref{fig:corrnoiseatten} that the vertical dispersion of the measurements is large for both panels. This could be accounted for by the diversity of the observing conditions (Table~\ref{tab:obs}). At larger separations, we retrieve similar correlations with the FWHM and the raw noise level, but they are less pronounced because the speckle noise is less and less dominant (results not shown). We do not determine correlations with the parameters estimated by the AO system.} {We discuss the errors on the speckle attenuation and FWHM in Sect.~\ref{sec:analysesdi}. The error on the normalized raw noise level is the same as the error on the PSF considered in the manuscript (20\%).}

\subsection{Dependencies of the ASDI self-subtraction}
\label{sec:corrselfsub}

This quantity depends on both observational and spectral factors (Eq.~(\ref{eq:fsdi})). We use the FWHM as the parameter that is representative of the observations and the companion flux ratio \fluxratio (Fig.~\ref{fig:rfluxteff}) as spectral factor. The \fluxratio ratio is calculated by interpolating the model relation on the mass measured at 5~$\sigma$. Because the self-subtraction is a function of the angular separation, we present the results for two separations in Fig.~\ref{fig:corrselfsub}. For the FWHM (left panel), we do not see any trends if we consider all the points for each separation.

In contrast, we find that the self-subtraction is strongly correlated with \fluxrationospace. For \fluxrationospace\,$\gtrsim$\,2 (T$_{\rm{eff}}$\,$\lesssim$\,800~K, Fig.~\ref{fig:rfluxteff}), it is less than 2. In this panel, we also represent the theoretical curves of the self-subtraction due to SDI alone measured for diffraction-limited PSF. Although the data points include both SDI and ADI attenuations, we note that they {follow quite closely} the theoretical relations. If we focus on the ``peak'' of the latter, watching out for the horizontal shift for the curve determined at 0.5$''$, we note that the corresponding \fluxratio decreases from a value of 1 at the image center as the separation increases. This feature is accounted for well by a polynomial law of degree 3. At 0.5$''$, most measurements are in the regime \fluxrationospace\,$<$\,1, which means that the companion is brighter at 1.625~$\muup$m than at 1.575~$\muup$m. This is the case for \teffnospace\,$\gtrsim$\,1\,000~K (Fig.~\ref{fig:rfluxteff}).

The self-subtraction does not show trends with the FWHM for a separation of 1$''${. This} seems to be inconsistent with Fig.~\ref{fig:phirseeingecmeant0mean}, in which we found a correlation of the geometric part of the self-subtraction $\phi(\vec{r})$ with this parameter. However, unlike $\phi(\vec{r})$, the ASDI self-subtraction is a degenerated quantity. Thus, we have to distinguish different regimes for the self-subtraction when analyzing possible trends. When \fluxratio is high ($\gtrsim$2), the PSF morphology (FWHM, asymmetries) and chromatic behavior (altered by different spectral filters and by the coronagraph chromaticity) will have little impact on the self-subtraction. We observe in the {right-hand} panel of Fig.~\ref{fig:corrselfsub} that the measurements are close to the theoretical predictions. In contrast, when \fluxrationospace\,$\simeq$\,1, the PSF properties will influence the self-subtraction value. This is confirmed in Fig.~\ref{fig:corrselfsub}, where the discrepancies between the measurements and the theoretical case are greater. If we examine the data measured at 1$''$ further in the {left-hand} panel of Fig.~\ref{fig:corrselfsub} and analyze the points with small and large self-subtractions separately, we find that the former are not correlated with the FWHM. However, for large self-subtractions, we retrieve a behavior similar to what we obtained for $\phi(\vec{r})$ (Fig.~\ref{fig:phirseeingecmeant0mean}). {The errors on the quantities represented in Fig.~\ref{fig:corrselfsub} are described in Sect.~\ref{sec:analysesdi}.}

\end{appendix}

\end{document}